\documentclass[12pt]{iopart}

\usepackage{graphicx}
\usepackage{placeins} 
\usepackage{shuffle}

\usepackage[utf8]{inputenc}
\usepackage{amssymb}
\expandafter\let\csname equation*\endcsname\relax
\expandafter\let\csname endequation*\endcsname\relax
\usepackage{bbold}
\usepackage{amsmath}
\usepackage{slashed}
\usepackage{nicefrac}
\usepackage{braket}
\usepackage{hyperref}
\usepackage[nosort]{cite}

\usepackage{etoolbox}

\makeatletter
\def\@mkboth#1#2{}
\newlength\appendixwidth
\preto\appendix{\addtocontents{toc}{\protect\patchl@section}}
\newcommand{\patchl@section}{%
  \settowidth{\appendixwidth}{\textbf{Appendix }}%
  \addtolength{\appendixwidth}{1.5em}%
  \patchcmd{\l@section}{1.5em}{\appendixwidth}{}{\ddt}%
}
\makeatother

\usepackage{tikz}

\usetikzlibrary{calc,arrows, decorations.markings, shapes.misc, decorations.pathmorphing}

\def\boxsize{0.7}
\def\blobsize{0.55cm}
\def\bwblobsize{0.35cm}

\tikzset{
  ->-/.style={
    decoration={
      markings,
      mark=at position #1 with {\arrow{latex}}},
    postaction={decorate}
  },
  ->-/.default=0.5
}

\tikzset{
    wavy/.style={decorate, decoration={snake}, draw=red},
}

\tikzset{VO/.style={cross out, draw, 
         minimum size=5pt, 
         inner sep=0pt, outer sep=0pt}}

\tikzset{VOline/.style={decorate, decoration={snake}},
}

\tikzset{
    partial ellipse/.style args={#1:#2:#3}{
        insert path={+ (#1:#3) arc (#1:#2:#3)}
    }
}

\newcommand{\whitedot}{\node[circle, fill=white, draw, inner sep=0pt, minimum size=\bwblobsize]}
\newcommand{\greyblob}{\node[circle, fill=black!20, draw, inner sep=2pt, minimum size=\blobsize]}

\newcommand{\drawLLwhite}{\whitedot (LL) at (-\boxsize, -\boxsize) {};}

\newcommand{\drawUL}[1]{\greyblob (UL) at (-\boxsize,  \boxsize) {#1};}
\newcommand{\drawUR}[1]{\greyblob (UR) at ( \boxsize,  \boxsize) {#1};}
\newcommand{\drawLR}[1]{\greyblob (LR) at ( \boxsize, -\boxsize) {#1};}

\newcommand{\drawboxinternallines}{
  \draw (UL) -- (UR) -- (LR) -- (LL) -- (UL);
}

\newcommand{\drawregionvariables}[5]{
  \draw ( 0.0, -1.4) node {#1};
  \draw (-1.4,  0.0) node {#2};
  \draw ( 0.0,  1.4) node {#3};
  \draw ( 1.4,  0.0) node {#4};
  \draw ( 0.0,  0.0) node {#5};
}

\newcommand{\cA}{\mathcal{A}}
\newcommand{\cC}{\mathcal{C}}
\newcommand{\cM}{\mathcal{M}}
\newcommand{\cN}{\mathcal{N}}
\newcommand{\cL}{\mathcal{L}}
\newcommand{\cO}{\mathcal{O}}
\newcommand{\cS}{\mathcal{S}}
\newcommand{\eps}{\epsilon}

\newcommand{\lt}{\tilde{\lambda}}
\newcommand{\lan}{\langle}
\newcommand{\ran}{\rangle}

\newcommand \vev [1] {\langle{#1}\rangle}
\newcommand \bev [1] {[{#1}]}

\newcommand{\black}{\color{black}}

\newcommand{\da}{{\dot{\alpha}}}
\newcommand{\db}{{\dot{\beta}}}
\newcommand{\alg}[1]{\mathfrak{#1}}

\newcommand{\tla}{\tilde{\lambda}}
\newcommand{\la}{\lambda}

\newcommand{\be}{\begin{equation}}
\newcommand{\ee}{\end{equation}}

\begin{document}

\begin{flushright}
	SAGEX-22-02\\
	HU-EP-22/06 \\
	QMUL-PH-22-01
\end{flushright}

\title[Modern Fundamentals of Amplitudes]{The SAGEX Review on Scattering Amplitudes\\
Chapter 1:  Modern Fundamentals of Amplitudes }

\author{Andreas~Brandhuber$^{1}$, Jan~Plefka$^{2}$ and Gabriele~Travaglini$^{1}$}

\address{$1$ Centre for Theoretical Physics, 
						Department  of Physics and Astronomy,\\
						Queen Mary University of London, 
						London E1 4NS, 
						United Kingdom}
						\vspace{4pt}
						
						\address{$2$ Institut f\"ur Physik und IRIS Adlershof, Humboldt-Universit\"at zu Berlin, \\
  Zum Gro{\ss}en Windkanal 2, D-12489 Berlin, Germany}
\ead{a.brandhuber@qmul.ac.uk, jan.plefka@hu-berlin.de, g.travaglini@qmul.ac.uk}
\vspace{10pt}

\begin{abstract}
This chapter  introduces  the  foundational concepts and techniques  for  scattering amplitudes. It is meant to be accessible to readers with 
only a basic understanding of
quantum field theory. 
Topics covered include: the four-dimensional  spinor-helicity formalism and the colour decomposition of Yang-Mills scattering amplitudes;   the study of soft and collinear  limits of Yang-Mills and gravity amplitudes; 
the BCFW recursion relation and generalised unitarity, also in the superamplitudes formalism of $\cN{=}4$ supersymmetric Yang-Mills;  an overview  of standard and hidden symmetries  of the \mbox{$S$-matrix} of $\cN{=}4$ supersymmetric Yang-Mills, such as the conformal,   dual conformal and Yangian symmetries; and a brief  excursus on form factors of protected and non-protected operators in Yang-Mills theory. Several examples and  explicit calculations are also provided. 

\end{abstract}

%
%
%
%
%


\newpage

\tableofcontents

\newpage


\setcounter{footnote}{0}


\section{Introduction}

The  most remarkable property of scattering amplitudes is their unexpected simplicity. Consider for example the scattering  of $2{\to}n$ gluons at tree level. Textbooks usually discuss the case  $n{=}2$, which requires the computation of four Feynman diagrams. They often fail to mention that, as $n$ grows, life is not so simple, as this table  shows:
\begin{table}[!ht]
\begin{center}
\begin{tabular}{||c||c|c|c|c|c|c|c|c||} \hline
$n$ & $2$ & $3$ & $4$ & $5$ & $6$ & $7$ & $8$ & $9$  
\\ \hline
$\#$ of diagrams & 
$4$ & $25$ & $220$ & $2,485$ & $34,300$ & $559,405$ & $10,525,900$ & $224,449,225$
\\
\hline
\end{tabular}
\caption{{\it The number of Feynman diagrams that contribute to  $2{\to} n$ gluon scattering at tree level \cite{Mangano:1990by}. 
This number grows factorially with $n$.}}
\label{table}
\end{center}
\end{table}
\FloatBarrier
\noindent
If  the result of a calculation as $n$ increases were to   grow in complexity in the way  the table above suggests, there would be no surprise. This is not the case. Indeed, there are  families of amplitudes for which  all-multiplicity expressions are available. The  most famous  one is the infinite sequence of  Maximally Helicity Violating gluon amplitudes, or MHV in short,  where all gluons have the same helicity except two, say $i$ and~$j$ (in a convention where the momenta of the $n$ particles  are all outgoing). For any $n$, these amplitudes  are expressed by the spectacularly beautiful Parke-Taylor formula 
 \cite{Parke:1986gb,Mangano:1987xk} 
\begin{align}
\label{one}
A_{n}^{\text{MHV}}(1^+, \ldots , i^-, \ldots j^- , \ldots ,  n^+) = i g^{n-2} \frac{ \vev{i j }^4}{\vev{12}\vev{23} \cdots \vev{n1}} \, . 
\end{align} 
One does not  need to understand the meaning of the symbols  in  \eqref{one} (which will be explained later) to appreciate that Feynman diagrams fail to account for  its simplicity, which is effectively independent of the number of gluons. In a landmark paper \cite{Witten:2003nn}, Witten related the simplicity of \eqref{one} to the fact that when transformed to Penrose's twistor space \cite{Penrose:1967wn,Penrose:1972ia}, MHV amplitudes have  support on  
the simplest curve in twistor space -- a (complex)~line.  
This result  led to  remarkable closed formulae for the tree-level $S$-matrix of $\cN{=}4$ super Yang-Mills (SYM) \cite{Roiban:2004vt, Berkovits:2004hg,Roiban:2004yf}, and a novel diagrammatic approach that uses MHV amplitudes as effective vertices \cite{Cachazo:2004kj}.

Two tasks are then ahead. The first  is to provide a framework, or choose coordinates,  that makes this simplicity manifest;  this is similar to picking polar coordinates to describe circular motion.  The second is to devise methods and find symmetries which can explain this simplicity, at the same time providing new,   powerful ways to  calculate amplitudes while avoiding Feynman diagrams. This chapter provides the beginning of  an answer to both tasks. 

The most economic language to describe the scattering of massless particles is  the spinor-helicity formalism, which we introduce in Section~\ref{sec:2}. It provides a parameterisation of the  momenta  and  polarisations of massless particles in terms of a set of variables which automatically satisfy the on-shell condition $p^2{=}0$ for lightlike momenta.  Section~\ref{sec:3} introduces colour decomposition, which leads to the concept of colour-ordered, or partial amplitudes in Yang-Mills theory -- quantities which depend only on kinematic data but not on colour, which will be one of the main  subjects of the rest of this article. With the aim of deriving amplitudes without ever looking at a Lagrangian, we discuss in Section~\ref{sec:4} the possible forms of the smallest scattering amplitudes of particle of spin~$s$, showing that they can be derived  from symmetry principles alone. Starting from these building blocks, in Section~\ref{sec:5} we introduce the BCFW recursion relation, one of the most efficient methods to derive the tree-level $S$-matrix of Yang-Mills theory and gravity. 
In Section~\ref{sec:6} we pause and consider the basic symmetry of scattering amplitudes --  the Poincar\'{e} group  (translations plus Lorentz)  -- and  the conformal group, which is an invariance of tree-level Yang-Mills amplitudes. 
Amplitudes are  singular in soft and collinear limits, with  a universal behaviour which is often very useful to constrain their form. The corresponding factorisation theorems are derived at tree level in Yang-Mills and gravity theories in Section~\ref{sec:7}, using a combination of  MHV diagrams  and recursion relations. Section~\ref{sec:8} introduces supersymmetry  and superamplitudes -- objects with package together amplitudes with a fixed total helicity, and  are invariant under supersymmetry transformations. Here we focus on maximally supersymmetric Yang-Mills theory, and  formulate supersymmetric BCFW recursion relations, also deriving MHV superamplitudes as an example. 
It has often been said that the scattering amplitudes in $\cN{=}4$ SYM are the ``hydrogen atom'' of four-dimensional relativistic scattering (see e.g.~\cite{Lance-talk}). 
This is due to the fact that they are very constrained: 
superamplitudes in $\cN{=}4$ SYM  enjoy the superconformal symmetry of the Lagrangian of the theory, as well as  certain  hidden symmetries of its $S$-matrix: the dual superconformal and Yangian symmetries.  
We review these  in Section~\ref{sec:9}, again focusing on the MHV superamplitude as a simple example. In Section~\ref{sec:10} we introduce the modern unitarity-based approach to compute loop amplitudes in theories with and without supersymmetry. In particular we review the computation of MHV (super)amplitudes both from two-particle and quadruple cuts, and of the all-plus four-point amplitude at one loop in pure Yang-Mills. Finally, Section~\ref{sec:11} serves as a taster of recent applications of on-shell techniques devised for amplitudes to form factors. These  are slightly off-shell quantities,  falling  in between amplitudes (fully on shell) and correlation functions (fully off shell). In \mbox{\ref{app:1}} we outline our conventions and the Lorentz transformation properties of the spinor variables introduced in Section~\ref{sec:2}.

\section{Spinor-helicity formalism} 
\label{sec:2}

\subsection{Massless particles and their helicity}
Elementary particles carry an internal angular momentum known as spin $\vec{S}$. The projection of the particle's spin
on the direction of motion is known as its {\it helicity} 
$h:= \dfrac{\vec{p}\cdot \vec{S}}{|\vec{p}\,|}$, where $\vec{p}$ denotes
the particle three-momentum. If the particle is massless, the helicity is a Lorentz-invariant quantity%
\footnote{This can be understood as follows: 
for a massive particle a Lorentz boost can be used to go to a frame in which the helicity is flipped,
however no boost can ``overtake'' a massless particle, which  moves at the speed of light.}.
Moreover, for massless particles of spin $s$ the helicity can only take the extremal values  $h{=}\pm s$. 
Scattering states of massless particles are therefore labeled by the on-shell momentum and helicity: $|p,h\rangle$.
Let us now take a look at the cases of spin $s=1/2$, $1$ and  $2$.

{\bfseries{$s=\nicefrac{1}{2}$}}. The momentum-space Dirac equation for positive- and    negative-energy solutions,    $u(p)$ and $v(p)$,   reads
\begin{equation}
(\slashed{p} -m )u(p)=0\, , \qquad    (\slashed{p} +m )v(p)=0\, .
\end{equation}
Clearly, they coincide in the massless case $\slashed{p} u=0=\slashed{p}v$. States of definite helicity are obtained via the projectors $\frac{1}{2}(1\pm \gamma_5)$,
\begin{equation}
    u_\pm=\frac{1}{2}(1\pm \gamma_5)u(p)\, , \qquad
      v_\mp=\frac{1}{2}(1\pm \gamma_5)v(p)\, , \qquad
\end{equation}
and in the massless case one can identify $u_\pm(p){=}v_\mp(p)$. Hence spin $\nicefrac{1}{2}$
states are labeled by $|p,\pm \nicefrac{1}{2}\rangle$.

{\bfseries{$s=1$}}. Gauge fields carry helicities $h=\pm 1$ described by polarisation vectors
$\epsilon_\mu^{(\pm)}(p)$ that obey the transversality condition 
\begin{align}
\label{polorth}
    p\cdot \epsilon^{(\pm)}(p)=0\, , 
\end{align}
as well as the relations
\begin{equation}\label{poldef}
    \epsilon^{(\pm)}(p)\cdot \epsilon^{(\pm)}(p)=0\, , 
    \quad  \epsilon^{(+)}(p)\cdot  \epsilon^{(-)}(p)=-1\, , \quad (\epsilon^{(\pm)}_\mu(p))^\ast = (\epsilon^{(\mp)}_\mu(p))\,.
\end{equation}
The corresponding  on-shell  states are labeled as $|p,\pm 1\rangle$.

{\bfseries{$s=2$}}. Gravitons come in two helicities $h=\pm 2$. Their
symmetric polarisation tensors $\epsilon_{\mu\nu}^{(\pm\pm)}(p)$ obey
$p^\mu \epsilon_{\mu\nu}^{(\pm\pm)}(p){=}0$,  and can be chosen to be traceless: $\epsilon^{(\pm\pm)\,\mu}{}_\mu=0$.
They can be represented as \emph{direct products} of gauge field polarisation vectors:
\begin{equation}
\label{gravpoldef}
\epsilon^{(++)}_{\mu\nu}(p)=\epsilon^{(+)}_\mu(p)\epsilon^{(+)}_\nu(p)\, , \qquad
\epsilon^{(--)}_{\mu\nu}(p)=\epsilon^{(-)}_\mu(p)\epsilon^{(-)}_\nu(p)\, .
\end{equation}
This representation automatically entails the above on-shell properties.

\subsection{Momenta and polarisations of massless particles}
\label{sec:2.2}

The key property of the spinor-helicity formalism is to provide a  representation  of   momenta and polarisations using one set of variables that automatically 
obey the on-shell constraint $p^2{=}0$ as well as the conditions on the polarisations, e.g.~$\slashed{p}u_\pm{=}0$ for spin-$\nicefrac{1}{2}$ particles, or \eqref{polorth} for gluons.
These variables ultimately lead to simpler final expression for the amplitudes of fermions, gluons,
photons and gravitons. The starting point is to 
rewrite $p^\mu$ as a Weyl bi-spinor:
\begin{equation}
\label{isomorphism}
    p^\mu \to p^{\dot\alpha\alpha} = \bar{\sigma}_\mu^{\dot\alpha\alpha} \, p^\mu 
    =\left ( \begin{matrix} p^0-p^3 & -p^1 +ip^2\\ -p^1-ip^2& p^0+p^3
    \end{matrix} \right )\, ,
\end{equation}
where $\bar{\sigma}_\mu^{\dot\alpha\alpha} {=}(\mathbb{1},-\vec{\sigma})$ and 
$\vec{\sigma}$ are  the Pauli matrices. This relation implements the isomorphism between the Lorentz group $SO(3,1)$ and $SL(2, \mathbb{C})$, as discussed  in \ref{app:1}.
The crucial observation is now that the on-shell condition for a massless particle $p^2{=}0$ is equivalent to
$
\text{det}\, p\!=\!0\, 
$,
and  the rank of the matrix $p^{\da\alpha}$ is thus equal to one.
Hence,  the four-momentum
can  be written~as
\be
\label{fundamental!}
p^{\da\alpha}=  \tla^{\da}\, \la^{\alpha}\, .
\ee
This is one of the most important formulae in this article.  $\la^{\alpha}$ and $\tla^{\da}$ are commuting Weyl spinors, known as  \emph{helicity spinors} \cite{Xu:1986xb,
    Gunion:1985vca,Kleiss:1985yh} (see \cite{Bjorken:1966kh,Henry:1967jm,DeCausmaecker:1981wzb,DeCausmaecker:1981jtq,Berends:1981uq,Berends:1983ez,Berends:1983ey,Berends:1984qe,Berends:1984qf} for a precursor formalism). 
For complexified momenta,  $\la$ and $\tla$ are independent
variables, and
importantly \eqref{fundamental!} is invariant under a {\it little-group} transformation 
\begin{align}
    \label{littlegroup}
\lambda \to z \lambda\, , \qquad
\lt \to z^{-1} \lt\, , \qquad  z\in \mathbb{C}^\ast\ . 
\end{align}
On the other hand
in real Minkowski space the four-momentum is real, which  translates into the condition 
$(\la^{\alpha})^{\ast}{=}\pm \tla^{\da}$, where the  sign is the same as that  of the 
energy $p^0$. That also reduces the little group to a $U(1)$ (since $|z|=1$ in this case), as expected for massless particles
\cite{wigner1, wigner2}.  
For real momenta, an explicit realisation of the spinors~is
\begin{align}
\la^{\alpha}&= \frac{1}{\sqrt{p^{0}-p^{3}}}\, \begin{pmatrix} p^{0}-p^{3}\cr -p^{1}-ip^{2}\cr
\end{pmatrix}\, , \qquad
\tla^{\da}= \frac{1}{\sqrt{p^{0}-p^{3}}}\, \begin{pmatrix} p^{0}-p^{3}\cr -p^{1}+ip^{2}\cr
\end{pmatrix}\, .
\end{align}
Since 
$|p^{0}|\geq|p^{3}|$, the quantity  $\sqrt{p^{0}+p^{3}}$ is real (imaginary) for positive (negative) $p^{0}$.

Spinor indices are  raised or lowered with the  Levi-Civita tensor:
\be
\la_{\alpha}:=\epsilon_{\alpha\beta}\, \la^{\beta}\, , \qquad
\tla_{\da}:=\epsilon_{\da\db}\, \tla^{\db}\, ,
\ee
which allows us to form  two basic Lorentz-invariant quantities
\begin{align}
\label{bracket-conv}
    \lan ij\ran := \lambda^\alpha_i\lambda_{j \alpha}\, , \qquad 
[ij] := \lt_{i \dot{\alpha}} \lt_j^{\dot{\alpha}}\, , 
\end{align}
introducing the NW-SE (SW-NE) contractions for the undotted and dotted Weyl indices and the  handy bracket notation (see  \ref{app:1} for a discussion of the 
Lorentz transformation properties of spinors). Here $i$ and $j$ denote the particles' labels.
 We can then write the product of two momenta $p_i$ and $p_j$ as
\begin{equation}
   \epsilon_{\alpha\beta}\epsilon_{\dot\alpha\dot\beta}\,  p_i^{\dot\alpha\alpha}\, p_j^{\dot\beta\beta} 
    =\epsilon_{\alpha\beta}\epsilon_{\dot\alpha\dot\beta} \, \bar{\sigma}^{\dot\alpha\alpha}_\mu\, \bar{\sigma}^{\dot\beta\beta}_\nu \, p_i^\mu\, p_j^\nu=2\, p_i\cdot p_j\, , 
\end{equation}
where $\epsilon_{\alpha\beta}\epsilon_{\dot\alpha\dot\beta} \, \bar{\sigma}^{\dot\alpha\alpha}_\mu\, \bar{\sigma}^{\dot\beta\beta}_\nu  {=} {\rm Tr} (\bar{\sigma}_\mu \sigma_\nu) {=} 2 \eta_{\mu \nu}$, and we have defined  $\sigma_{\mu\, \alpha\dot\alpha}{:=}\epsilon_{\alpha\beta} \epsilon_{\dot\alpha\dot\beta}
\bar{\sigma}_\mu^{\dot\beta\beta}{=}(\mathbb{1},\vec\sigma) $.
 Mandelstam invariants also have a very simple representation in spinor variables:
\begin{equation}
    s_{ij}= (p_i+p_j)^2= 2\, p_i\cdot p_j = \lan ij \ran [ji]\, .
\end{equation}
We have seen that  spinor-helicity variables are useful to describe the momenta of massless on-shell particles
as the mass-shell condition is automatically met, but what is their relation to the helicity of the
on-shell states? One quickly sees that they solve the massless Dirac equation and can
be identified with the helicity states $u_\pm(p)$ and $v_\pm(p)$. Indeed, using the chiral representation of  the Dirac matrices $\gamma^\mu$, one has
\begin{align}
\label{pslash}
\slashed{p}  = p_{\mu}\, \gamma^{\mu} &=    
\begin{pmatrix} 
      0 & p_{\alpha\da} \\
      p^{\db\beta} & 0 \\
   \end{pmatrix}
   =
    \begin{pmatrix} 
      0 & \la_{\alpha}\tla_{\da} \\
      \tla^{\db}\la^\beta & 0 \\
   \end{pmatrix}\, .
\end{align}
Now writing 
\be
u_{+}(p)= v_{-}(p)=    
\begin{pmatrix} 
      \la_{\alpha} \\
      0 \\
   \end{pmatrix}
   := |p\rangle\, ,\qquad
u_{-}(p)= v_{+}(p)=    
\begin{pmatrix} 
      0 \\
      \tla^{\da} \\
   \end{pmatrix}
   := |p]\, ,
\ee
using the convenient bra-ket notation $|\, \bullet\, \rangle$ and $|\, \bullet\, ]$, we see that the massless
Dirac equation  $\slashed{p} |p\rangle {=} \slashed{p}\, |p]{=} 0$ is satisfied since 
$\lan\la\, \la\ran{=}[\tla\, \tla]{=}0$. 
Hence, the helicity states of massless spin-$\nicefrac{1}{2}$ fermions are captured by $\la$ and $\tla$.
For negative momenta, we  will  define 
\begin{align}
\label{analytic-cont}
    |-p\ran := i\, |p\ran\, , \qquad |-p] := i\, |p]\, .
\end{align}
Moving on to massless spin-$1$ states,  we can re-express the polarisation vectors $\epsilon_\mu^{(\pm)}(p)$ as bi-spinors via $\epsilon^{(\pm)}_{\alpha\dot\alpha}=\sigma_{\alpha\dot\alpha}^\mu \epsilon^{(\pm)}_{\mu} $, with 
\begin{align}
\label{epsmp}
    \eps^{(-)}_{\alpha \dot{\alpha}} =\sqrt{2} \frac{\lambda_\alpha  \tilde{\xi}_{\dot{\alpha}}}{[\lambda \xi]}\, , \qquad \eps^{(+)}_{\alpha \dot{\alpha}} =\sqrt{2} \frac{\xi_\alpha  \lt_{\dot{\alpha}} }{\lan \xi \lambda \ran}\, .
\end{align}
Here  $\xi$ and $\tilde\xi$ are arbitrary reference spinors that will drop out of
any final expression for a scattering amplitude. The only condition  is that they are not parallel to 
$\la$ and $\tla$, e.g.~$\xi \neq c\, \la$. In fact the freedom in choosing a reference spinor in
the polarisation bi-spinors can be attributed to gauge transformations, since
\be
\epsilon^{(+)}_{\alpha\dot\alpha}(\xi+\delta\xi)= \epsilon^{(+)}_{\alpha\dot\alpha}(\xi)
+ p_{\alpha\da} \sqrt{2} \frac{\lan \la \, \delta\xi\ran}{\lan \la\,  \xi\ran^2}\, .
\ee
We also note the completeness relation 
$
\sum_{h=\pm}(\epsilon^{(h)})_\mu(\epsilon^{(h)})_\nu^*=-\eta_{\mu\nu}+\frac{p_{\mu}q_{\nu}+p_{\nu}q_{\mu}}{p\cdot q}
$, 
where $q_{\alpha\dot\alpha}=\xi_\alpha \tilde{\xi}_{\dot\alpha}$.
Graviton polarisations then follow from \eqref{gravpoldef} as products of the $ \eps^{(\pm)}_{\alpha \dot{\alpha}}$.

\subsection{Massive particles}

We can  also introduce on-shell variables for \emph{massive} momenta \cite{Arkani-Hamed:2017jhn}. In this case the on-shell condition $\text{det}\, p{=}p^2{ =} m^2$ implies that $p_{\alpha\dot\alpha}$ has rank two and can be expressed in terms of a pair of spinor variables $\la^I$ and $\tla_I$ with $I=1,2$.
The bi-spinor representation of a four-dimensional massive momentum then becomes
\begin{align}
\label{mastermassive}
p_\mu\sigma^\mu_{\alpha\da}&:=p_{\alpha\da}= \lambda_{\alpha}^I\tilde{\lambda}_{\da \, I} \, .
\end{align}
Also note that the on-shell condition 
becomes 
\begin{align}
  \text{det} \, p\,  = \text{det} \lambda \times 
\text{det} \tilde\lambda 
= m^2 \, .
\end{align}
For real momenta, \eqref{mastermassive} is invariant under $SU(2)$ transformations $L$ acting on the $I$~indices: 
$\lambda^I {\to}  \lambda^J {L_J}^I$, $\tilde\lambda_I {\to} {(L^{-1})_I}^J \tilde\lambda_J$,
which are naturally  identified with the little group transformations of massive particles 
\cite{wigner1, wigner2}.


\subsection{Useful formulae}
We close this section with two useful  formulae for our helicity spinors.
The first is the Schouten identity
$
\vev{\la_{1}\,\la_{2}}\, \la_{3}^{\alpha} + \vev{\la_{2}\,\la_{3}}\, \la_{1}^{\alpha} + 
\vev{\la_{3}\,\la_{1}}\, \la_{2}^{\alpha} = 0$, 
or, contracting with an arbitrary spinor $\lambda_a$,  
\begin{align}
\vev{12}\, \vev{3a} + \vev{23}\, \vev{1a} + \vev{31}\, \vev{2a} =0\ , 
\end{align}
and similarly for the conjugate spinors.
It  reflects the fact that there is no completely anti-symmetric three-tensor $\Omega^{\alpha\beta\gamma}$.
A second identity descends from   momentum conservation $\sum_{i=1}^{n} \la_{i}^{\alpha}\, \tla_{i}^{\da} =0 $, from which it follows that 
$
\sum_{i=1}^{n}\, \vev{a\, i }\, [i\, b] =0
$,
for arbitrary $\la_{a}$ and $\tla_{b}$. Finally we quote the two useful relations
    \begin{align}
\begin{split}
\lan ab\ran[bc]\lan cd\ran[da]  \\
[ab]\lan bc\ran[cd]\lan da\ran  
\end{split}
\ \bigg\}
\!=\! {\rm Tr} \Big( \frac{1\mp\gamma^5}{2}\slashed{a}\slashed{b} \slashed{c}\slashed{d}\Big)  =2\Big[  (a{\cdot}b)  (c{\cdot}d) -  (a{\cdot}c)(b{\cdot}d) + (a{\cdot}d)(b{\cdot}c) \mp i \epsilon (abcd)\Big],
    \end{align}
   where $\eps(abcd) := \epsilon_{\mu \nu \rho \sigma}a^\mu b^\nu c^\rho d^\sigma$.

\section{Colour decomposition}
\label{sec:3}

We now turn our attention to gauge field theories. Having introduced helicity spinors as efficient variables to describe the kinematics, we now introduce a formalism that allows to disentangle the colour degrees of freedom from the
kinematic ones. There are two such formalisms for an efficient colour management: the trace-based and the
structure constant based (or DDM) formalism. In $SU(N_c)$ gauge theories coupled to matter, one mostly 
encounters two representations of
the gauge group: 
\begin{itemize}
    \item {Adjoint representation:} gluons $A^a_\mu$ and their superpartners (gluinos and scalars) carry
    adjoint indices $a-1,\ldots,N^2_c-1$.
    \item {Fundamental \& anti-fundamental representation:} quarks and anti-quarks carry (anti)-fundamental indices
    $i=1,\ldots, N_c$ and $\bar i=1,\ldots , N_c$.
\end{itemize}
The $SU(N_c)$ algebra is represented by  fundamental generators $(T^a)_i{}^{\bar{j}}$ which are $N_c\!\times\!N_c$ hermitian, traceless
 matrices. In our conventions the structure constants take the form 
\begin{equation}
\label{fasTr}
    f^{abc}=-\frac{i}{\sqrt{2}} \, \Tr (T^a [T^b, T^c])\, , 
\end{equation}
or $[T^a, T^b] {=} i\sqrt{2} f^{abc} T^c$, with  $\Tr(T^a T^b){=}\delta^{ab}$. 
Moreover, the $SU(N_c)$ Fierz-type~identity
\begin{equation}
\label{SUNFierz}
 (T^a)_{i_1}^{\bar{j}_1}\,  (T^a)_{i_2}^{\bar{j}_2}= \delta_{i_1}^{\bar{j}_2}\,
\delta_{i_2}^{\bar{j}_1} - \frac{1}{N_c}\delta_{i_1}^{\bar{j}_1}\,
\delta_{i_2}^{\bar{j}_2} \, ,
\end{equation}
is important for the colour decomposition of
amplitudes and
can be understood as a completeness relation for a basis of Hermitian matrices spanned
by $\{\mathbb{1}, T^a\}$.

\subsection{Trace basis}

\begin{figure}[tt]
\begin{center}
\scalebox{0.8}{\includegraphics{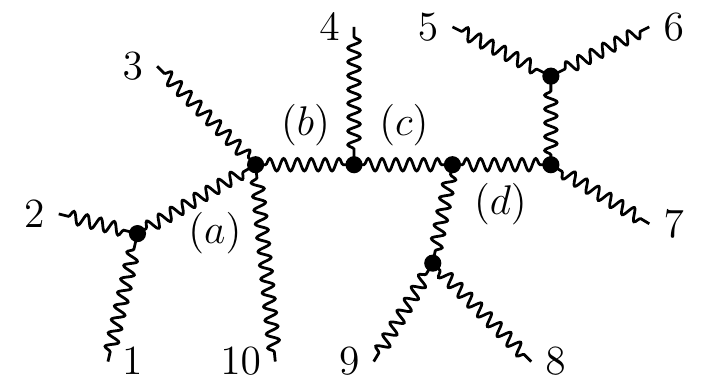}}
\begin{align*}
&(a):  (p_{1}+p_{2})^{2}\to 0\, , \qquad \qquad \qquad \qquad  (b):  (p_{10}+ p_{1}+p_{2}+p_{3})^{2}\to 0\, \\
&(c):  (p_{5}+p_{6}+p_{7}+p_{8}+p_{9})^{2}\to 0\, , \qquad (d):  (p_{5}+p_{6}+p_{7})^{2}\to 0
\end{align*}
\end{center}
\caption{\it 
Possible poles in a colour-ordered Feynman diagram. 
}
\label{Fig:regionmomenta}
\end{figure}

The colour dependence of a given Feynman graph arises from its vertices. The three-gluon vertex carries one structure constant $f^{abc}$, the four-gluon interaction a product of two $f^{abc}$, while the gluon-quark-anti-quark
interaction comes with a generator $(T^a)_i{}^{\bar j}$. 
In order to work out the colour dependence of a given Feynman diagram,  imagine replacing all structure constants appearing in it  by the trace formula \eqref{fasTr}. 
This transforms the expression to products
of $T^{a}$ generators with contracted and open indices. Open fundamental indices correspond to quark lines in the diagram, open adjoint indices to the external gluon states.
Contracted adjoint indices can be used to merge traces and products of generators 
by  repeatedly applying the Fierz-type identity~\eqref{SUNFierz}:
\begin{align}
(A\, T^{a} \, B)_{i}^{\bar j}\, 
(C\, T^{a} \, D)_{k}^{\bar l}
= (A \, D)_{i}^{\bar l}\, (C\, B)_{k}^{\bar j}
-\frac{1}{N_{c}} (A\, B)_{i}^{\bar j}\, 
(C\, D)_{k}^{\bar l} \, .
\end{align}
In  the end we will arrive at an expression of traces and strings 
of $T^a$'s with only open indices corresponding to external states of the form
\begin{align}
\Tr(T^{a_{1}}\cdots T^{a_{n}})\ldots
\Tr(T^{b_{1}}\cdots T^{b_{m}})\, (T^{c_{1}}
\cdots T^{c_{p}})_{i_{1}}^{\bar j_{1}}
\ldots
 (T^{d_{1}} \cdots
T^{d_{p}})_{i_{s}}^{\bar j_{s}}\, .
\end{align}
For pure gluon amplitudes,  things are even  simpler: in pure Yang-Mills theory the interaction
vertices of $SU(N_{c})$ and $U(N_{c})$ gauge groups are identical, as $f^{0bc}\!=\!0$ by virtue of \eqref{fasTr}
where 
$T^{0}= \frac{\mathbb{1}}{\sqrt{N_{c}}}$ is the $U(1)$ generator (this leads to the photon decoupling
theorem discussed in Section~\ref{ampprop}). Hence, the $\frac{1}{N_{c}}$ part of \eqref{SUNFierz} is not active here.
In conclusion, tree-level  gluon amplitudes reduce to a \emph{single}-trace structure and can
be brought into the \emph{colour}-decomposed form 
\begin{align}
\label{I.26}
\cA_{n}^{\text{tree}}(\{ a_{i}, h_{i},p_{i}\}) 
=  
\sum_{\sigma\in S_{n}/\mathbb{Z}_{n}}\, 
\Tr(T^{a_{\sigma_{1}}}\,  T^{a_{\sigma_{2}}}\cdots T^{a_{\sigma_{n}}} )\, 
A_{n}^{\text{tree}}({\sigma_{1}}, {\sigma_{2}}, 
\ldots , \sigma_{n}) \, .
\end{align}
Here $h_{i}$ denote the helicities and $a_{i}$ the adjoint colour indices of the external states, 
and we use the notation $\sigma = \{p_{\sigma},h_{\sigma}\}$.
Moreover, $S_{n}/\mathbb{Z}_{n}$ is the set of all non-cyclic permutations of $n$ elements,  which is
equivalent to $S_{n-1}$. The $A_{n}$ are called  \emph{partial} or \emph{colour-ordered} amplitudes
and carry all kinematic information that is now separated from the colour degrees of freedom.
Partial amplitudes $A_{n}$ are simpler than the full amplitudes $\cA_{n}$ as they are individually gauge invariant
and  exhibit poles only in channels of cyclically adjacent momenta
$(p_{i}+p_{i+1}+\cdots + p_{i+s})^{2}\to 0$, see 
Figure~\ref{Fig:regionmomenta}.

For tree-level gluon-quark-anti-quark amplitudes with a single quark line one has
\begin{align}
\cA_{n,q\bar q}^{\text{tree}} (\{ a_{i}, h_{i},p_{i}\}|\{i,q_{1}^{h_{q_{1}}},
\bar j,q_{2}^{h_{q_{2}}}\}) = 
 \sum_{\sigma\in S_{n-2}}   
(T^{a_{\sigma_{1}}}\, \cdots T^{a_{\sigma_{n-2}}})_{i}^{\bar j}\, 
 A_{n,q\bar q}^{\text{tree}}(\sigma_{1},
\ldots, {\sigma_{n-2}}| q_{1}^{h_{q_{1}}} , q_{2}^{h_{q_{2}}}) \, . 
\end{align}
Increasing the number of quark lines yields a more involved structure as more strings and $\frac{1}{N_{c}}$
factors appear, see \cite{Mangano:1990by,Ita:2011ar,Reuschle:2013qna} for details. 

At  loop level,  pure gluon amplitudes contain  also multi-trace  contributions  arising   from 
the merging performed  using  \eqref{SUNFierz}.   For example, at one loop  one has
\begin{align}
\begin{split}
\label{leading-colour-one-loop}
\cA_{n}^{\text{1-loop}}(\{ a_{i}, h_{i},p_{i}\}) &=  N_c\,
 \sum_{\sigma\in S_{n}/\mathbb{Z}_{n}}\,
\Tr(T^{a_{\sigma_{1}}}\, T^{a_{\sigma_{2}}}\cdots T^{a_{\sigma_{n}}} )\, 
A_{n;1}^{(1)}(\sigma_{1}, 
\ldots , \sigma_{n} ) \\
+ \sum_{i=2}^{[n/2]+1}\, \sum_{\sigma\in S_n/\mathbb{Z}_n} &
\Tr(T^{a_{\sigma_{1}}} \cdots T^{a_{\sigma_{i-1}}} )\,
\Tr(T^{a_{\sigma_{i}}}\cdots T^{a_{\sigma_{n}}})\, 
A_{n;i}^{(1)}(\sigma_{1}, \ldots , \sigma_{n}  )\, ,
\end{split}
\end{align}
where the $A_{n;1}^{(1)}$ are called the primitive (colour-ordered) amplitudes, and the 
$A_{n;c>1}^{(1)}$ are the higher primitive amplitudes. The latter can be expressed as linear
combinations of the primitive ones \cite{Bern:1990ux}. 
In the large-$N_{c}$ limit the single-trace 
contributions are enhanced. In colour-summed cross sections, which are of interest in 
applications, the contribution of the higher primitive amplitudes is suppressed by $\frac{1}{N_{c}^{2}}$.

\subsection{DDM basis}

\begin{figure}[tt]
\begin{center}
\scalebox{0.6}{\includegraphics{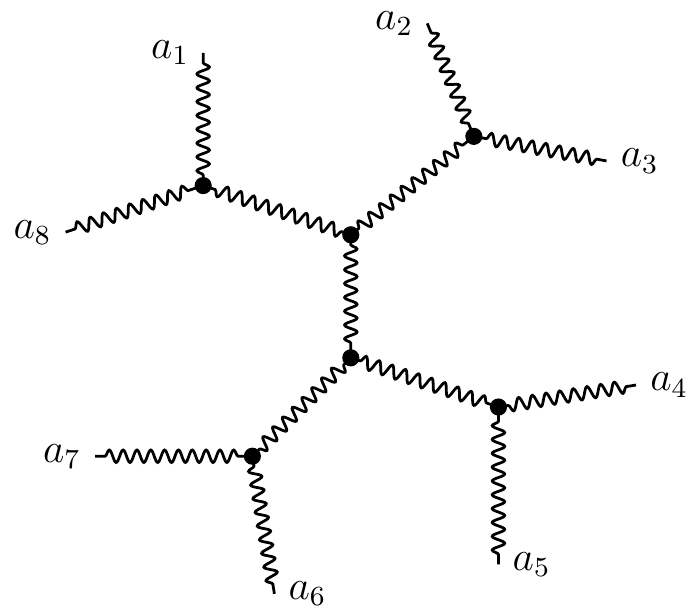}}
\end{center}
\caption{\it 
Typical colour tree in a structure constant based (DDM) colour expansion. 
}
\label{Fig:tree}
\end{figure}

An alternative basis for the colour decomposition of  pure-gluon (or purely adjoint particles) 
amplitudes makes use of the structure constants $f^{abc}$ and is due to Del Duca, Dixon and Maltoni (DDM) 
\cite{DelDuca:1999rs}. Consider the colour dependence of an $n$-gluon tree  amplitude. This can
be represented as a sum over tri-valent graphs with vertices linear in $f^{abc}$, see Figure~\ref{Fig:tree}.
In this process we artificially ``blow'' up a four-valent gluon vertex to sums of products
of tri-valent vertices by multiplying it by $1=q^{2}/q^{2}$ where $\frac{i}{q^{2}}$ is the propagator
of the ``blown up'' leg. One then uses the Jacobi identity 
\begin{align}
\begin{split}
& \quad \, \, \, \, f^{abe}\, f^{cde} =\ \ 
f^{dae}\, f^{bce}\ \,  -\ \ \  f^{dbe} \, f^{ace} \\
&\scalebox{0.7}{\includegraphics{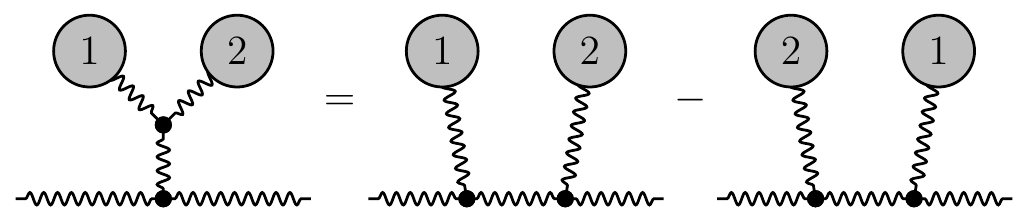}}
\end{split}
\end{align}
successively in order to shrink branched trees to branchless ones resulting in
a ``half-ladder'' expression.
In this way we can completely reduce a coloured amplitude to a half-ladder basis
in colour space:
\begin{align}
\label{DDM}
\begin{split}
\cA_{n}^{\text{tree}}(\{ a_{i}, h_{i},p_{i}\}) 
\!=\!  
\sum_{\sigma\in S_{n-2}}\, & 
f^{a_{1}a_{\sigma_{2}}e_{1}}\, f^{e_{1}a_{\sigma_{3}}e_{2}} \, f^{e_{2}a_{\sigma_{4}}e_{3}} 
\cdots f^{e_{n-3}a_{\sigma_{n-1}}a_{n}} 
A_{n}^{\text{tree}}(1, {\sigma_{2}}, 
\ldots , {\sigma_{n-1}}, n) \, ,
\end{split}
\end{align}
%
%
%
where we now sum over permutations $\sigma$ of the $n{-}2$ elements $\{2,3,\ldots, n-1\}$.
The half-ladder colour basis fixes two (arbitrary) legs, here 1 and  $n$:
$$
f^{a_{1}a_{\sigma_{2}}e_{1}}\, f^{e_{1}a_{\sigma_{3}}e_{2}} \, f^{e_{2}a_{\sigma_{4}}e_{3}} 
\cdots f^{e_{n-3}a_{\sigma_{n-1}}a_{n}}  = 
\raisebox{-0.7cm}{\scalebox{0.9}{\includegraphics{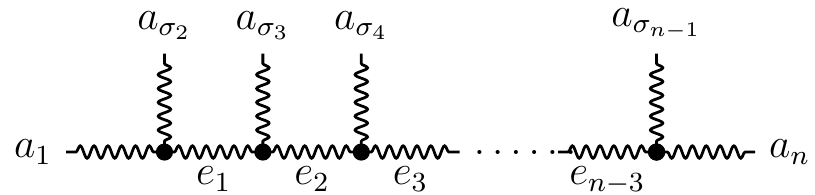}}}
$$
therefore  the DDM  basis consists of $(n-2)!$ independent partial amplitudes.
This is to be contrasted with the  $(n-1)!$ partial amplitudes in the trace basis.
Hence, there must exist non-trivial identities between  partial amplitudes
allowing one to reduce the basis accordingly. These are known as  Kleiss-Kuijf
relations \cite{Kleiss:1988ne} and take the form
\be
A_{n}^{\text{tree}}(1,\{\alpha\}, n, \{\beta\}) = (-1)^{n_{\beta}}
\sum_{\sigma\in \alpha \shuffle \beta^{T}} A_{n}^{\text{tree}}(1,\sigma, n) \, , 
\ee
where $n_{\beta}$ denotes the number of elements in the set $\beta$ and $\beta^{T}$ is the
set $\beta$ with reversed ordering. The shuffle or ordered permutation
$\displaystyle \alpha  \shuffle \beta^{T}$ means to merge $\alpha$ and $\beta^{T}$ while preserving the individual orderings of $\alpha$ and $\beta^{T}$. The Kleiss-Kuijf relations can be proven by
rewriting the DDM basis in terms of the trace basis discussed above.

It turns out that there exists yet another non-trivial identity between partial amplitudes
allowing one to further reduce the basis of primitive amplitudes
to $(n-3)!$ independent elements. This is due to the  Bern-Carrasco-Johansson relation  \cite{Bern:2008qj,Bern:2010ue},  discussed in Chapter~2 of this
review \cite{Bern:2022wqg}. It takes the schematic form
\be
A_{n}^{\text{tree}}(\sigma_{1},\ldots,\sigma_{n}) = 
\sum_{\rho\in S_{n-3}} 
K^{(\sigma)}_{\rho}
A_{n}^{\text{tree}}(1,2,{\rho_{3}},\ldots, \rho_{n-1},n)\, , 
\ee
with kinematic-dependent coefficients $K^{(\sigma)}_{\rho}$.
Finally, we note that there is also a generalisation of the DDM basis to include fundamental matter \cite{Johansson:2015oia,Melia:2015ika}.

\subsection{Colour-ordered Feynman rules}

One can establish colour-ordered Feynman rules that generate the partial (colour-ordered) amplitudes by stripping off
the colour factors from the usual Feynman rules. This is particularly easy for the gluon
and quark propagators,
\be
\raisebox{-0.3cm}{\scalebox{0.9}{\includegraphics{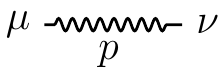}}} = -\frac{i}{p^{2}+i\varepsilon}\eta_{\mu\nu}\, ,\qquad \qquad
\raisebox{-0.3cm}{\scalebox{0.9}{\includegraphics{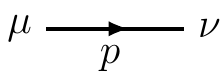}}} = \frac{i\slashed{p}}{p^{2}+i\varepsilon}\eta_{\mu\nu}\, ,
\ee
while for the vertices one finds
\begin{align}
\label{Feynrulveert}
\begin{split}
\raisebox{-0.8cm}{\scalebox{0.8}{\includegraphics{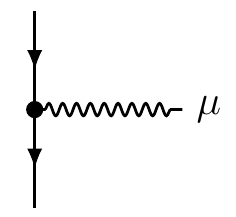}}} &= -\frac{i}{\sqrt{2}} g\, \gamma^{\mu}
\, , \\
\raisebox{-1cm}{\scalebox{0.8}{\includegraphics{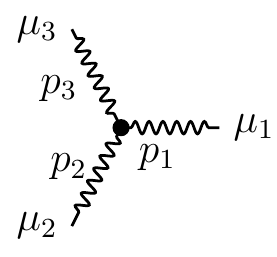}}} 
&= \frac{i}{\sqrt{2}} g \Big[ (p_1-p_2)^{\mu_3}\eta^{\mu_1\mu_2}+(p_2-p_3)^{\mu_1}\eta^{\mu_2\mu_3}+(p_3-p_1)^{\mu_2}\eta^{\mu_3\mu_1}\Big]\, ,\\
\raisebox{-0.8cm}{\scalebox{0.8}{\includegraphics{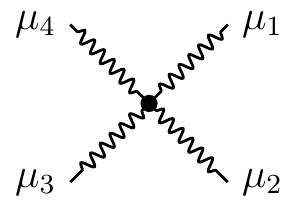}}} &= \frac{i}{2}g^{2} \Bigl [ 
2\eta_{\mu_{1}\mu_{3}}\eta_{\mu_{2}\mu_{4}} -\eta_{\mu_{1}\mu_{2}}\eta_{\mu_{3}\mu_{4}} -
\eta_{\mu_{2}\mu_{3}}\eta_{\mu_{4}\mu_{1}}\Big ]\, .
\end{split}
\end{align}

\subsection{General properties of colour-ordered amplitudes}
\label{ampprop}
Colour-ordered amplitudes are gauge invariant and obey general properties which  
reduce considerably the number of independent structures:
\begin{itemize}
\item[{\bf 1.}]~Cyclicity: 
\be A(1,2,\ldots, n) = A(2,\ldots, n,1)
\, , 
\ee
which follows from consistency with the definition \eqref{I.26} and cyclicity of the trace. 
\item[{\bf 2.}]~Parity: \be\left[ A(1,2,\ldots, n)\right]^{*} = A(\bar 1, \bar 2,\ldots,\bar n)\, . 
\ee
Here $\bar i$ denotes the inversion of the helicity of particle $i$. 
\item[{\bf 3.}]~Charge conjugation: 
\be A(1_q,2_{\bar q},3,\ldots, n) = -
A(1_{\bar q},2_{q},3,\ldots, n)\ , 
\ee
that is, flipping the helicity of a quark line changes the sign of the amplitude. This descends from 
the colour-ordered quark-quark-anti-quark vertex above.
\item[{\bf 4.}]~Reflection: \be A(1,2,\ldots, n) = (-1)^n\, A(n,n-1,\ldots,1)\, . 
\ee
It follows from the anti-symmetry of the colour-ordered gluon vertices
under reflection of all legs. It also holds in the presence of quark lines but only at tree~level.
\item[{\bf 5.}]~Photon decoupling, or dual Ward identity:
\begin{align}
\sum_{\sigma\in \mathbb{Z}_{n-1}} A(\sigma_1, \ldots , \sigma_{n-1}, n) =0\, , 
\end{align}
where $\sigma = \{\sigma_1, \ldots , \sigma_{n-1}\}$ are cyclic permutations of $\{1, 2,  \ldots, n-1\}$. 
It follows from  \eqref{I.26} and the fact that a gluon  amplitude with a single  photon vanishes since~$f^{0bc}{=}0$.
\end{itemize}

\section{Three-point amplitudes}
\label{sec:4}

\subsection{From symmetries}
\label{sec:4.1}
Scattering amplitudes are  covariant under  little group transformations of massless momenta \eqref{littlegroup}.
This  is encoded in the following relation \cite{Witten:2003nn}:
\begin{align}
\label{lg}
    -\frac{1}{2} \Big( \lambda_i^\alpha \ \frac{\partial}{\partial \lambda_i^\alpha} - \tilde{\lambda}_i^{\dot{\alpha}}\frac{\partial}{\partial \tilde{\lambda}_i^{\dot{\alpha}}}\Big) A \, = \, h_i \, A\, , 
\end{align}
where $h_i$ is the helicity of particle $i$. It is an immediate consequence of  how  the wavefunction of a particle of helicity $h_i$  scales under \eqref{littlegroup}.
Combined with Lorentz invariance, \eqref{lg} can be used to determine the functional form of three-point amplitudes of particles of any spin, without ever looking at a Lagrangian, as we now show.  

We begin by noting that momentum conservation $p_1 + p_2 + p_3=0$ implies $p_i\cdot p_j{=}0$ for $i, j{=}1,2,3$. In real Minkowski space this means that 
$\langle i\, j\rangle {=}0$ and $[ i\, j] {=}0$ for all particles: simply there is no scattering! Life is less constrained in complexified Minkowski space, where the spinors $\lambda$ and $\lt$ become independent, and  two  solutions are possible: 
\begin{align}
\langle i\, j\rangle =0\, \ \text{and} \ \, [i\, j]\neq 0\, , \qquad \text{or} \qquad \langle i\, j\rangle \neq 0\, \ \text{and} \ \, [i\, j] = 0\, , \qquad \forall \, i, j\, .
    \end{align}
Looking for instance at the helicity assignment $1^{-s}, 2^{-s}, 3^{+s}$, one can immediately see, using \eqref{lg}, that the answer must have the form  
\begin{align}
        A(1^{-s}, 2^{-s}, 3^{+s}) \sim  \big[ A(1^{-}, 2^{-}, 3^{+})\big]^s\,  
\end{align}
where for the amplitude with $s{=}1$ two options arise: $A(1^-, 2^-, 3^+) \!\sim\!{\langle 1\, 2\rangle^3}/ ({\langle 2\, 3\rangle\langle 3\, 1\rangle})$ 
    or $ A(1^-, 2^-, 3^+) \!\sim \![ 2\, 3][ 3\, 1]/  [ 1\, 2]^3$.
It turns out that Nature has chosen 
the first one, and we will set 
\begin{align}
\label{3pag}
    A(1^-, 2^-, 3^+) \, = \, i g \,  \frac{\langle 1\, 2\rangle^3}{\langle 2\, 3\rangle\langle 3\, 1\rangle}\, ,  
    \end{align}
where $g$ is the  Yang-Mills coupling constant.%
\footnote{The factor of $i$ comes from the Dyson expansion of the $S$-matrix, and in our conventions, scattering amplitudes are the elements of the matrix $iT$ where $S=\mathbb{1} + i T$.}
There are several reasons to see why this is the correct choice. First, an $n$-point amplitude has dimension $4{-}n$. With the choice of \eqref{3pag}, the coupling constant $g$ is dimensionless, as the Yang-Mills coupling should be, while  the other option  requires a dimensionful coupling. This would also imply that the corresponding interaction  in the Lagrangian is non-local. 

The amplitude in \eqref{3pag} is the first  in the MHV family \eqref{one}.  We now quote  the  $\overline{\text{MHV}}$ three-point amplitude, which is obtained from \eqref{3pag}  by replacing
$\lan a  b \ran \to - [a\, b]$:%
\footnote{Flipping the helicity sends $\lambda_\alpha\to \lt_{\dot{\alpha}}$, and a minus sign  arises from the different convention in defining the angle and square  brackets as in \eqref{bracket-conv}. 
}
\begin{align}
\label{3bpag}
    A(1^+, 2^+, 3^-) \, = \,- i g \,  \frac{[ 1\, 2]^3}{[ 2\, 3][ 3\, 1]}\, .  
    \end{align}
Little group scaling also fixes the possible form of the all-minus and all-plus three-point amplitudes: 
$A(1^-, 2^-, 3^-) {\sim} \langle 1\, 2\rangle \langle 2\, 3\rangle \langle 3\, 1\rangle$ and $A(1^+, 2^+, 3^+) {\sim} [1\, 2] [2\, 3] [3\, 1]$, but in Yang-Mills theory the proportionality constant is zero. These amplitudes  can be generated in a theory with a higher-dimensional, non-renormalisable interaction of the form $\text{Tr}\,  F^3$, where $F$ is the field strength \cite{Dixon:1993xd,Dixon:2004za,Broedel:2012rc}. 

Conceptually it is very important that we can determine three-point amplitudes just from symmetry considerations. These amplitudes will be the seeds of the BCFW recursion relation, discussed in Section~\ref{sec:5}.

\subsection{From Feynman diagrams}
As a useful exercise in spinor gymnastics, we will now 
derive \eqref{3pag} from  QCD Feynman rules.
Using the colour-ordered three-point vertex in \eqref{Feynrulveert} we find  (with all momenta taken as outgoing)
\begin{align}
   \hspace{-0.2cm} A(1^- 2^- 3^+) 
   \!=\!i\frac{g}{\sqrt{2}}\Big[ (p_1{-}p_2)\!\cdot\!\eps^{(+)}_3  \eps^{(-)}_1\!\cdot\!\eps^{(-)}_2+(p_2{-}p_3)\!\cdot\!\eps^{(-)}_1  \eps^{(-)}_2\!\cdot\!\eps^{(+)}_3 + (p_3{-}p_1)\!\cdot\!\eps^{(-)}_2  \eps^{(+)}_3\!\cdot\!\eps^{(-)}_1\Big] , 
\end{align}
where the polarisation vectors are given in \eqref{epsmp}. Choosing  the same reference spinor for the two negative-helicity gluons we can set $\eps_1^{(-)} {\cdot} \eps_2^{(-)} =0$, and using momentum conservation and the transversality condition $p_i\cdot \eps_i=0$ 
we can write this as
\begin{align}
   \hspace{-0.3cm} A(1^- 2^- 3^+) = ig \sqrt{2} 
   \Big[ p_2\!\cdot\!\eps^{(-)}_1 \, \eps^{(-)}_2\!\cdot\!\eps^{(+)}_3 -p_1\!\cdot\!\eps^{(-)}_2\,  \eps^{(+)}_3\!\cdot\!\eps^{(-)}_1\Big]\, . 
\end{align}
One can easily work out the various dot products
\begin{align}
\begin{split}
   \eps^{(-)}_2\!\cdot\!\eps^{(+)}_3&= -\frac{\lan 2 \xi\ran[3\xi ] }{[ 2 \xi] \lan 3 \xi\ran} \, , \qquad \qquad\, 
   \eps^{(-)}_1\!\cdot\!\eps^{(+)}_3= -\frac{\lan 1 \xi\ran[3\xi ] }{[ 1 \xi] \lan 3 \xi\ran} \, , 
   \\
     p_2\cdot \eps_1^{(-)}&= \frac{1}{\sqrt{2}}\frac{\lan 12 \ran[2 \xi]}{[1 \xi]}\, , \qquad \quad\ \ \,
     p_1\cdot \eps_2^{(-)}= -\frac{1}{\sqrt{2}}\frac{\lan 12 \ran[1 \xi]}{[2 \xi]}\, ,
     \end{split}
\end{align}
and therefore, 
\begin{align}
\begin{split}
    A(1^- 2^- 3^+) &= -ig \lan 12\ran \frac{[3\xi]}{\lan 3\xi\ran}
   \Big(  \frac{\lan 2\xi\ran}{[1\xi]} +\frac{\lan 1\xi\ran}{[2\xi]}  \Big) = 
    ig \lan 12\ran \frac{[3\xi]}{\lan 3\xi\ran}
   \frac{\lan \xi | p_1 + p_2 |\xi] 
  }{[1\xi][2\xi]}
  =
  ig \frac{\lan 12\ran [3\xi]^2 }{[1\xi][2\xi]}
   \, . 
   \end{split}
\end{align}
Finally we use three-point momentum conservation to simplify
\begin{align}
\frac{[3\xi] }{[1\xi]}=\frac{\lan 23\ran [3\xi] }{\lan 23\ran[1\xi]}=\frac{\lan12\ran}{\lan23\ran}\, , \qquad \quad 
\frac{[3\xi] }{[2\xi]}=\frac{\lan 13\ran [3\xi] }{\lan 13\ran[2\xi]}=\frac{\lan12\ran}{\lan31\ran}\, ,
    \end{align}
    thus arriving at the result \eqref{3pag}. One could repeat this calculation for the scattering of three gravitons, this time using the three-point vertex of \cite{DeWitt:1967uc}, arriving at a result proportional to  $\big[ A(1^{-}, 2^{-}, 3^{+})\big]^2$. The expression for the vertex in that paper contains at least 171 terms, which gives no   hints of such a  remarkable squaring relation!%
    \footnote{The reader is not encouraged to try.}

\section{BCFW recursion relation}
\label{sec:5}

\subsection{Derivation of the recursion}
\label{sec:der-BCFW}

It was long believed  that amplitudes may be determined from their analytic properties. The route followed in  \cite{Eden:1966dnq} was to complexify Mandelstam invariants and study  amplitudes as a  function of these. Unfortunately, complex analysis in many variables is complex! 
The Britto-Cachazo-Feng-Witten (BCFW) recursion relation
\cite{Britto:2004ap,Britto:2005fq} avoids this problem by mapping the 
singularities of tree-level amplitudes into poles in a {\it  single} complex variable $z$. 
To see this at work,  consider a tree-level $n$-gluon amplitude $A_{n}( p_{1}, \ldots p_{n})$, and introduce the following deformation of the  spinors
of  two adjacent particles $1$ and $n$, often indicated as $[n\, 1\ran$:
\begin{align} \label{bcfw-n1}
\begin{split}
&\lambda_1 \rightarrow \hat{\lambda}_{1}(z) = \lambda_{1} - z \lambda_{n} \,, \qquad  \tla_{1}\rightarrow
\tla_{1} \, ,  \\
&\la_{n}\rightarrow
\la_{n} \, , \qquad\qquad \qquad \qquad 
\tilde{\lambda}_n \rightarrow \hat{\tilde\lambda}_{n}(z) = \tilde{\lambda}_{n} + z \tilde{\lambda}_{1} \, ,  
\end{split}
\end{align}
with $z \in {\mathbb{C}}$. We denote the shifted, $z$-dependent quantities by a hat.
The corresponding  deformation of the  momenta, 
\begin{align}
\label{complexmamienta}
p_{1}^{\dot\alpha {\alpha}} \to \hat{p}_{1}^{\dot\alpha {\alpha}} (z) = 
\tilde{\lambda}_{1}^{\dot\alpha}\, (\lambda_{1} -z \lambda_{n})^{\alpha}  \,,\qquad
p_{n}^{\dot\alpha {\alpha}} \to \hat{p}_{n}^{\dot\alpha {\alpha}} (z) = (\tilde{\lambda}_{n} + z  \tilde{\lambda}_{1} )^{\dot\alpha}\, \lambda_{n}^{\alpha}  \,,
\end{align}
 preserves both overall
 momentum conservation and the on-shell conditions,
\begin{equation}
\hat{p}_{1}(z) + \hat{p}_{n}(z) = p_{1} + p_{n}\, , \qquad 
\hat{p}_{1}^2(z) = 0\,,\qquad \hat{p}_{n}^{2}(z) = 0\, ,  
\end{equation}
so that 
$A_n(z){=}A_n \big(\hat{p}_1 (z), p_2, \ldots, p_{n-1}, \hat{p}_n(z)\big)$ 
is a  one-parameter family of amplitudes. Note that $\hat{p}_1$ and $\hat{p}_2$ in \eqref{complexmamienta} are now complex --  we are now  working in complexified Minkowski space. This  makes  the three-point amplitudes of Section~\ref{sec:4} non-vanishing, which will then   become the seeds of the recursion.

What are the analytic properties of  $A_{n}(z)$?
It is well known that tree amplitudes have simple poles in multi-particle channels. 
This can be seen from the Feynman diagrammatic expansion: pick  all  diagrams which have a propagator $i / P^2$, where  $P$ is a sum of momenta (which will be adjacent for colour-ordered amplitudes, or generic in gravity). 
As $P$ goes on shell, the singular  diagrams in this class combine into  the product of  an amplitude to the left and one to the right of this propagator. 
This implies   that the deformed amplitude $A_{n}(z)$  has precisely
$n{-}3$ \emph{simple} poles in $z$: with $P_{i}:=\sum_{j=1}^{i-1}p_{j}$, these have the form 
\begin{align}
\label{poleargumentbcfw}
\begin{split}
\frac{i}{\hat{P}_{i}^2(z)} &
:= 
\frac{i}{P_{i}^2 - z \langle n | P_{i} | 1 \rbrack} = -\frac{1}{\langle n | P_{i} | 1 \rbrack} 
\frac{i}{z - z_{P_{i}} }
\, , 
\end{split}
\end{align}
where $\hat{P}_{i}(z) = \hat{p}_{1}(z) + p_{2} + \cdots p_{i-1}$, and 
\be
\label{zPidef}
z_{P_{i}}=\frac{P_{i}^{2}}{\lan n|P_{i}|1]}\, , \qquad \quad \forall i\in[3,n-1]\, .
\ee
\begin{figure}[tt]
\begin{center}
\begin{equation*}
\raisebox{-1.9cm}{\scalebox{0.8}{\includegraphics{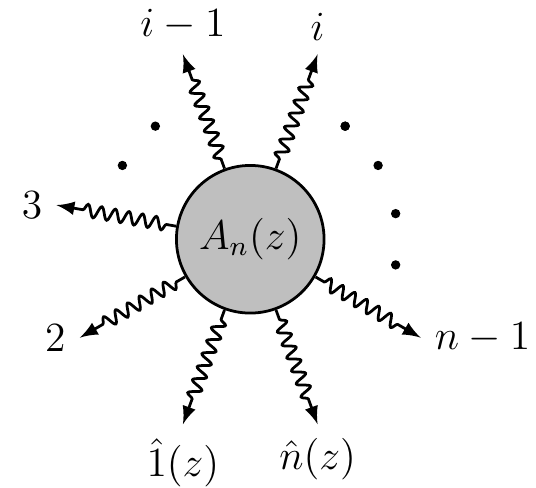}}}\,
\stackrel{z\to z_{P_{i}}}{\sim}
\frac{1}{z-z_{Pi}}\, \sum_h
\raisebox{-1.8cm}{\scalebox{0.8}{\includegraphics{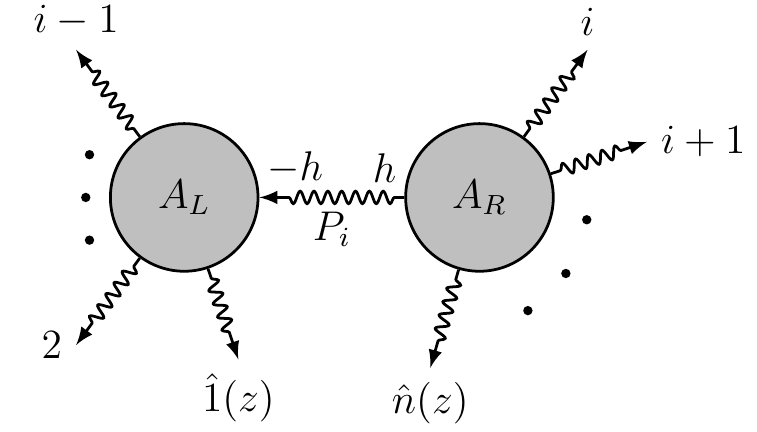}}}
\end{equation*}
\end{center}
\vspace*{-2mm}
\caption{\it 
Factorisation of the $z$-deformed amplitude $\cA_{n}(z)$.
}
\label{Fig:BCFW}
\end{figure}
It follows that, as  $z\to z_{P_{i}}$,  the amplitude $A_{n}(z)$ factorises as 
\begin{align}\label{limitpolezPi}
A_{n}(z) {\buildrel z\to z_{P_i} \over
{\relbar\mskip-1mu\joinrel\rightarrow}} \frac{i}{\hat{P}_{i}^2(z)}   \!\sum_{h=\pm} A_{L}(\hat{1}(z_{P_{i}}),2,\ldots, i-1,-\hat{P}^{-h}(z_{P_{i}})) \, A_{R}( \hat{P}^{h}(z_{P_{i}}) ,i ,\ldots,n-1, \hat{n}(z_{P_{i}}) )  ,  
\end{align} 
see Figure~\ref{Fig:BCFW}. 
The sum over $s$ in \eqref{limitpolezPi} runs over all possible
 states propagating between $A^{L}$ and $A^{R}$, and is theory dependent.  For gluons it is a sum over  $h= \{+,-\}$.

We are  only interested in the original amplitude, i.e.~$A_n (z{=}0)$, and   we can use complex analysis to construct it  from the knowledge of the residues of~$A_{n}(z)$:
\begin{align}
\label{BCFWrec}
\begin{split}
A_{n}  &= A_{n}(z{=}0) = \frac{1}{2\pi i}\oint_{C_{0}}\!\frac{dz}{z } \, A_n(z) 
= \sum_{i=2}^{n-1} \sum_{h=\pm} A_{L}^{-h}(z_{P_{i}}) \frac{i}{P_{i}^2} A_{R}^h(z_{P_{i}}) + {\rm Res}(z=\infty)\,.
\end{split}
\end{align}
Here $C_{0}$ is a  small circle around  $z{=}0$ that only contains  the pole around the origin. To obtain \eqref{BCFWrec} we have deformed  this into a large circle at  infinity, 
now   encircling all the poles $z_{P_{i}}$ in the complex plane but  with  an opposite
orientation. If $A_{n}(z) \to 0$ as $z\to \infty$ we can drop the boundary term
${\rm Res}(z=\infty)$.
As we shall argue in a moment, this is the case for gauge theories under certain conditions.
With this assumption, we arrive at the celebrated BCFW recursion relation \cite{Britto:2005fq}:
\begin{align}\label{eq-bcfw}
A_{n} =  
 \sum_{i=3}^{n-1} \sum_{h=\pm} & A_{i}\Bigl(\hat{1}(z_{P_{i}}),2,\ldots, 
-\hat{P}_{i}^{-h}(z_{P_{i}})\Bigr) \frac{i}{P_{i}^2}\,  A_{n+2-i}\Bigl (\hat{P}^{h}_{i}(z_{P_{i}}),i,\ldots, n-1, \hat{n}(z_{P_{i}}) \Bigr ) \, , 
\end{align}
with $z_{P_{i}}$  defined in \eqref{zPidef} and $P_{i}=p_{1}+p_{2}+\cdots + p_{i-1}$.
This relation is constructive: the amplitudes appearing on the right-hand side have lower 
multiplicity than $A_n$. Hence, with the seed three-gluon amplitudes
\eqref{3pag} and \eqref{3bpag},  we can use this relation to construct \emph{all} $n$-gluon
trees without using Feynman diagrams! In this derivation we chose to shift two neighbouring legs
$\hat{1}$ and $\hat{n}$. In fact, one can also shift non-neighbouring legs or even more than two legs
to obtain alternative recursion relations \cite{Risager:2005vk,Elvang:2008vz}.

An open issue is the vanishing of the boundary term in \eqref{BCFWrec}. For this we need that 
\be
\frac{1}{2\pi i}\oint_{\infty}\frac{dz}{z} \, {A_{n}(z)} =0\, , \ee
which in turns requires  a large-$z$ falloff of the amplitude as $A_{n}(z) {\sim} z^{-1}$.
In fact, the large-$z$ behaviour depends on the helicities of the shifted legs, and one can show that
\begin{align}
A(\hat{1}^{+},\hat{n}^{-})\stackrel{z\to \infty }{\sim} \frac{1}{z} \, , \qquad
A(\hat{1}^{+},\hat{n}^{+})\stackrel{z\to \infty }{\sim} \frac{1}{z} \, , \qquad
A(\hat{1}^{-},\hat{n}^{-})\stackrel{z\to \infty }{\sim} \frac{1}{z} \, , \qquad
\end{align}
yet $A(\hat{1}^{-},\hat{n}^{+})\!\stackrel{z\to \infty }{\sim}\!z^{3}$, which is then a forbidden $[n1\rangle$ shift.
It is  straightforward to show the first  relation by analysing  the
colour-ordered Feynman rules; the other scalings are more technical to derive
\cite{Arkani-Hamed:2008bsc}, see \cite{Henn:2014yza} for a pedagogical discussion.

\subsection{Gravity and other theories}

Can we generalise the BCFW recursion to other \emph{massless} quantum field theories? If we recap its
derivation, only two ingredients were  needed to establish it:
\begin{enumerate}
\item Tree-level amplitudes factorise on simple poles,
when the square of the sum of a subset of external momenta vanishes. Note that for colour-ordered amplitudes we only needed to consider adjacent channels but this was not essential, factorisation is a completely general property of unitary theories!
\item The deformed amplitude $A_{n}(z)$ falls off as $1/z$ at infinity. This depends on the theory and is related to its ultraviolet  behaviour.
\end{enumerate}
So in order to reconstruct 
tree amplitudes we need to
consider all multi-particle channels
\black 
\be
P_{I}^{\mu}:=\sum_{i\in I}p_{i}^{\mu}\, , \qquad \text{with}\quad
I\in\{\text{any subset of the momenta } p_{1}, \ldots, p_{n}\}\, .
\ee
Whenever  $P_{I}^{2}=0$ we have a pole, and if a two-particle BCFW shift is used the set $I$ must  contain only one of the shifted momenta so that $P_{I}^{2}$ becomes $z$-dependent.
Concretely, the BCFW recursion
for a shift of legs $1$ and $n$ as in \eqref{bcfw-n1}
in gravity takes the form \cite{Bedford:2005yy, Cachazo:2005ca}
\begin{align}\label{eq-bcfw-grav}
M_{n} =  
 \sum_Q \sum_{h=\pm\pm} 
 M_L\Bigl(\hat{1}(z_{P_{Q}}),Q, 
-\hat{P}_{Q}^{-h}(z_{P_{Q}})\Bigr)
\frac{i}{P_{Q}^2}\,  M_{R}\Bigl (\hat{P}^{h}_{Q}(z_{P_{Q}}),\bar{Q}, \hat{n}(z_{P_{Q}}) \Bigr ) \,,
\end{align}
where $Q$ denotes \emph{all} subsets of  momenta in  $\{p_{2},\ldots, p_{n-1}\}$, $\bar{Q}$ its complement and
$P_{Q}=p_{1}+\sum_{i\in Q}P_{i}$.
Finally, we note that the BCFW recursion can be  generalised to massive theories  \cite{Badger:2005zh,Badger:2005jv}, to rational parts of one-loop amplitudes in QCD and gravity 
\cite{Bern:2005hs,Bern:2005ji,Brandhuber:2007up,Dunbar:2010xk,Alston:2015gea},  form factors \cite{Brandhuber:2010ad,Brandhuber:2011tv}, non-linear sigma models and effective field theories \cite{Kampf:2012fn,Cheung:2014dqa,Cheung:2015ota,Mojahed:2021sxy}. Supersymmetric  recursion relations \cite{Brandhuber:2008pf,Arkani-Hamed:2008owk} are reviewed in Section~\ref{sec:8}. 
In the maximally supersymmetric case, that is in  $\mathcal{N}{=}4$ super Yang-Mills theory, a generalisation of the BCFW 
recursion to loop-level planar amplitudes was achieved using the formalism of on-shell diagrams and positive Grassmannians of \cite{Arkani-Hamed:2012zlh}. These important developments connecting to the Amplituhedron approach are reviewed in Chapter~7  of this review~\cite{Herrmann:2022nkh}. 
Criteria to construct recursion relations in general field theories were studied in \cite{Cheung:2015cba}, also making use of multi-line shifts \cite{Risager:2005vk,Cohen:2010mi}.

\subsection{The MHV amplitude from the BCFW recursion relation}
\label{sect:MHVfrom BCFW}

As an application, we  now derive by induction the   Parke-Taylor formula \eqref{one}.  
We already know from  Section~\ref{sec:4} that it  is true for $n\!=\!3$. 
Therefore we only need to prove  recursively that the formula is correct. 
We will focus on the case where particles $n$ and $1$ have negative helicity, and choose our $[n1\rangle$ shifts of \eqref{bcfw-n1}. The MHV amplitude has no multi-particle factorisation. In fact, only one
BCFW  diagram contributes, where $A_L$ in  Figure~\ref{Fig:BCFW} is a three-point $\overline{\rm MHV}$ amplitude \eqref{3bpag} and  $A_R$ is an $(n-1)$-point MHV amplitude.
From \eqref{zPidef}, the position of the pole is 
$
z_{P} = \frac{ (p_1 + p_2 )^2 }{ \langle n | P|1 \rbrack} = \frac{ \vev{12}\lbrack2 1 \rbrack}{\vev{n 2} \lbrack2 1\rbrack} = \frac{\vev{12}}{\vev{n 2}} \, .
$
The amplitudes $A_{L}$ and $A_{R}$ are then
\begin{align} 
\label{62}
\begin{split}
A_{L} =& A^{\overline{\text{MHV}}}_{3}( \hat{1}^{-},2^{+},-\hat{P}^{+} ) = -  i g  
\frac{ \lbrack 2 (-
\hat{P}) \rbrack^3}{\lbrack 12 \rbrack \lbrack (-\hat{P}) 1 \rbrack} \,, \\
A_{R} =& A^{\text{MHV}}_{n-1}(\hat{P}^- , 3^+ , 4^+, \ldots (n-1)^{+} , \hat{n}^{-} ) =   i g^{n-3}  \frac{ \vev{\hat{n} \hat{P}}^3 }{\vev{\hat{P} 3} \vev{34} \cdots \vev{(n-1) \hat{n}}}\,.
\end{split}
\end{align}
Using \eqref{analytic-cont}, the fact that $\lambda_n$ and $\tilde{\lambda}_1$ are not shifted in  our $[n1\rangle$ shift of \eqref{bcfw-n1}, as well~as 
\begin{align}
\vev{\hat{n} \hat{P}} \bev{\hat P 2} &= \vev{n\hat 1}\, \bev{12} = \vev{n1}\, \bev{12}\, , \qquad
\vev{3\hat P}  \,\bev{\hat P 1} 
=  \vev{32}\, \bev{21}\, , 
\end{align}
we find
\begin{align}
\hat{A}_{L}  \frac{i}{(p_1 + p_2)^2} \hat{A}_{R} &=
-i g^{n-2}\frac{\vev{n1}^{3}
\, \bev{12}^{3}}{\bev{12}\bev{21}\vev{32}\bev{21}\, \vev{12}\vev{34}\cdots
\vev{(n-1)\, n}}
= ig^{n-2} \frac{ \vev{n 1}^4}{\vev{12} \cdots \vev{n1}} \ , 
\end{align}
in agreement with  \eqref{one} for the chosen helicities. MHV amplitudes with different helicity assignments can easily be obtained using the same strategy as above.

\subsection{What's special about   Yang-Mills MHV amplitudes?}
\label{sec:2.4}

The MHV amplitude \eqref{one} derived in the last section is special in many ways. First, it does not have any multi-particle poles -- a fact that follows from the vanishing of the amplitudes%
\footnote{A proof that $A_n(1^\pm, 2^+, \ldots , n^+)=0$ for $n>3$ is provided in Section~\ref{sec:vanishing}.}
$A_n(1^\pm, 2^+, \ldots , n^+)$. Second, it  is a holomorphic function of the spinor variables $\lambda$. As anticipated in the introduction, Witten was able to relate  this   to the property that MHV  amplitudes have support on a complex line in twistor space 
\cite{Witten:2003nn}. This is  easy to show:  reintroducing the momentum conservation delta function $(2\pi)^{4}\delta^{(4)}(p) {=} \int\!d^4x \, e^{i p\cdot x}$, the  amplitude in twistor space  is obtained by performing a half-Fourier transform from spinor variables $(\lambda, \lt)$ to twistor variables  $(\lambda, \mu)$:
\begin{align}
\label{FTTTS}
\int\!\prod_{i=1}^n \frac{d^2 \lt_i}{(2\pi)^2} \, e^{i [\mu_i \lt_i] } A_n^{\rm MHV } \, (2\pi)^4\delta^{(4)} \Big( \sum \lambda_i \lt_i\Big) =
A_n^{\rm MHV } \int\!\!d^4x \, \prod_{i=1}^n \delta^{(2)} (\mu^{\dot{\alpha}}_i+ x^{\dot{\alpha} \alpha } \lambda_{i\alpha})  
.
\end{align}
Hence the transformed amplitude vanishes unless the gluon twistor space coordinates   $(\lambda_i, \mu_i)$ 
satisfy  
$
\mu^{\dot{\alpha}} + x^{\dot{\alpha} \alpha} \lambda_\alpha {=} 0$,  
$\dot{\alpha}{=}1,2$, which is the equation of a (complex) line in twistor space. As shown in \cite{Witten:2003nn},   amplitudes with $q$ negative-helicity gluons, which we call  ${\rm N}^{q-2}{\rm MHV}$, are supported on algebraic curves in twistor space of degree $q{-}1{+}L$, where $L$ is the number of loops. 
The case of disconnected curves leads to the so-called MHV diagram method \cite{Cachazo:2004kj}, while  connected prescriptions were developed in \cite{Roiban:2004vt, Berkovits:2004hg,Roiban:2004yf}.
\black

\section{Symmetries of scattering amplitudes}
\label{sec:6}

\subsection{Poincar\'e and conformal symmetry}
\label{sect:bosonicsym}

Let us now discuss the symmetry properties of scattering amplitudes. These
can be obvious (Poincar\'e), less obvious (conformal) or  hidden (dual conformal
or Yangian), as we will discuss in the following.%
\footnote{Hidden symmetries are not invariances of the action.}
In  relativistic quantum field theory, amplitudes  are
 Poincar\'e  invariant by construction. To see this, we seek a representation of the
Poincar\'e symmetry generators --  translation and Lorentz generators --  in spinor-helicity variables \cite{Witten:2003nn}. 
Translations $p^{\alpha\da}$ are realised  as a multiplicative operator 
\begin{align}
p^{\alpha\da} & = \sum_{i=1}^n \lambda_i^\alpha \, \tla_i^{\da}\, , 
\end{align}
and the  corresponding  invariance 
$p^{\alpha\da}\, \cA_n(\lambda_i,\tla_i) {=}
p^{\alpha\da}\,\delta^{(4)}(p^{\alpha\da}) \, A_n(\lambda_i,\tla_i) {=} 0$ 
is  manifest by virtue of the  total momentum conservation
delta function.
The  Lorentz generators 
are symmetric bi-spinors, 
$m_{\alpha\beta}$ and ${\overline m}_{\da\db}$, realised as  first-order differential operators,
\be
m_{\alpha\beta} = \sum_{i=1}^n \lambda_{i\, (\alpha}\, \partial_{i\,\beta)}\, , \qquad
{{\overline m}}_{\da\db} = \sum_{i=1}^n \tla_{i\, (\da}\, \partial_{i\, \db)}\, ,
\ee
with $\partial_{i\alpha} := \frac{\partial}{\partial \lambda_i^\alpha}$, 
$\partial_{i\da} := \frac{\partial}{\partial \tla_i^\da}$  and
$r_{(\alpha\beta)}:= \frac{1}{2}\, (r_{\alpha\beta} + r_{\beta\alpha})$ denotes symmetrisation.
Lorentz invariance  of $\cA_n(\lambda_i,\tla_i)$, that is  
$
m_{\alpha\beta}\, \cA_n(\lambda_i,\tla_i)  {=} {{\overline m}}_{\da\db}\, \cA_n(\lambda_i,\tla_i){=} 0
$
is manifest, as the spinor brackets $\vev{ij}$ and $\bev{ij}$ are invariant under $m_{\alpha\beta}$
and ${{\overline m}}_{\da\db}$, e.g.
\be
m_{\alpha\beta}\, \vev{jk} =
\sum_{i=1}^n \lambda_{i\, (\alpha}\, \partial_{i\,\beta)}\, \lambda_{j}^{\gamma}\lambda_{k\, \gamma}=
\lambda_{j\, \alpha}\lambda_{k\, \beta} -\lambda_{j\, \beta}\lambda_{k\,\alpha} + (\alpha\leftrightarrow \beta) =0\,.
\ee
Classical Yang-Mills theory is invariant under an additional,  less obvious symmetry: conformal
symmetry. It originates from the fact that  pure
Yang-Mills theory and massless QCD do not carry  any dimensionful parameter and  are thus  invariant under scale transformations (or dilatations)
$
x^\mu {\to} \kappa^{-1}\, x^\mu
$, or,  in momentum space
$
p^\mu {\to} \kappa\, p^\mu
$.
The dilatation generator in spinor-helicity variables acting on $n$-point amplitudes reads~\cite{Witten:2003nn}
\be
d= \sum_{i=1}^n\, \Big( \frac{1}{2} \lambda^\alpha_i\, \partial_{i\,\alpha} + 
 \frac{1}{2} \tla^\da_i\, \partial_{i\,\da} + 1\Big) \,, 
\label{d_def}
\ee
reflecting the mass dimensions $\nicefrac{1}{2}$ of the  spinors, 
i.e.~$[d,\lambda_i]=\frac{1}{2}\, \lambda_i$ and $[d,\tla_i]=\frac{1}{2}\, \tla_i$.

\subsection{Example: the MHV amplitude}

As an example, we now wish to    check the invariance
of the MHV amplitudes 
$
\cA_n^\text{MHV} {=}  \delta^{(4)}(\sum_i p_i) A_n^\text{MHV} $ with $A_n^\text{MHV}$ given in \eqref{one}.
The dilatation operator $d$ in  \eqref{d_def} simply measures the  mass
dimension of the object it acts on.
We note the mass dimensions $[\delta^{(4)}(p)]=-4$, $[\vev{i\, j }^4]=4$ and
$[(\vev{12}\cdots\vev{n1})^{-1}]=-n$, hence
\be
d\, \cA_n^\text{MHV} = (-4 + 4-n +n) \, \cA_n^\text{MHV} =0\, ,
\ee
as required. Relativistic scale-invariant quantum field theories are conformal, i.e.~the dilatation symmetry is accompanied by invariance under so-called special conformal transformations $k_{\alpha\da}$. 
This  symmetry generator is 
realised in terms of a second-order differential operator in  spinor-helicity variables
\cite{Witten:2003nn},
\be
k_{\alpha\da}= \sum_{i=1}^n \partial_{i\, \alpha}\, \partial_{i\, \da} \, .
\ee
Checking this symmetry for MHV amplitudes is instructive yet requires a little bit of algebra \cite{Witten:2003nn},
see \cite{Henn:2014yza} for a pedagogical exposition.

In summary, together with the Poincar\'e and dilatation
generators,  the set of operators $\{p_{\alpha\da}, k_{\alpha\da}, m_{\alpha\beta}, {{\overline m}}_{\da\db}, d\}$ generate
the four-dimensional conformal group  $SO(2,4)$ which leave tree-level pure Yang-Mills
and massless QCD amplitudes invariant.

\section{Collinear and soft limits in gauge theory and gravity}
\label{sec:7}

\subsection{Yang-Mills theory}

\subsubsection{Collinear limits}

Scattering amplitudes in Yang-Mills theories have a universal behaviour when two (or more) particle momenta become collinear, which in turn can be used to constrain their form, or check the correctness of a calculation. In the following we  discuss the case of two gluons with  momenta $p_1$ and $p_2$ becoming collinear. This is described by setting
 $p_1 {=} z P$ and $p_2 {=} (1-z) P$, where $P:=p_1 + p_2$ and $P^2\to 0$ in the collinear limit. The universal behaviour of  tree-level  amplitudes can then be described~as 
 \begin{align}
    A_n (1,  \ldots, n) 
    \ {\buildrel p_1 \parallel p_2 \over
{\relbar\mskip-1mu\joinrel\longrightarrow}} \ 
\sum_{h = \pm} {\rm Split}_{-h}(1,2)
\, 
A_{n-1} (P^h, 3, \ldots, n)
\ . 
\end{align}
The  {\it splitting amplitudes}  ${\rm Split}_{- \lambda}(1,2)$ diverge in the collinear limit, and are given by
\begin{align}
\begin{split}
\label{collsplit}
{\rm Split}_{-}(1^-, 2^-) & =0\ , \qquad \qquad \qquad \qquad
        {\rm Split}_{-}(1^+, 2^+)  =\frac{1}{\sqrt{z(1-z)}} \frac{1}{\lan 12\ran}\, , \\ 
        {\rm Split}_{+}(1^+, 2^-) & = \frac{(1-z)^2}{\sqrt{z(1-z)}} \frac{1}{\lan 12\ran}
        \ , \qquad  {\rm Split}_{-}(1^+, 2^-)  = - \frac{z^2}{\sqrt{z(1-z)}} \frac{1}{[ 12]}
        \ .
\end{split}
\end{align}
An elegant way to derive this universal behaviour at tree level is based on  the  MHV diagram method%
\footnote{MHV diagrams can also be understood as multi-line BCFW recursion relations,  where one shifts the $\lt$ spinors of {\it all} the negative-helicity gluons \cite{Risager:2005vk}.}
\cite{Cachazo:2004kj}, later extended to  to loop amplitudes  in \cite{Brandhuber:2004yw, Brandhuber:2005kd}. 
While we will not review it here  (see e.g.~\cite{Brandhuber:2011ke} for details), the basic rules are very easy to explain: MHV amplitudes are continued off shell  and used as vertices;   to  an internal leg whose momentum $P$  is  a sum of several external momenta, we associate the spinor 
\begin{align}
\label{MHV-prescr}
    \lambda_{\widehat{P}}^\alpha \to P^{\dot{\alpha}\alpha} \tilde{\xi}_{\dot\alpha}\, , 
\end{align}
where $|\xi]$ is a  reference spinor (this is often called an {\it off-shell continuation} of the spinor);  and  MHV vertices are joined using scalar propagators $i/P^2$. Finally, by counting negative helicities, one can immediately see that MHV diagrams contributing to an $\mathrm{N^{\it k}MHV}$ amplitude must contain $k{+}1$ MHV vertices. 
To make contact with  Section~\ref{sec:2.4},  note that  this corresponds to a disconnected curve of degree $k{+}1$ in twistor space -- the union of $k{+}1$ complex lines.

MHV diagrams  treat positive and negative helicities on different footing,  hence we need to distinguish two types of collinear limits: those  where the number of negative helicities is unchanged, that is 
$++ \to +$ and $+- \to -$; and those where 
this number
is reduced by one, that is 
$-- \to -$ and $+- \to +$. 
In both cases, the MHV diagrams that contribute  in the  limit have the two legs that become collinear
attached to the same MHV vertex \cite{Cachazo:2004kj}. In  the first case, corresponding to the  second and third splitting amplitudes  in \eqref{collsplit}, 
the collinear behaviour descends directly from the single MHV vertex containing the two momenta that are becoming  collinear; while in the second, it  arises from an MHV diagram where the two particles going collinear  belong to a three-point MHV vertex, connected to another MHV vertex with the usual 
scalar propagator of the  MHV diagrammatic approach. 
\begin{figure}[ht]
\begin{center}
\scalebox{0.45}{\includegraphics{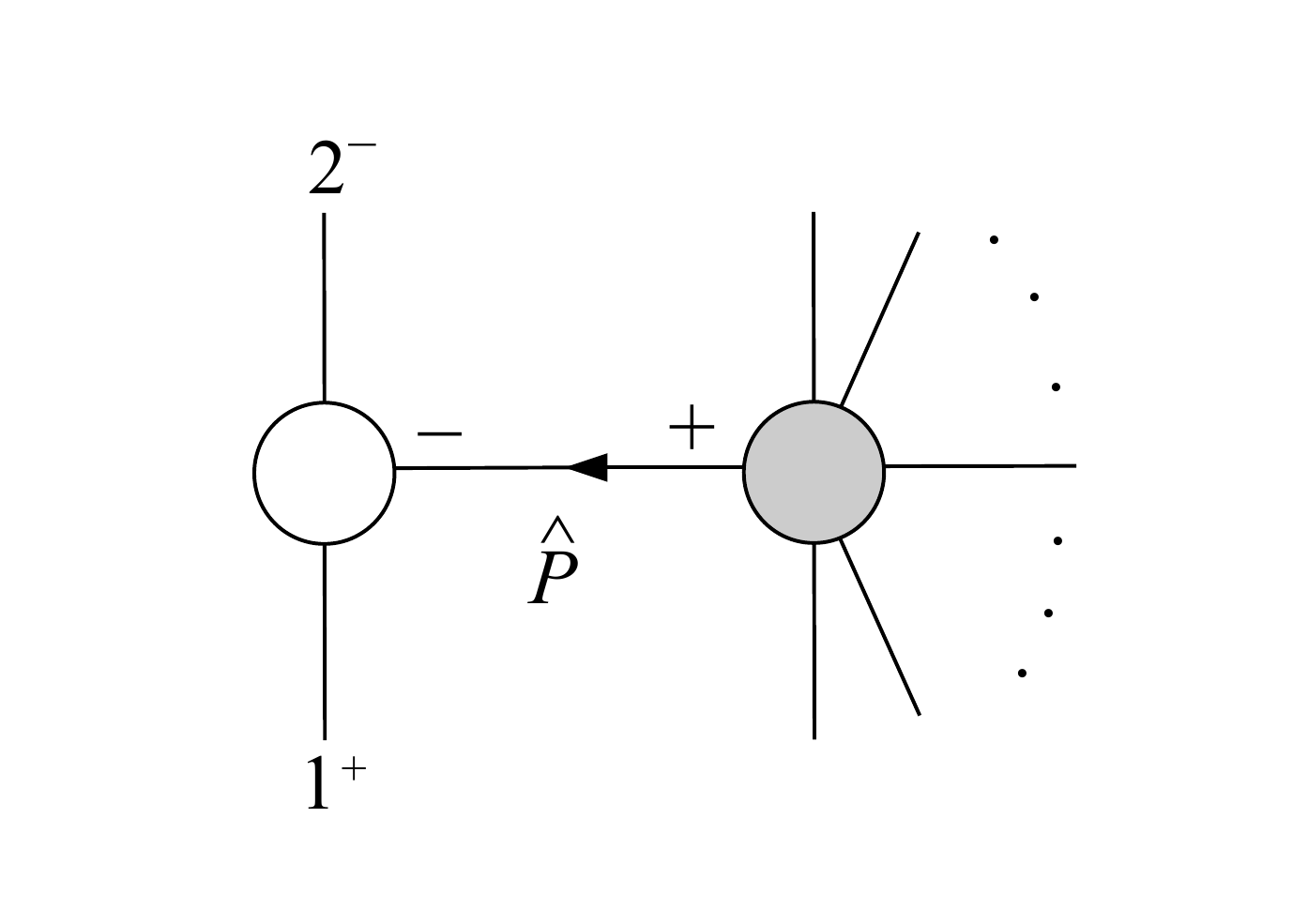}}
\end{center}
\vspace*{-10mm}
\caption{\it 
The MHV diagram contributing to the collinear limit $--\to -$. The grey (white) amplitudes is (anti)MHV. 
}
\label{fig:coll-limit}
\end{figure}
As an example we now derive collinear factorisation in  the case $+-\to +$. The relevant  MHV diagram is shown in Figure~\ref{fig:coll-limit}, and gives 
\begin{align}
    i \frac{\lan 2\,  {-\hat{P}}\ran^3}{\lan -\hat{P} \, 1\ran \lan 12 \ran}\frac{i}{\lan 12\ran [21] } A_{n-1} (  \hat{P}^+ , \ldots )\ . \end{align}
Following \eqref{MHV-prescr}, the spinor $\lambda_{\hat{P}}$ is given by  $\lambda_{\hat{P}} = (p_1+p_2) |\xi] / [ \hat{P} \xi]$, where $|{\xi}]$ is the reference spinor. Using this and \eqref{analytic-cont}, we get
\begin{align}
 \lan 2 \, {-\hat{P}}\ran = i \frac{\lan 21\ran [1\xi]}{[\hat{P} \xi]}\ , 
 \qquad 
 \lan 1 \, {-\hat{P}}\ran = i \frac{ \lan 1 2\ran [2 \xi]}{[\hat{P} \xi]}\ , 
\end{align}
and hence 
\begin{align}
    A_n  \ {\buildrel p_1 \parallel p_2 \over
{\relbar\mskip-1mu\joinrel\longrightarrow}} \ 
-  \frac{1}{[12]} \, \frac{[1  \xi]^3}{[\hat{P}\xi ]^2 [2 \xi  ]}
A_{n-1} ( \hat{P}^+,  \ldots )\ .  
\end{align}
Replacing $\lt_1 \to \sqrt{z} \lt_{\hat{P}}$, $\lt_2 \to \sqrt{1-z} \lt_{\hat{P}}$, we arrive at 
\begin{align}
  A_n  \ {\buildrel p_1 \parallel p_2 \over
{\relbar\mskip-1mu\joinrel\longrightarrow}} \ 
-\frac{z^2}{\sqrt{z (1-z)}}\frac{1}{[12]} \, 
A_{n-1} (P^+,  \ldots )\ ,     
\end{align}
where $P= p_1+p_2$, thus reproducing the last splitting amplitude in \eqref{collsplit}.

We conclude by  mentioning that collinear behaviour at one loop \cite{Kosower:1999rx,Bern:1999ry}
can also be studied \cite{Brandhuber:2005kd} using quantum MHV diagrams \cite{Brandhuber:2004yw,Bedford:2004py,Bedford:2004nh,Quigley:2004pw}.

\subsubsection{Soft limits}

\begin{figure}[ht]
\begin{center}
\scalebox{0.45}{\includegraphics{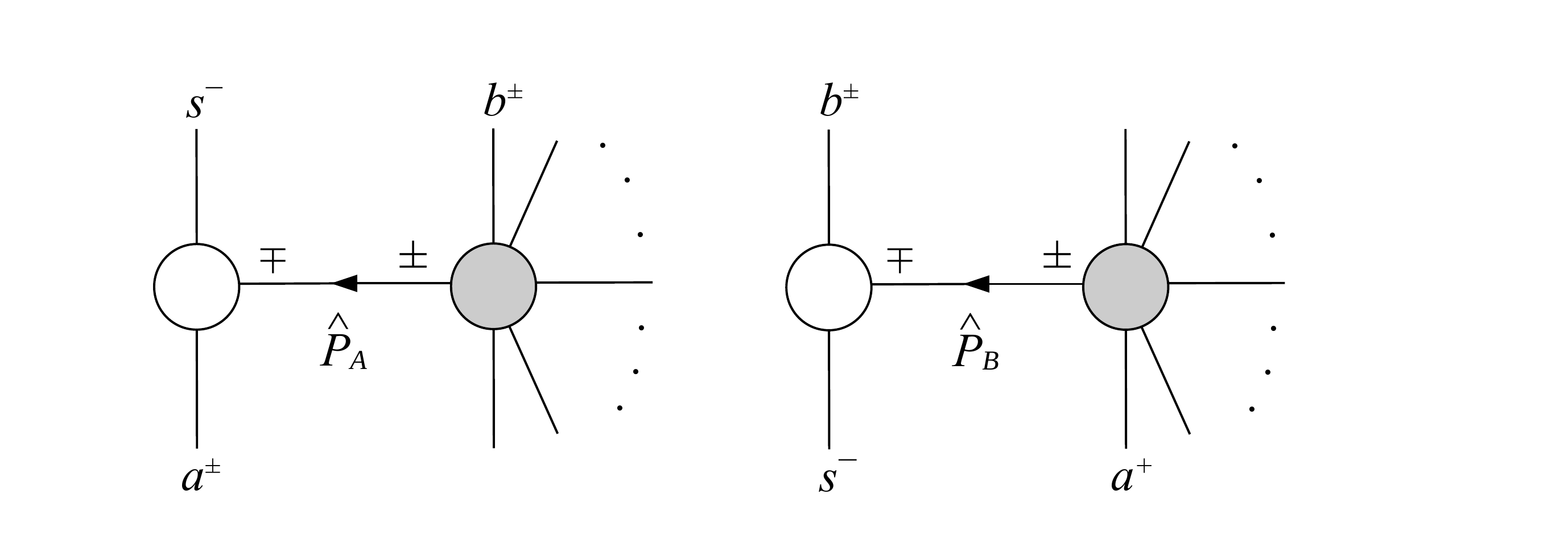}}
\end{center}
\vspace*{-10mm}
\caption{\it 
The two MHV diagrams contributing to the soft limit $p_s\to 0$ for the helicities $(a^+, s^-, b^+)$. 
}
\label{fig:soft-limits}
\end{figure}
Amplitudes have a universal behaviour also in soft limits, where the momentum of a particle becomes small.%
\footnote{An extensive discussion of soft limits can be found in Chapter~11 of this review \cite{McLoughlin:2022ljp}.}
At tree level, 
\begin{align}
\label{soft-tree}
A_n
(1,\ldots, a, s, b, \ldots, n)
\ 
{\buildrel p_s \to 0 \over
{\relbar\mskip-1mu\joinrel\longrightarrow}}
\ 
\mathcal{S}^{(0)} ( a, s, b)
\ A_{n-1}
(1,\ldots, a , b, \ldots, n)
\ , 
\end{align}
where $\mathcal{S}^{\rm (0)} ( a, s, b)$ is a 
tree-level soft (or eikonal) factor, 
\begin{align}
\label{soft-funct-tree}
\mathcal{S}^{(0)} ( a, s^+, b) \ = \ 
\frac{\lan a \, b \ran }{ 
\lan a \, s \ran \, \lan s \, b \ran }
\ , \qquad 
\mathcal{S}^{(0)} ( a, s^-, b) \ = \ 
- \,
\frac{ [ a \, b ] }{ 
[ a \, s ]  \, [ s \, b ]  }
\ . 
\end{align}
Note the dependence on the helicity of the soft particle (but not on the helicities of the particles adjacent to it in colour space). 
The derivation from MHV diagrams is straightforward for the first case. The second case, where the gluon becoming soft has negative helicity, 
is  special since 
MHV vertices have precisely two negative helicities and a generic MHV vertex would simply vanish in the limit. Two diagrams have to be considered in this case, shown in Figure~\ref{fig:soft-limits}:   
in the  first one,   an 
MHV three-point vertex with  external gluons  $a$ and $s$
($s$ is the leg whose momentum is becoming soft)
is joined to an MHV vertex 
to which the leg $b$ belongs,  maintaining  the colour
ordering $a$, $s$, $b$; in the second, 
$s$ and $b$ belong  to a three-point MHV vertex, which is then 
linked to a second MHV vertex containing the gluon  $a$.  Focusing on the case where particles $a$ and $b$ have positive helicities as an example, the  first diagram gives
\begin{align}
\begin{split}
\label{uno}
    i \frac{\lan s\, {-\hat{P}_A} \ran^3}{\lan -\hat{P}_ Aa \ran \lan as \ran} \frac{i}{\lan sa\ran [ as]}A_{n-1} (\hat{P}_A, b, \ldots ) & = \  \frac{[a \xi]^3}{[s \xi][sa]}\frac{1}{[\hat{P}_A\xi]^2}A_{n-1} (\hat{P}_A, b, \ldots ) \\
    & \to  \frac{1}{[s \xi]} \frac{[a\xi]}{[sa]}A_{n-1} (\hat{P}_A, b, \ldots )\ ,
\end{split}
\end{align}
while the second evaluates to
\begin{align}
\begin{split}
\label{due}
    i \frac{\lan \, {-\hat{P}_B} s\ran^3}{\lan s b\ran\lan b\, {-\hat{P}_ B}\ran} \frac{i}{\lan sb\ran [ bs]}A_{n-1} (\hat{P}_B, b, \ldots ) & = - \ \frac{[b \xi]^3}{[s \xi][sb]}\frac{1}{[\hat{P}_B\xi]^2}A_{n-1} (a^+, \hat{P}_B, \ldots ) \\
    & \to - \frac{1}{[s \xi]} \frac{[b\xi]}{[sb]}A_{n-1} (a^+, \hat{P}_B, \ldots )\ ,
\end{split}
\end{align}
where $|\xi]$ is the usual MHV-diagram reference spinor, and 
\begin{align}
|\hat{P}_A\ran =     \frac{(p_a + p_s) |\xi]}{[\hat{P}_A\xi]}\ , \qquad 
|\hat{P}_B\ran =     \frac{(p_b + p_s) |\xi]}{[\hat{P}_B\xi]}\ .
\end{align}
We also used   \eqref{analytic-cont}, and 
\begin{align}
\lan s \hat{P}_A\ran &= \frac{\lan sa\ran [a \xi]}{[\hat{P}_A \xi]}\ , \quad     
\lan a \hat{P}_A\ran = \frac{\lan as\ran [s \xi]}{[\hat{P}_A \xi]}\ , \quad 
\lan s \hat{P}_B\ran &= \frac{\lan sb\ran [b \xi]}{[\hat{P}_B \xi]}\ , \quad     
\, \lan b \hat{P}_B\ran = \frac{\lan bs\ran [s \xi]}{[\hat{P}_B \xi]}\ .
    \end{align}
Summing  the contributions in \eqref{uno} and \eqref{due}, and taking the soft limit (with $\hat{P}_A \to p_a$, $\hat{P}_B \to p_b$) we   obtain
\begin{align}
A_n(a, s^-, b, \ldots) {\buildrel p_s \to 0 \over
{\relbar\mskip-1mu\joinrel\longrightarrow}} - \frac{[ab]}{[as][sb]}A_{n-1} (a, b, \ldots )\ ,
\end{align}
in agreement with $\mathcal{S}^{(0)}(a, s^-, b)$  in \eqref{soft-funct-tree}. Similar derivations can be carried out for  the other possible  helicities of particles $a$ and $b$.

\subsubsection{Soft limits from recursion relations}
\label{sec:slYMBCFW}
There is an alternative, powerful way to derive soft theorems  from  the BCFW recursion relation. It was originally proposed in \cite{Cachazo:2014fwa},  where it was found that not only the leading but also the subleading soft behaviour of graviton amplitudes is universal.  A similar approach was devised in Yang-Mills theory in \cite{Casali:2014xpa}, as we now briefly review. 
Choosing to shift the momenta of particles $s$ and $b$,  a single diagram contributes in the soft limit, which is identical to that   on the left-hand side of Figure~\ref{fig:soft-limits} now to be interpreted as a BCFW diagram. For concreteness we carry out the computation for the case that legs $a$ and $s$ carry helicity $+1$,  however the result is  independent of the helicity of particle $a$, hence   we will drop its helicity label. With the shifts $\hat{\lambda}_s = \lambda_s + z \lambda_b$, 
$\hat{\lt}_b = \lt_b - z \lt_s$, the recursive diagram evaluates to \begin{align}
A_n(a, s^+, b, \ldots) \to -i \frac{[as]^3}{[-\hat{P}_Aa][ s \, {-\hat{P}_A}]} \frac{i}{(p_a + p_s)^2 }A_{n-1} (z^\ast)\ , 
\end{align}
and  $z^\ast = - \lan as\ran / \lan ab\ran$ is the position of the pole for this BCFW diagram. The internal momentum evaluated at this pole can be written as $\hat{P}_A = \lambda_a \big[\lt_a + \lt_s (\lan s b\ran / \lan a b \ran)\big]$ (after using the Schouten identity), and using this  one quickly arrives at
\begin{align}
A_n(a^+, s^+, b, \ldots) \, {\buildrel p_s \to 0 \over
{\relbar\mskip-1mu\joinrel\longrightarrow}} \,  \frac{\lan ab\ran }{\lan as\ran \lan sb\ran }A_{n-1} (z^\ast)\ , 
\end{align}
where $A_{n-1} (z^\ast) = A_{n-1} ( \{\lambda_a,\lt_a + 
\frac{\lan s b\ran}{ \lan a b \ran} \lt_s  \}, \{ \lambda_b, \lt_b + \frac{\lan a s\ran}{ \lan a b \ran} \lt_s \}, \ldots) $. 
To leading order in the soft limit, one  simply replaces $A_{n-1} (z^\ast)\to A_{n-1} (a, b, \ldots )$ thus reproducing the soft factor in \eqref{soft-funct-tree}.  One can also be more ambitious and  keep subleading terms in the limit. Rescaling the soft momentum as $p_s \to \delta \, p_s$ to keep track of terms, one  finds that 
\begin{align}
    A_{n} (a, s^+, b, \ldots ) {\buildrel p_s \to 0 \over
{\relbar\mskip-1mu\joinrel\longrightarrow}} \left( \frac{1}{\delta^2} \cS^{(0)} + \frac{1}{\delta} \cS^{(1)} \right) A_{n-1} (a, b, \ldots )+ \cO(\delta)\ , 
\end{align}
where the subleading soft factor $\cS^{(1)}$ is
\begin{align}
\cS^{(1)}=  \frac{1}{\lan as\ran} \lt_s \frac{\partial}{\partial \lt_a} + \frac{1}{\lan sb \ran } \lt_s \frac{\partial}{\partial \lt_b} 
\ .  
\end{align}

\subsection{Gravity}

\subsubsection{Collinear limits}

Unlike Yang-Mills amplitudes,  gravity amplitudes in real Minkowski space are non-singular in  collinear limits, more precisely  they only have phase singularities, which become simple poles in complex Minkowski space. Concretely \cite{Bern:1998sv}, if we send $p_i \to z P$ and $p_j \to (1-z) P$
as $P^2 = (p_i+p_j)^2 \to 0$, we have 
\begin{align}
    \cM_{n} (i^{h_i},j^{h_j},\ldots)
    {\buildrel p_i || p_j \over
{\relbar\mskip-1mu\joinrel\longrightarrow}}
\sum_{h=\pm\pm} {\rm Split}^{\rm GR}_{-h}(i^{h_i},j^ {h_j})
\cM_{n-1} (P^{h},\ldots) + R_n \, ,
\end{align}
where $h$ and  $\sigma$ denote the helicities of the gravitons. The remainder $R_n$ is free of phase singularities/poles and the splitting amplitudes are given by
\begin{align}
\begin{split}
    {\rm Split}^{\rm GR}_{--}(i^{++},j^{++}) & = - \frac{1}{z(1-z)}\frac{[i\,j]}{\langle i \,j \rangle}  ,
    \ \ 
   {\rm Split}^{\rm GR}_{++}(i^{--},j^{++}) = - \frac{z^3}{(1-z)} \frac{[i\,j]}{\langle i \,j \rangle} ,  \\
  {\rm Split}^{\rm GR}_{++}(i^{++},j^{++}) &= 0  ,
\end{split}
\end{align}
where the missing cases can be obtained from parity, or simply vanish. The ratio of spinor brackets appearing in the splitting amplitudes is manifestly a phase in real Minkowski space, but in complex Minkowski space the brackets are independent and if the collinear limit is taken as $\langle ij \rangle \to 0$,  the ratio becomes singular.

The gravity splitting amplitudes can be derived easily using the  fact that the three-graviton amplitudes are simply  squares of the corresponding three-gluon amplitudes leading to a simple relation between graviton and gluon splitting amplitudes \cite{Bern:1998sv}
\begin{align}
{\rm Split}_{\pm\pm}^{\rm GR}(i^{2 h_i},j^{2 h_j}) = s_{ij} \left[{\rm Split}_{\pm}(i^{h_i},j^{h_j})\right]^2
\ ,
\end{align}
where the Yang-Mills splitting amplitudes are given in \eqref{collsplit}.

\subsubsection{Soft limits}
As already mentioned in Section~\ref{sec:slYMBCFW}, the leading \cite{Weinberg:1965nx},   subleading and sub-subleading \cite{Cachazo:2014fwa} soft limits of gravity amplitudes are universal.%
\footnote{The sub-subleading soft factor quoted here is for Einstein-Hilbert theory. In general it can receive additional, theory-dependent corrections  \cite{Laddha:2017ygw}.}
These   can be obtained using the four-dimensional BCFW recursion relation \cite{Cachazo:2014fwa}, with the result 
\begin{align}
\cM_{n} {\buildrel p_s \to 0 \over
{\relbar\mskip-1mu\joinrel\longrightarrow}}  \left(\frac 1 \delta \, \cS^{(0)}_{\text{grav}}(q)+ \cS^{(1)}_{\text{grav}}(q)
+ \delta\,  \cS^{(2)}_{\text{grav}}(q)\right)\, \cM_{n-1}
+ \cO(\delta^{2}) \, .
\end{align}
where, for a positive-helicity soft graviton  $s^+$,
\begin{align}
\begin{split}
\label{singlesoftgrp}
\cS^{(0)}_{\rm grav} (s^{+}) &= \sum_{a}\frac{[s a]}{\langle s a\rangle}\frac{\langle x a\rangle}{\langle x s \rangle} \frac{\langle y a\rangle }{\langle y s \rangle}\ , \qquad 
\cS^{(1)}_{\rm grav}(s^{+}) = \frac{1}{2}\sum_{a}\frac{[s a]}{\langle s a\rangle}\left(\frac{\langle x a\rangle}{\langle x s \rangle} + \frac{\langle y a\rangle }{\langle y s \rangle}\right){\tilde \lambda}^{\dot \alpha}_{s} \frac{\partial}{\partial{\tilde \lambda}_a^{\dot \alpha}}\, .
\\ 
\cS^{(2)}_{\rm grav}(s^{+}) &= \frac{1}{2} \sum_{a} \frac{[s a]}{\langle s a\rangle} {\tilde \lambda}_s^{\dot \alpha} {\tilde \lambda}_s^{\dot \beta} \frac{\partial^2}{\partial {\tilde \lambda}_a^{\dot \alpha} \partial {\tilde \lambda}_a^{\dot \beta}} \ . 
\end{split}
\end{align}
The sum over $a$ is over the remaining $n{-}1$ particles, and 
 $|x\ran $ and $|y\ran$ are  reference spinors. 
 The soft factors for the case where $s$ has negative helicity can be found by conjugation. 
 $\cS^{(0)} $ is the famous Weinberg soft factor \cite{Weinberg:1965nx}, and we also quote below  expressions for the soft factors
 valid in any dimension in terms of  polarisation tensors:  %
 \begin{align}
 \begin{split}
\cS^{(0) }  & =      \sum_a  \frac{k^\mu_a \eps_{\mu \nu}(s) k^\nu_a }{k_a \cdot p_s}
\ ,      \qquad \qquad \quad 
\cS^{(1)} = - i \sum_a 
\frac{k^\mu_a \eps_{\mu \nu}(s) J^{\nu \rho}_a k_{s\rho} }{k_a \cdot p_s}\ , 
\\
\cS^{(2)} &  = - \frac{1}{2} 
\sum_a 
\frac{ \eps_{\mu \nu}(s) k_{s \rho} J^{\mu \rho}_a k_{s\sigma}J^{\nu \sigma}_s }{k_a \cdot p_s}\ , 
\end{split}
 \end{align}
 where $J^{\mu \nu}_a = L^{\mu \nu}_a + \Sigma^{\mu \nu}_a $, and 
 $
     L^{\mu \nu}_a = i \Big( k^\mu_a \frac{\partial}{\partial k_{a \nu}} - k^\nu_a \frac{\partial}{\partial k_{a \mu}}
     \Big)$, 
$     \Sigma^{\mu \nu}_a = i \Big( \eps^\mu_a \frac{\partial}{\partial \eps_{a \nu}} - \eps^\nu_a \frac{\partial}{\partial \eps_{a \mu}}
     \Big) $. 
We also mention that soft theorems beyond leading order can be elegantly derived  from gauge  invariance 
\cite{Broedel:2014fsa,Bern:2014vva}. Finally, it is interesting to note that 
double soft limits are also universal, and corresponding theorems can be established, with the simultaneous and consecutive limits leading to  different types of  universal behaviour \cite{Klose:2015xoa,Volovich:2015yoa}.



\setcounter{footnote}{0}


\section{Supersymmetric amplitudes }
\label{sec:8}

\subsection{Generalities}

\label{sec:generalities-super}

The spectrum of maximally supersymmetric  $\cN{=}4$ SYM  theory contains the following states:%
\footnote{See e.g.~\cite{DHoker:2002nbb} for a review.} 
\begin{itemize}
    \item 
two gluons $G^{\pm} (p)$ with helicities $1, -1$, 
\item
four Weyl  fermions $\psi_A$ with  helicity $+1/2$,  transforming in the fundamental of the $R$-symmetry group $SU(4)_{R}$, and  four Weyl fermions $\bar\psi^A$ with helicity $-1/2$  in the anti-fundamental representation,  with 
$A=1,\ldots , 4$,
and 
\item 
six real scalar fields (corresponding to particles of zero helicity)  $\phi_{[AB]}$ in the antisymmetric tensor representation of the $R$-symmetry group ($A, B=1,\ldots , 4$). \end{itemize}
One can then   combine the states into  an on-shell superfield \cite{Nair:1988bq}
\begin{align}
\label{superNair} 
\Phi (\eta, p) \, := \, G^+(p) + \eta^A \psi_A (p)  + \frac{\eta^A \eta^B}{ 2!} \phi_{[AB] } (p) + 
\eps_{ABCD}  \frac{\eta^A \eta^B \eta^C}{ 3!}  \bar{\psi}^D(p)  + \eta^1\eta^2 \eta^3 \eta^4
 G^{-} (p) 
 \ ,  
 \end{align}
 where the $\eta^A$ are 
 four auxiliary  Gra{\ss}mann variables. For each particle, the coordinates $(\lambda, \lt, \eta)$ parameterise an on-shell superspace \cite{Ferber:1977qx}. 
The supersymmetry generators  $q^A$ and  $\bar{q}_A$ satisfy the algebra $\{ q^A_\alpha, \bar{q}_{B \dot{\alpha}} \} =  \lambda_\alpha \lt_{\dot \alpha}\, \delta^A_{\, B}$, and have a natural realisation on this superspace  as $q^A_\alpha = \lambda_{\alpha} \eta^A$, 
$\bar{q}_{A \dot{\alpha}}  = 
\lt_{\dot{\alpha}} \frac{\partial}{\partial \eta^A} $, or, for $n$ particles, 
\begin{align}
\label{qqbardef}
q^A_\alpha \ = \ \sum_{i=1}^n \lambda_{i\alpha} \eta^A_i
\ , 
\qquad 
\bar{q}_{A \dot{\alpha}}  \ = \ 
\sum_{i=1}^n \lt_{i\dot{\alpha}} \frac{\partial}{\partial \eta^A_i} 
\ . 
\end{align}
The next step is to combine all amplitudes with a   given  number of particles $n$ and fixed total helicity   into a   {\it superamplitude}%
\footnote{Not to be confused with the complete amplitudes of Section~\ref{sec:3}, traditionally denoted in the same way.}
$\cA$. This superamplitude  can  then 
 be expanded in powers of the $\eta^{A}_i$s, with each  coefficient of the expansion being a   component amplitude. A term containing $k_i$  powers of $\eta_i$  corresponds  to an amplitude  where the $i^\mathrm{th}$ particle has helicity  $h_i = 1 - k_i/2$, with the  total helicity  being $\sum_{i=1}^n h_i$. In other words, to get an amplitude  with helicity $h_i$ for particle $i$ we need to pick the term containing $2 - 2h_i$ powers of $\eta_i$ in the superamplitude.


\setcounter{footnote}{0}


Superamplitudes  are invariant under the $q$ and $\bar{q}$ supersymmetries, in addition to being invariant under translations. The latter symmetry is implemented by pulling out a  $\delta$-function of total momentum conservation $\delta^{(4)}(p)$, with $p{:=}\sum_{i=1}^n \lambda_i \lt_i$, and similarly we can 
realise the $q$-supersymmetry manifestly   via  a $\delta$-function of supermomentum conservation.%
\footnote{The three-point case is special and will be discussed  in \eqref{3ptmhvbar}.}
Summarising, we will set 
\begin{align} 
\label{superAdef}
\cA_n  :=  
\delta^{(4)} (p) \delta^{(8)}(q) \ A_n \ ,  
\end{align}
where 
$q{=}\sum_{i=1}^n \eta_i \lambda_i$ is  the total   supermomentum. It is then easily checked that invariance under  $\bar{q}$ supersymmetry implies  that 
$\bar{q} A_n {=}0$ on the support of the two $\delta$-functions.

\subsection{MHV and NMHV superamplitudes}
\label{sec:ex-super-ampl}

Our first example is  the MHV superamplitude. Its elegant  expression was  given in \cite{Nair:1988bq}:
\begin{align}
\label{superMHV}
\cA_n^{\mathrm{MHV}}(1, \ldots, n)  \ = \ i\, g^{n-2}\,   \frac{ \delta^{(4)} (p) \, \delta^{(8)} (q)}{\lan 12\ran \lan 23\ran \cdots \lan n1\ran }
\ .  
\end{align}
From this it is easy to extract component amplitudes as outlined in the previous section.
For instance, 
the  MHV amplitude  with negative helicity gluons $i^-$ and $j^-$ can be extracted  as  the coefficient of 
$\eta_i^4 \eta_j^4$ in the expansion of 
\eqref{superMHV}, 
leading to%
\footnote{A useful formula is $\delta^{(8)} (\lambda_1 \eta_1 + \lambda_2 \eta_2 + \cdots) = \lan 12\ran^4 \prod_{A=1}^4 \eta_1^A \eta_2^A + \cdots$.}
 \begin{align}
A_n^{\rm MHV}(1^+, \dots, i^-, \dots, j^-, \dots, n^+)=ig^{n-2}\, \frac{\langle{ij\rangle}^4}{\langle 12\rangle\langle 23\rangle \cdots  \langle n1\rangle}\, .
\nonumber
 \end{align}
 Recall that we derived this for neighboring $\{i,j\}=\{n,1\}$ in Section~\ref{sect:MHVfrom BCFW}.
 
Next we  consider the  NMHV superamplitudes. These have the form 
 \cite{Drummond:2008vq, Drummond:2008bq}
\begin{align}
\label{superNMHV}
  \cA_n^{\mathrm{NMHV}} =    \cA_n^{\mathrm{MHV}} \sum_{u,v=i+2}^{i+n-1} R_{i u v}
  \ , 
\end{align}
where the functions   $ R_{rst}$ are  defined as
\begin{align}
\label{Rinv}
\begin{tikzpicture}[thick, scale=0.7]
 \node[left] at (-2.5,0) {$R_{rst}:=$};
  \drawLLwhite
  \drawUL{}
  \drawUR{}
  \drawLR{}
   \drawregionvariables{$x_r$}{$x_{r+1}$}{$x_s$}{$x_t$}{}
  \drawboxinternallines
  \draw (UL) -- ++( 180:0.8) node[anchor=east] {$r+1$};
  \draw[dotted] (UL)+( 170:0.6) to [bend left=45] ++( 100:0.6);
  \draw (UL) -- ++(  90:0.8) node[anchor=south] {$s-1$};
  \draw (UR) -- ++( 0:0.8) node[anchor= west] {$t-1$};
  \draw[dotted] (UR)+(  90:0.6) to [bend left=45] ++(  0:0.6);
  \draw (UR) -- ++(  90:0.8) node[anchor=south] {$s$};
  \draw (LR) -- ++(   0:0.8) node[anchor=west] {$t$};
  \draw[dotted] (LR)+( -10:0.6) to [bend left=45] ++( -80:0.6);
  \draw (LR) -- ++( -90:0.8) node[anchor=north] {$r-1$}; 
  \draw (LL) -- ++(-135:0.8) node[anchor=north east] {$r$};
  \node[right] at (2.5,0) {$\displaystyle=\frac{\langle s-1\,s\rangle \langle t-1\,t\rangle \, \delta^{(4)}\big(
  \Xi_{r st}
  \big)
  }{x_{st}^2 \langle r|x_{rt}x_{ts}|s-1\rangle  \langle r|x_{rt}x_{ts}|s\rangle  \langle r|x_{rs}x_{st}|t-1\rangle  \langle r|x_{rs}x_{st}|t\rangle }$}; 
\end{tikzpicture}
\end{align}
and 
$\Xi_{r st}\!:=\!\langle r|x_{rs}x_{st}|\theta_{tr}\rangle  + \langle r|x_{rt}x_{ts}|\theta_{sr}\rangle$. 
Here we have introduced the so-called  dual, or region (super)momenta%
\footnote{See Section~\ref{sec:DSI} for a discussion of such quantities in the context of dual superconformal invariance.} 
$x_i$  and $\theta_i$, defined  $\lambda_i \lt_i:= x_i - x_{i+1} $, $\lambda_i \eta_i := \theta_i - \theta_{i+1}$, so that $x_{ij} = \sum_{k=i}^{j-1} \lambda_k\lt_k $, $\theta_{ij} = \sum_{k=1}^{j-1} \lambda_k \eta_k$, with $x_{n+1} = x_1$, 
$\theta_{n+1} = \theta_1$. We also showed a convenient diagrammatic notation for the invariants introduced in \cite{Bern:2004bt}.
In Section~\ref{sec:dss-ex} we will prove that the NMHV is dual superconformal covariant.

\subsection{Supersymmetric BCFW recursion relation}
\label{sec:susyBCFW}

We now discuss how to supersymmetrise the BCFW recursion relation
of Section~\ref{sec:der-BCFW} \cite{Brandhuber:2008pf,Arkani-Hamed:2008owk}. 
As in the non-supersymmetric case, we construct amplitudes recursively starting from 
 two three-point superamplitudes: the first one has the total MHV helicity, and is given by \eqref{superMHV} for $n\!=\!3$, 
while the  three-point  ${\rm \overline{MHV}}$ superamplitude is  \cite{Brandhuber:2008pf, Arkani-Hamed:2008owk}
\begin{align}
\label{3ptmhvbar}
\mathcal{A}^{\overline{\mathrm{MHV}}}_3  \ = \ -i\, g\,  
\delta^{(4)} (p_1 + p_2 + p_3 )\, 
\frac{ \delta^{(4)} ( \eta_1 [23] \, + \, \eta_2 [ 31] \, + \, \eta_3 [12] ) }{ [12] \, [23]\, [31] } 
\ . 
\end{align} 
It was shown in  \cite{Brandhuber:2008pf} that,  despite its slightly unusual supersymmetric delta function,  the  ${\rm \overline{MHV}}$ superamplitude is  invariant under supersymmetry, as well as covariant under the  dual superconformal symmetry of  \cite{Drummond:2008vq}.  

 Similarly to the discussion of Section~\ref{sec:4.1},  three-point superamplitudes    can  be determined from symmetry considerations alone up to an overall normalisation. 
For instance, the form of the three-point MHV superamplitude
 can be fixed   by requiring that it depends only on the holomorphic spinors $\lambda_1, \lambda_2, \lambda_3$ and satisfies the relations
$
\hat{h}_ i\,  \cA^{{\mathrm{MHV}}}_3  {=} \cA^{{\mathrm{MHV}}}_3$,   
$i = 1, 2, 3$, 
where 
\begin{align}
\hat{h}_i :=   \frac{1}{ 2}  
\left( 
-  \lambda_{i}^{\alpha} \frac{\partial}{ \partial \lambda_{i}^{ \alpha}}  +  \lt_{i}^ { \dot{\alpha}}    \frac{\partial}{  \partial \lt_{i}^{\dot{\alpha}}} + 
\eta^A_i \frac{\partial}{ \partial \eta^A_i}\right)
\ ,
\end{align}  
which express the fact that the on-shell superfield \eqref{superNair} has helicity $+1$.

\subsubsection{Derivation}

We now derive the supersymmetric recursion relation. We begin by observing that in order to maintain supersymmetry we must accompany the momentum shifts by a   supermomentum shift. 
The  following (super)shifts 
\begin{align}
\label{supershift}
\hat{\lt}_1 (z) := \lt_1 + z \lt_2 \ , \quad
\hat{\lambda}_2 (z) = \lambda_2 - z \lambda_1 \ , \quad 
 \hat{\eta}_1 (z) = \eta_1+ z \eta_2 \ , 
 \end{align}
manifestly preserve (super)momentum conservations and  the on-shell conditions. As in the non-supersymmetric case, we define a one-parameter family of superamplitudes, 
\begin{align}
\label{superaofz}
\cA_n (z) := \cA_n (\{\lambda_1, \hat{\lt}_1, \hat{\eta}_1\}, \{\hat{\lambda}_2, \lt_2, \eta_2\}, \ldots )
\ , 
\end{align}
where the dots denote the unshifted (super)momenta of  the remaining $n{-}2$ particles. 
The derivation of the  recursion relation parallels  that of its non-supersymmetric cousin, with the result  \cite{Brandhuber:2008pf,Arkani-Hamed:2008owk}
\begin{align}
\label{superampli}
\cA_n \ = \ 
\sum_{P} 
\int\!\!d^4\eta_{\hat{P}} \  \cA_L(z_{P})
 \frac{i}{P^2} \cA_R(z_{P})
 \ , 
\end{align}
where $\eta_{\hat{P}}$ is the Gra{\ss}mann  coordinate associated to the internal particle with momentum $\hat{P}$. The  sum is over  all diagrams where   such that  the shifted momenta belong to different  superamplitudes. The two superamplitudes in \eqref{superampli} are  computed  on the solution $z_P$ of  $\hat{P}^2 (z)=0$, with  $\hat{P} (z) := P + z \lambda_1 \lt_2$. 
Note that in  \eqref{superampli}
 the
total helicities  of $\cA_L$ and  $\cA_R$ must sum to the total helicity of $\cA$.

The derivation of  \eqref{superampli} rests on the important  fact that $\cA_n(z) \to 0$ as $z\to \infty$ 
\cite{Brandhuber:2008pf, Arkani-Hamed:2008owk}. Specifically, we will now show that 
\begin{align} 
\label{behsup}
\mathcal{A}_n^{\cN=4} (z) \stackrel{z\to \infty }{\sim}  \frac{1}{z} \ , \qquad 
\cA_n^{\cN=8} (z)\stackrel{z\to \infty }{\sim}  \frac{1}{z^2} 
\ .
\end{align}
To do so, we  note that in the maximally supersymmetric $\cN{=}4$ SYM or $\cN{=}8$ supergravity theories,  we have enough   supersymmetry transformations 
to set to zero {\it two} of the $\eta^A$ variables in the superamplitude $\cA_n( \lambda_1, \lt_1, \eta_1; \lambda_2, \lt_2, \eta_2; \ldots ; \lambda_n, \lt_n, \eta_n)$, for instance 
$\eta_1$ and $\eta_2$. 
We can  then  determine the $2 \cN$ parameters $ \zeta^{\dot{\alpha}}_B$
in a    generic   $\bar{q}$ supersymmetry transformation  $\bar{q}_\zeta := \zeta^{\dot{\alpha}}_B \bar{q}^B_{\dot{\alpha}}$, with  $B=1, \ldots , \cN$,  in such a way that 
$
e^{\bar{q}_\zeta} \eta_1^A = e^{\bar{q}_\zeta} \eta_2^A = 0
$,
that is 
$\bar{q}_\zeta \eta_{1,2}^A = - \eta_{1,2}^A $.
The solution is 
\begin{align}
\label{solsp}
\zeta_{\dot{\alpha}}^A \ = \frac{1}{ [12]} \big(- \lt_{1 \dot{\alpha}} \eta_2^A +  \lt_{2 \dot{\alpha}} \eta_1^A\big)
\ , 
\end{align}
and  the action on the remaining $n-2$ Gra{\ss}mann variables  $\eta_i$ is 
\begin{align}
\label{etap}
e^{\bar{q}_\zeta}  \eta_i \ := \ \eta_i^\prime \ = \ \eta_i - \eta_1 
\frac{ [i\, 2]}{[1\, 2]} + \eta_2 \frac{[i\, 1]}{ [1\, 2]}
\ .
 \end{align}
As we have seen in Section~\ref{sec:generalities-super}, supersymmetry invariance of a  superamplitude
implies that $\delta^{(4)} (p) \delta^{(2\cN)} (q) \, [\bar{q}^B_{\dot\beta} A_n] = 0 $, hence 
 $e^{\bar{q}_\zeta} A_n = A_n$ on the support of the delta functions. Acting with the $\bar{q}$ operator explicitly, we get
 \begin{align}
 \label{remove}
& \, \cA_n (\lambda_1, \lt_1, 0; \lambda_2, \lt_2, 0; \lambda_3 , \lt_3, \eta_3^\prime; \ldots ) 
\, = \, \cA_n (\lambda_1, \lt_1, \eta_1; \lambda_2, \lt_2, \eta_2; 
\lambda_3 , \lt_3, \eta_3; \ldots )
 \ , 
 \end{align}
with  $\eta_i^\prime$ defined as in \eqref{etap} (for all $i{=}3, \ldots , n)$. 
We can now use \eqref{remove} to  prove that our superamplitudes $\cA_n (z)$ defined in \eqref{superaofz}  have the large-$z$ behaviour advertised in   \eqref{behsup}.    
The key observation is that the supersymmetry transformation that sets $\eta_1 (z) $ and $\eta_2$ to zero is  $z$-independent: indeed, using \eqref{solsp} and \eqref{supershift} we see that  
\begin{align}
\zeta_{\dot{\alpha}}^A =   \frac{- \hat{\lt}_{1 \dot{\alpha}} \eta_2^A +  \lt_{2 \dot{\alpha}} \hat{\eta}_1^A}{ [12]}  \ = \ 
  \frac{- {\lt}_{1 \dot{\alpha}} \eta_2^A +  \lt_{2 \dot{\alpha}} {\eta}_1^A}{ [12]} 
\ . 
 \end{align}
As a result  $
\cA_n(z) {=} \cA_n( \lambda_1, \hat{\lt}_1, 0; \hat{\lambda}_2, \lt_2, 0; \ldots ; \lambda_i, \lt_i, \eta_i^\prime; \ldots ; \lambda_n, \lt_n, \eta_n^\prime)$, 
where crucially  none of the  $\eta_i^\prime$   contain $z$:  the only $z$-dependence  occurs through $\hat{\lt}_1$ and $\hat{\lambda}_2$. The large-$z$  behaviour of  $\mathcal{A}_n(z)$ is then identical to that of  a gluon (or graviton)  amplitude where particles  1 and 2 have   positive helicity.  Such  amplitudes  fall off as $1/z$ at large $z$  for Yang-Mills  theory \cite{Britto:2005fq}, or   $1/z^2$ in gravity  \cite{Arkani-Hamed:2008bsc}, thus proving~\eqref{behsup}.

 \subsubsection{Application to  MHV superamplitudes}
 
\begin{figure}[ht]
\begin{center}
\scalebox{0.45}{\includegraphics{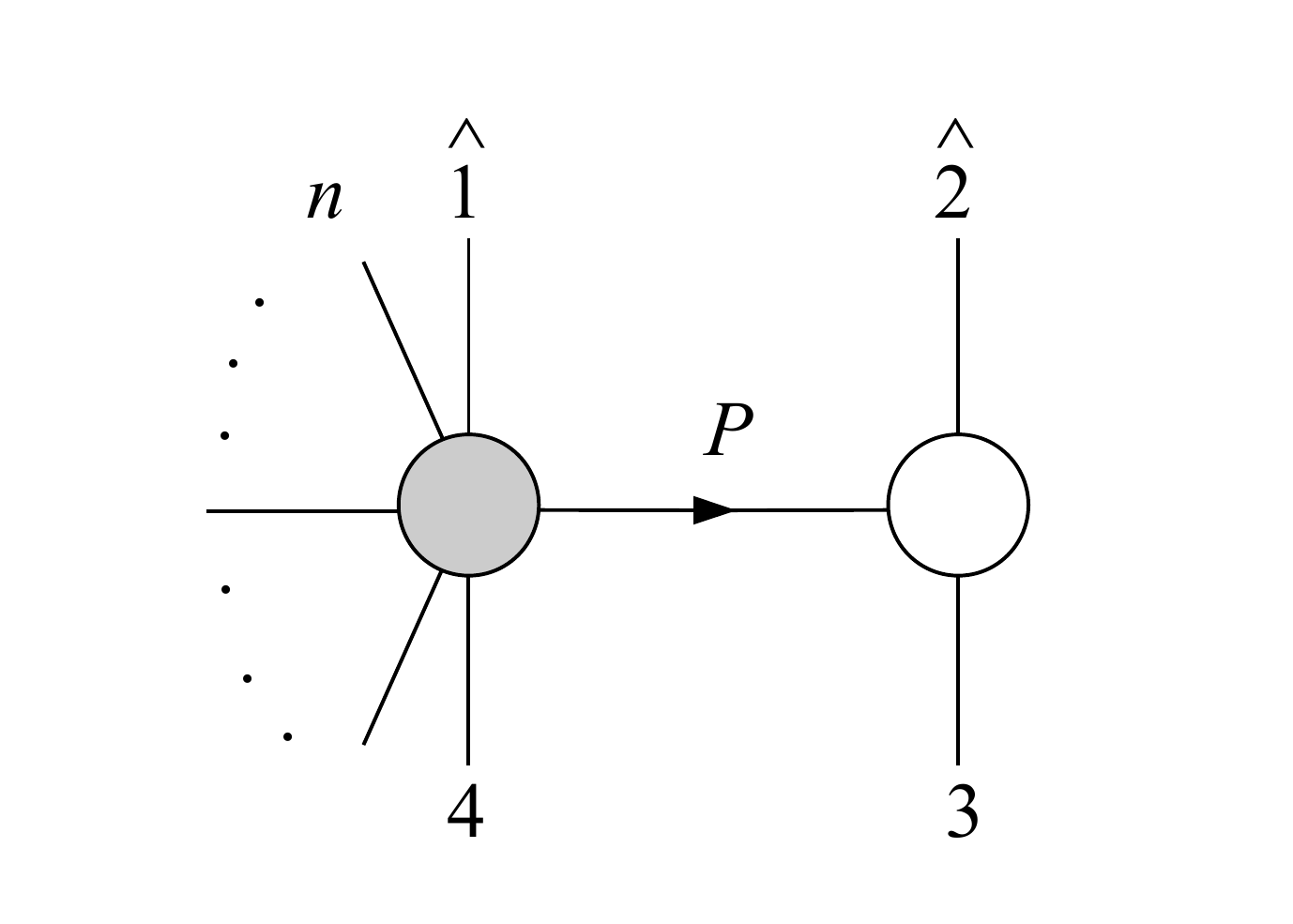}}
\end{center}
\vspace*{-10mm}
\caption{\it 
BCFW diagram  
for the  $n$-point 
MHV  recursion. In the derivation we  use a $[12\ran$ shift.
}
\label{fig:MHVsuper}
\end{figure}

 We now use the supersymmetric recursion relation of   \cite{Brandhuber:2008pf,Arkani-Hamed:2008owk}  to derive the MHV superamplitude
 \eqref{superMHV}.
With  the supershifts in \eqref{supershift},  there is a single recursive diagram to consider, shown   in Figure~\ref{fig:MHVsuper}.
 The right-hand side is always a three-point ${\rm \overline{MHV}}$ superamplitude, whereas that  on the left-hand side is an $(n{-}1)$-point  {\rm MHV} superamplitude. We will assume that the latter has the form given in   \eqref{superMHV}, and  then derive it for $n$ points using the recursion. Starting from $n{=}4$ this gives of course a derivation of the superamplitude at any $n$. 
The building blocks in the  supersymmetric recursion  \eqref{superampli} are then  
 \begin{align}
\cA_L & = i g^{n-3} \, \delta^{(4)} \Big (\sum_{i=4}^n p_i+  \hat{p}_1 + \hat{P} \Big) \, \frac{
\delta^{(8)} (   \sum_{i=4}^n q_i +  \lambda_1 \hat{\eta}_1  +  \eta_{\hat{P}} \lambda_{\hat{P}}  ) 
}{ 
\lan 1 \hat{P} \ran \lan \hat{P}  4 \ran \cdots \lan n 1 \ran 
} 
\ , 
\\ \nonumber
\cA_R & =   -ig\, \frac{\delta^{(4)} ( \hat{p}_2 + p_3 - \hat{P} ) \, \delta^{(4)} ( \eta_{-\hat{P}} [23] + \eta_2 
[3  -\!\!\hat{P}] + \eta_3 [-\!\hat{P} \, 2]   ) }{
[-\!\hat{P}\, 2] [23] [3\, -\!\hat{P}]
} 
\ .
\end{align}
Using   the identity
\begin{align}
&\ \delta^{(8)} \big( \hat{\eta}_1 \lambda_1 + \sum_{i=4}^n \eta_i \lambda_i + \eta_{\hat{P}} \lambda_{\hat{P}}  \big) \, 
\delta^{(4)} ( \eta_{-\hat{P}} [23] + \eta_2 
[3  -\!\!\hat{P}] + \eta_3 [-\!\hat{P} \, 2]   )
\nonumber \\ 
&= \ 
\delta^{(8)} \big( \sum_{i\in L, R} \hat{\eta}_i \hat{\lambda}_i \big) \, 
\delta^{(4)} ( \eta_{-\hat{P}} [23] + \eta_2 
[3  -\!\!\hat{P}] + \eta_3 [-\!\hat{P} \, 2]   ) \ , 
\end{align}
and (super)momentum conservation  
$
\sum_i \hat{\eta}_i \hat{\lambda}_i {=} \sum_i \eta_i \lambda_i$, $\sum_i \hat{p}_i {=} \sum_i p_i$, 
we arrive at the result 
$
\cA_n =
 \delta^{(4)} \big(\sum_{i\in L,R} p_i\big) \,
\delta^{(8)} \big( \sum_{i\in L, R} \eta_i \lambda_i \big)  \, A_n
$,
where 
\begin{align}
A_n =
\frac{ig^{n-2}}{ P_{23}^2} \, \frac{1}{ \lan45\ran \cdots \lan n1\ran  [23] \, 
\lan 1 \hat{P}\ran \lan \hat{P} 4\ran [ {-\hat{P}} 2 ] [ 3 \,{-\hat{P}}] 
} \, 
\int\!d^4\eta_{\hat{P}} \, 
\delta^{(4)} ( \eta_{-\hat{P}} [23] + \eta_2 
[3 \hat{P}] + \eta_3 [\hat{P} 2]   ). 
\end{align}
It is straightforward  to see that 
$\lan 1 \hat{P}\ran \lan \hat{P} 4\ran [ -\hat{P} 2 ] [ 3\, {-\hat{P}}] {=}-\lan1|2|3] \lan 4|3|2]{=}-
\lan12\ran \lan34\ran [23]^2$, finally obtaining 
\begin{align}
A_n \ = \  \frac{ ig^{n-2} }{\lan 1 2\ran \lan  23\ran \cdots \lan n1\ran}
\ . 
\end{align}
 We also note that  the supersymmetric recursion relation was  solved in closed form in  \cite{Drummond:2008cr}. 

\subsection{Vanishing Yang-Mills amplitudes}
\label{sec:vanishing}
$n$-gluon amplitudes with $n{>}3$  where all or all but one of the gluons have the same helicity are zero at tree level in any theory.%
\footnote{And to all loops in the presence of supersymmetry, see e.g.~\cite{Dixon:1996wi} for a proof.} Intriguingly, one can derive this fact using supersymmetry: at tree level $\cN\!=\!4$ SYM  has the same gluon  amplitudes of pure Yang-Mills; because of the $\delta^{(8)}$ of supermomentum conservation, the first non-vanishing amplitudes must have at least two negative-helicity gluons (providing  each  four powers of $\eta$), except for the three-point case \eqref{3ptmhvbar} which is quartic in $\eta$. Pleasingly,  supersymmetry can be used to make powerful statements on non-supersymmetric amplitudes!

\section{Superconformal, dual superconformal and Yangian symmetries}
\label{sec:9}

As mentioned in the introduction,  scattering amplitudes in $\mathcal{N}{=}4$ SYM are remarkably  simple. Thanks to the finiteness of the theory \cite{Mandelstam:1982cb} they are ultraviolet finite, and furthermore they are constrained by several symmetries. Some of these are symmetries of the  Lagrangian  -- the standard superconformal symmetry group -- but in addition there are symmetries which are visible only in the $S$-matrix of the theory: the dual superconformal and Yangian symmetries. In the next sections we present a snapshot of these symmetries, and describe some of their consequences on the $S$-matrix of $\mathcal{N}{=}4$~SYM.

\subsection{Superconformal symmetry}

We introduced the supersymmetry generators $q^{A}_{\alpha}$ and $\bar{q}_{A \dot{\alpha}}$ of
 $\mathcal{N}{=}4$ SYM  
in \eqref{qqbardef}, where we saw that  they leave the superamplitude invariant by virtue of the
supermomentum conserving delta function $\delta^{(8)}(q)$ of \eqref{superAdef}.
In  the presence of conformal symmetry, the commutator of a special conformal
and the supersymmetry generators introduces a set of new Gra{\ss}mann-odd generators known as superconformal  generators,  $s$ and $\bar{s}$:
\begin{align}
[k_{\alpha\da} , q^{\beta\, A} ] &= \delta_\alpha^\beta\, \bar s_\da^A\, , \quad\qquad
\bar s_\da^A =\eta^A\, \tilde\partial_\da \, , &\nonumber\\
[k_{\alpha\da} , \bar q^{\db}_{A} ] &= \delta_\da^\db\, s_{\alpha\, A} \, ,  \qquad
s_{\alpha\, A} = \partial_\alpha\, \partial_A\, .&
\end{align}
The complete $\mathcal{N}\!=\!4$ superconformal symmetry algebra finally takes the form 
\begin{align}
\begin{split}
&\{ q^{\alpha\, A}, \bar q^\da_B\}  = \delta^A_B\, p^{\alpha\da}\, , \qquad\qquad
\{ s_{\alpha\, A}, \bar s_\da^B\}  = \delta^A_B\, k_{\alpha\da} \\
&\{ q^{\alpha\, A}, s_{\beta\, B} \} = m^\alpha{}_\beta\, \delta^A_B +\delta^\alpha_\beta\, r^A{}_B + \frac 1 2 \,
\delta^\alpha_\beta\, \delta^A_B\, (d+c) \\
&\{ \bar q^{\da}_A, \bar s_{\db}^B \} = {\overline m}^\da{}_\db\, \delta^A_B -\delta^\da_\db\, r^B{}_A 
+ \frac 1 2 \,\delta^\da_\db\, \delta^B_A\, (d-c) 
\\
&[p^{\alpha\da}, s_{\beta\, A}] = \delta^\alpha_\beta\, \bar q^\da_A\, ,\qquad \qquad \quad [p^{\alpha\da}, \bar s^A_\db ]
= \delta^\da_\db\, q^{\alpha\, A}\, ,
\label{psu224alg}
\end{split}
\end{align}
with the central charge 
$
c=1+\frac 1 2\, ( \la^\alpha\partial_\da - \tla^\da\partial_\da -\eta^A\partial_A) = 1-h
$
as well as an additional
global $\alg{su}(4)$ $R$-symmetry generator $r^A{}_B$ %
\begin{align}
r^A{}_B&=\eta^A\, \partial_B -\frac 1 4 \delta^A_B\, \eta^C\,\partial_C\, , \qquad
\partial_A:=\frac{\partial}{\partial \eta^A}\, ,
\end{align}
which acts as an internal rotation in $\eta^A$-space. This superalgebra is known as $\alg{psu}(2,2|4)$.

\subsection{Dual superconformal symmetry}

\label{sec:DSI}

Remarkably, the $\mathcal{N}{=}4$ SYM theory enjoys an additional hidden invariance  known as dual superconformal symmetry.
To make this symmetry  manifest, one has to parameterise the momenta and supermomenta of the scattered particles in terms of  dual momenta $x_i$ and supermomenta $\theta_i$. These are defined as  
\begin{align}
\label{constraintssuperspace}
p_{i \alpha \dot{\alpha}} = \lambda_{i \alpha}\lt_{i \dot{\alpha}} =   (x_{i}- x_{i+1})_{\alpha \dot{\alpha}}
\ , \qquad 
\eta_i^A \lambda_{i\alpha} = \theta_{i\alpha}^{A} - \theta_{i+1\, \alpha}^{A} \ ,
\end{align}
and we require that $x_{n+1} {=} x_1$ and $\theta_{n+1} {=} \theta_1$.
Note that one can 
make consistent assignments for the region momenta only for   planar diagrams. An advantage of this parameterisation is that  momentum conservation is  automatic: the only constraint on the $x_i$s is  the on-shell conditions $(x_i {-} x_{i+1})^2{=}0$, while the fermionic variables $\theta_i$ must also satisfy the on-shell condition  
$
(\theta_i - \theta_{i+1} ) \lambda_i {=} 0 
$.
Momentum and supermomentum conservation are then implemented with the delta functions $\delta^{(4)} (x_1 - x_{n+1}) \, \delta^{(8)} (\theta_1 - \theta_{n+1})$. 

Without spoiling momentum and supermomentum conservation, we can then act with inversions on the dual momenta and supermomenta~\cite{Drummond:2008vq}: 
\begin{align}
\label{LTBV}
x_{\alpha \dot{\beta}}  & \to \ I[ x_{\alpha \dot{\beta}}] = \frac{x_{\beta \dot{\alpha}} }{ x^2} := x_{\beta \dot{\alpha}}^{-1}
\ ,  \qquad 
\theta^{A \alpha}  \to \ I[\theta^{A \alpha}] =  (x^{-1})^{\dot{\alpha} \beta} \theta_{\beta}^{A}
\ . 
\end{align}
This transformation makes sense since  dual momenta, unlike the momenta,  are unconstrained. It is also important  that  dual conformal inversions do not change the lightlike nature of a momentum --  this is indeed one of the claims to fame of the conformal group: $(\frac{x^\mu}{x^2} - \frac{y^{\mu}}{y^2})^2 =0$ if $(x-y)^2=0$. Note that \eqref{LTBV} implies that 
\begin{align}
\label{trxij}
    I[ (x_{ij})_{\alpha \dot{\beta}}] = - (x_j^{-1} x_{ij} x_i^{-1})_{\beta \dot{\alpha}} \ ,  
\end{align}
with $x_{ij} {:=} x_i {-}x_j$, and in particular $I[ x_{ii{+}1}] = - x_{i{+}1}^{-1} x_{ii{+}1} x_i^{-1}$.  In order to determine what is $I[\lambda^{\alpha}]$, we note that we want to preserve  the constraint $\lambda^{\beta} (x_{i i+1})_{\beta \dot{\alpha}}=0 $. It then follows that  $  (x_{i{+}1}^{-1} x_{ii{+}1} x_i^{-1})_{\alpha \dot{\beta}}\, I[\lambda^{\beta}]=0 $, 
which can be solved by choosing  
\begin{align}
\label{spintran}
\lambda_{i}^{\beta}  \to I[\lambda_{i}^{\beta}] = (x_i^{-1})^{\dot{\beta}\alpha}  \lambda_{i\alpha}\, .  
\end{align}
This  also implies that 
\begin{align}
\label{trbrij}
    \lan i\,  i{+}1\ran \to \frac{\lan i\,  i+1 \ran }{ x_{i}^2}\ ,  \end{align}
as it can be seen after using $ x_{i i{+}1} | i\ran =0$. 
The transformation  of $\lt_i$ under an inversion can be found by noticing that  from  $\lambda_i \lt_i {=} x_{i i{+}1}$ it follows that  $\lt_i^{\dot{\alpha}} \!=\!x_{i i{+}1}^{\dot{\alpha} \beta}\lambda_{i{+}1 \beta}/ \lan i i{+1}\ran $. Using then  \eqref{trxij}, \eqref{spintran} and \eqref{trbrij} one quickly arrives at 
\begin{align}
    \lt_i^{\dot{\alpha}} \to I[\lt_i^{\dot{\alpha}}] = \lt_{i \dot{\beta}} (x_{i+1}^{-1})^{\dot{\beta} \alpha}\ . 
\end{align}
Special conformal transformations are then obtained by performing  an inversion followed by a translation and  another inversion. Combined  with supersymmetry, this covers all  superconformal transformations. 

The  dual 
supersymmetries are either manifest or  related to ordinary special superconformal symmetry \cite{Drummond:2008vq},
which is an invariance of the $\cN{=}4$ theory. Hence the invariance of the $S$-matrix under the full dual
superconformal symmetry only  requires that we  prove invariance under dual
inversions. This was achieved  
in \cite{Brandhuber:2008pf}, by constructing the  supersymmetric BCFW recursion relation reviewed in Section~\ref{sec:susyBCFW}. 
In a nutshell, the proof relies on the fact that the  building blocks of each  recursive diagram respect  dual superconformal symmetry, hence guaranteeing the covariance of the final answer.

\begin{figure}
\begin{center}
\scalebox{0.55}{\includegraphics{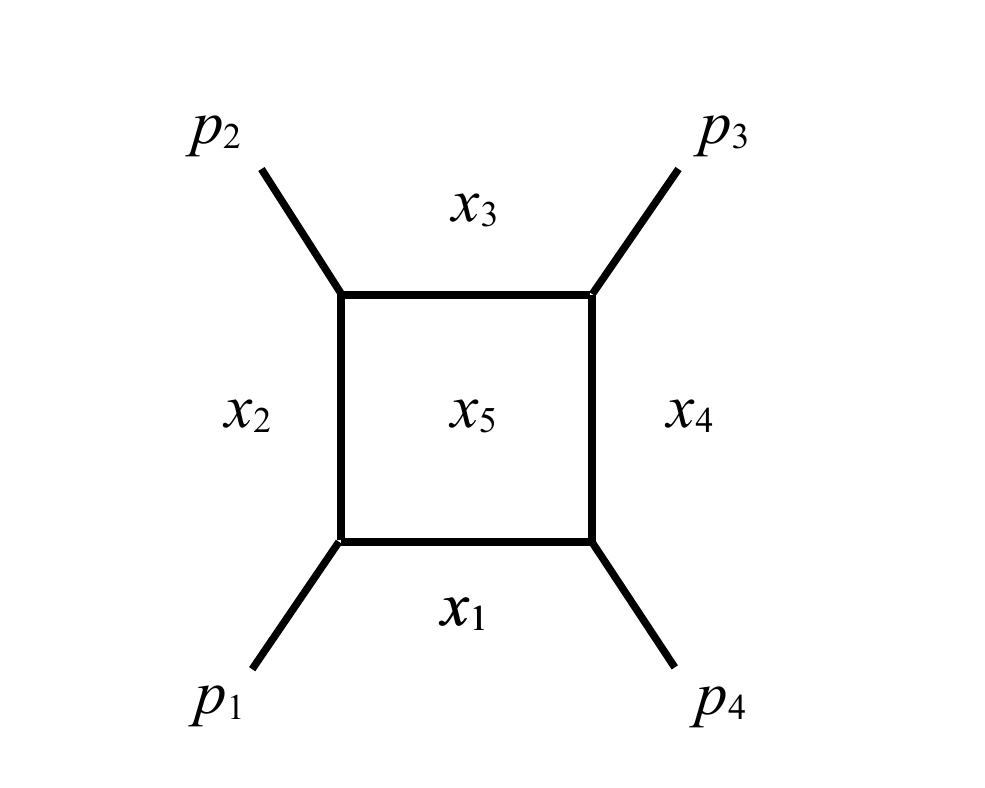}}
\end{center}
\vspace*{-10mm}
\caption{\it 
A one-loop box function. Here $p_1{:=} x_1 {-} x_{2},\ldots , p_4 {:=} x_4{-}x_1 $.
}
\label{fig:box-function}
\end{figure}
Finally, we mention that the first
strong hint of  dual conformal symmetry  was observed at loop level rather than at tree level \cite{Drummond:2006rz}, as we now outline. It is well known that all one-loop amplitudes in the maximally supersymmetric theory can be written in terms of box integrals \cite{Bern:1994zx}, such as the one shown in Figure~\ref{fig:box-function}. A box integral can be defined as  
\begin{align}
\label{bilm}
I (x_1, \ldots , x_4)  \ = \ 
\int\!\!{d^4x_5 \over (2\pi)^4}\, {1 \over x_{51}^2 x_{52}^2 x_{53}^2 x_{54}^2} \ , 
\end{align}
where we have introduced dual momenta as  $p_1{:=} x_1 {-} x_{2},\ldots ,   p_4 {:=} x_4{-}x_1 $, and the momenta of the internal legs are $x_{51}, \ldots , x_{54}$.
The advantage of this expression is that the loop measure is simply $d^4x_5$, and there is no need to pick a particular internal leg as the integration variable. Note that we have written the integration measure in four dimensions; this is allowed only when the integral does not require (infrared) regularisation, which is the case when all the $p_i$ are massive,%
\footnote{The corresponding so-called ``four-mass''  box has been evaluated in \cite{tHooft:1978jhc}, see also \cite{Denner:1991qq,Bourjaily:2019jrk,Duhr:2019tlz} for more recent calculations of the same quantity.}
otherwise we can simply replace  $d^4x_5\to d^Dx_5$ with $D{=}4{-}2\epsilon$, and choose $\epsilon{<}0$.  Leaving momentarily this fact aside, 
 let us study the transformation properties of \eqref{bilm} under dual conformal symmetry. Under the inversion \eqref{LTBV},  we simply have
\begin{align}
    x_{ij}^2 \to \frac{x_{ij}^2}{x_i^2 x_j^2} \ ,  
\end{align}
so that introducing $x_5^\prime = (x_5)^{-1}$, and using 
$x_{5^\prime i^\prime}^2 {=} \frac{x_{5 i}^2}{ x_5^2 x_i^2}$ as well as $d^4x_5^\prime  = \frac{d^4x_5}{(x_5^2)^4}$, we find that 
\begin{align}
I (x_1^\prime, \ldots , x_4^\prime) = (x_1^2\cdots x_4^2) I (x_1, \ldots , x_4)\ . \end{align}
Hence the box integrals are covariant under inversions, and since   integrals are invariant under translations of the dual momenta, it follows that all box integrals, if evaluated strictly in four dimensions, are dual conformal covariant \cite{Drummond:2006rz}. 

Usually one encounters box integrals where at least one of the external momenta is massless, in which case they are infrared divergent and have an  anomaly (computed in \cite{Brandhuber:2009xz}); these integrals are usually called ``pseudo-conformal''. Dual conformal symmetry is then 
 anomalous at loop level 
\cite{Drummond:2008vq,Brandhuber:2009kh}, and  the anomaly of the amplitudes turns out to be  closely related to that of the polygonal lightlike Wilson loop   \cite{Drummond:2007cf,Drummond:2007au} dual to the amplitude \cite{Alday:2007hr,Drummond:2007aua,Brandhuber:2007yx}.
Using this anomaly one can find useful  
 constraints on   supercoefficients in the expansion of  superamplitudes in an integral basis~\cite{Brandhuber:2009xz,Elvang:2009ya}. We also mention that pseudo-conformality of the integrals has been used to write the  four-point MHV amplitude up to five  \cite{Bern:2007ct}, six and seven loops \cite{Bourjaily:2011hi}, following the remarkable direct calculations at three \cite{Bern:2005iz} and four loops  \cite{Bern:2006ew,Cachazo:2006az}.  We will come back to loop amplitudes in Section~\ref{sec:10}.

\subsection{Dual superconformal covariance of the MHV and NMHV superamplitudes}
\label{sec:dss-ex}

We begin by showing that the tree-level MHV
superamplitude \eqref{superMHV} is covariant under dual conformal symmetry \cite{Drummond:2008bq}. As explained earlier, it is sufficient to consider dual inversions. Using \eqref{trbrij}, we  see that $\prod_{i=1}^n \lan i i+1\ran \to \prod_{i=1}^n 
x_i^2  \lan i i+1\ran$. One then  observes that the combination of delta functions $\delta^{(4)} (x_i - x_{n+1}) \, \delta^{(8)} (\theta_1 - \theta_{n+1})$ is invariant under inversions. Hence
the MHV superamplitude  transforms covariantly under inversions:
\begin{align}\label{MHVtrans}
  \cA_n^{\rm MHV}(1,2,\dots,n) \rightarrow \cA_n^{\rm MHV}(1,2,\dots,n) \  \prod_{k=1}^n
  {x_k^2} \ .
\end{align}
Next we discuss the NMHV superamplitudes, whose explicit expression is shown in \eqref{superNMHV}. To prove  that they transform covariantly, we need  to show that the   $R$-functions  
in \eqref{Rinv}  are dual superconformal invariant. It is convenient to define the~four-bracket 
\begin{align}\label{fourbrack}
 \braket{i,j-1,j,k}:=\langle i|x_{ij}x_{jk}|k-1\rangle \langle j-1 j\rangle \ ,
\end{align}
whose usefulness arises from the fact that it is  a dual conformal invariant. 
An elegant way to see this is to introduce  momentum twistors \cite{Hodges:2009hk,Mason:2009qx}
\begin{align}
 Z^{\hat{A}}_i&=\begin{pmatrix} \lambda_i^\alpha\\ \mu_i^{\dot \alpha}\end{pmatrix} \, , \qquad  \mu_i^{\dot \alpha}=x_i^{\dot \alpha \alpha} \lambda_{i \alpha}\, , 
\end{align}
on which  conformal transformations act linearly --   they are realised as  $SL(4)$ transformations on the index $\hat{A}$. 
The four-bracket \eqref{fourbrack} can then be recast as
\begin{equation}
  \braket{i,j-1,j,k}= \epsilon_{\hat A \hat B \hat C \hat D} Z^{\hat A}_i Z^{\hat B}_{j-1} Z^{\hat C}_j Z^{\hat D}_k\, , 
\end{equation}
which is  manifestly invariant under $SL(4)$ transformations. To address dual  superconformal transformations, it  is  then convenient to introduce  supertwistor variables 
\begin{align}
\label{supertwistors}
 \mathcal{Z}_i^M&=\begin{pmatrix} Z_i^{\hat A} \\ \chi_i^{A}\end{pmatrix} \ ,\qquad   \chi^A_i=\theta_i^{A\, \alpha} \lambda_{i \alpha}\ . 
\end{align}
These  transform in the fundamental representation of the supergroup $SL(4|4)$, whose  projective real section $PSU(2,2|4)$ is precisely the (dual)  superconformal group of  $\mathcal{N}\!=\!4$ SYM.
Given five arbitrary superstwistors $\mathcal{Z}_a,\ldots, \mathcal{Z}_e$, it is  straightforward to show that 
\begin{align}\label{fivebrack}
[a,b,c,d,e]= \frac{\delta^{(4)}(\braket{a,b,c,d} \chi_{e}+\text{cyclic})}{\braket{a,b,c,d}\braket{b,c,d,e}\braket{c,d,e,a}\braket{d,e,a,b}\braket{e,a,b,c}}
\end{align}
is an $SL(4|4)$ invariant.%
\footnote{Note that \eqref{fivebrack} is invariant under 
$
\mathcal{Z}^{M}_i\to \zeta_i \mathcal{Z}^{M}_i$, 
in other words  these  are projective coordinates in super twistor space. 
This transformation   is related to  little group scaling of the spinor-helicity variables.}
This is useful since one can prove that   
$
 R_{rst}=[s-1,s,t-1,t,r] $ \cite{Mason:2009qx}, 
from which dual superconformal invariance of the $R$-functions follows.

\subsection{Yangian symmetry}

The generators of the dual superconformal symmetry algebra $\{ P, K,S, \bar{S}, Q, \bar{Q} \}$ 
are most naturally  written  in an extended superspace given by the set of  variables  
$\{ \lambda^{\alpha}, \tilde{\lambda}^{\da}, \eta^{A}, x^{\alpha\da}, \theta^{\alpha A}\}$ that are subject to the constraints of \eqref{constraintssuperspace}.
Explicitly, the dual superconformal generators $K$ and $S$ take the form
\begin{align}
\begin{split}
\label{KSdef}
K^{\alpha \dot{\alpha}} &= \sum_{i=1}^{n} \biggl[ x_i^{\alpha \dot{\beta}} x_i^{\dot{\alpha} \beta} \frac{\partial}{\partial x_i^{\beta \dot{\beta}}}
+ x_i^{\dot{\alpha} \beta}  \theta_i^{\alpha B} \frac{\partial}{\partial \theta_i^{\beta B}} +
  x_i^{\dot{\alpha} \beta} \lambda_i^{\alpha} \frac{\partial}{\partial \lambda_i^{\beta}}
 + x_{i+1}^{\alpha \dot{\beta}} \tilde{\lambda}_i^{\dot{\alpha}}
  \frac{\partial}{\partial \tilde{\lambda}_{i}^{\dot{\beta}}} + \tilde{\lambda}_i^{\dot{\alpha}} \theta_{i+1}^{\alpha B} \frac{\partial}{\partial \eta_i^B} \biggr] \,, \\ 
S_{\alpha}^A &= \sum_{i=1}^{n} \biggl[ -\theta_{i \alpha}^{B} \theta_i^{\beta A}
  \frac{\partial}{\partial \theta_i^{\beta B}} + x_{i \alpha}{}^{\dot{\beta}} \theta_i^{\beta
    A} \frac{\partial}{\partial x_i^{\beta \dot{\beta}}} + \lambda_{i \alpha}
  \theta_{i}^{\gamma A} \frac{\partial}{\partial \lambda_i^{\gamma}} 
  + x_{i+1\,
    \alpha}{}^{\dot{\beta}} \eta_i^A \frac{\partial}{\partial \tilde{\lambda}_i^{\dot{\beta}}} -
  \theta_{i+1\, \alpha}^B \eta_i^A \frac{\partial}{\partial \eta_i^{B}} \biggr] \,,
\end{split}
\end{align}
and can be shown to commute with the constraints \eqref{constraintssuperspace}.
An interesting question  is what algebraic structure emerges if one commutes the 
superconformal and dual superconformal generators with one another, i.e.~studies the closure of
the two algebras. It turns out that this induces an 
infinite-dimensional symmetry algebra known as the Yangian $Y[\alg{psu}(2,2|4)]$ \cite{Drummond:2009fd}.
A Yangian algebra $Y(\alg{g})$ built upon a simple Lie algebra $\alg{g}$ 
is a deformation of the loop algebra realised by  generators $J_{a}^{(n)}$  with levels $n\in\mathbb{N}$ \cite{Drinfeld:1985rx,Drinfeld:1986in}. The  
level-zero and level-one generators obey the commutation relations 
\be
\label{comrels}
[ J_{a}^{(0)}, J_{b}^{(0)}\} = f_{ab}^{c}\, J_{c}^{(0)} \, ,\qquad
[ J_{a}^{(0)}, J_{b}^{(1)}\} = f_{ab}^{c}\, J_{c}^{(1)} \, ,
\ee
where  $[.,.\}$ denotes a graded 
commutator.
The higher-level generators follow from commutators of the level-one generators. In addition, there are Serre relations \cite{Drinfeld:1985rx,Drinfeld:1986in} which generalise the usual 
Jacobi identities. 
The co-products of the level-zero and  level-one Yangian generators express the action
 on two-particle states, and read
\begin{align}
\begin{split}
\Delta(J^{(0)}_{a}) &= J^{(0)}_{a} \otimes \mathbb{1} + \mathbb{1} \otimes J^{(0)}_{a} \, ,\\
\Delta(J^{(1)}_{a}) &= J^{(1)}_{a} \otimes \mathbb{1} + \mathbb{1} \otimes J^{(1)}_{a} 
+ f_{a}{}^{bc}\, J^{(0)}_{b} \otimes J^{(0)}_{c}\, . 
\label{coproduct2}
\end{split}
\end{align}
In the last term above, the adjoint indices of the structure constant are
raised and lowered with the group metric 
$\Tr(J^{(0)}_{R\, a}\, J^{(0)}_{R\,b})$ with $J^{(0)}_{R, a}$  
in the defining representation of~$\alg{g}$.
The  level-one generators  are then given by
\begin{align}
\label{J1explicitdef}
J_{a}^{(1)} = \sum_{i=1}^{n} J^{(1)}_{ia} + f_{a}{}^{cb} \sum_{i\le 1 < j \le n} J_{ib}^{(0)} J_{jc}^{(0)} \,.
\end{align}
In the problem at hand, the level-zero generators $J^{(0)}_{a}$  coincide with the 
generators of the superconformal algebra $\alg{psu}(2,2|4)$ of \eqref{psu224alg}.
Interestingly, the dual superconformal  generators $K$ and $S$ of \eqref{KSdef}
can be identified with the level-one Yangian generators of $Y[\alg{psu}(2,2|4)]$.
In order to see this one solves the constraints (\ref{constraintssuperspace}) 
via
\be
x_i^{\alpha \dot\alpha} = x_1^{\alpha \dot\alpha} - \sum_{j<i} \lambda_j^{\alpha}\tilde \lambda_j^{\dot\alpha}\,, \qquad \theta_i^{\alpha A} = \theta_1^{\alpha A} - \sum_{j<i} \lambda_j^{\alpha} \eta_j^A\, \hspace{10pt} \text{ for } \hspace{10pt} 2\leq i \leq n+1
\, , 
\ee
eliminating $x^{\alpha\dot\alpha}_{i}$ and $\theta^{\alpha A}_{i}$,
and expresses the dual superconformal generators 
in the original superspace variables
$\{\lambda^{\alpha}_{i}, \tilde{\lambda}^{\da}_{i}, \eta^{A}_{i} \}$ to
discover that some of the generators become trivial, namely $P$ and $Q$, 
while others overlap with the original superconformal ones, namely 
$\bar{S}$ and $\bar{Q}$. The non-trivial generators turn out to be
$K$ and $S$. One can  show
that   $S$ is explicitly given,    up to a  term $\Delta S$ that trivially annihilates the  amplitudes,
 by
\begin{align}\label{dualS}
S^A_{\alpha} + \Delta S^A_{\alpha} = -\frac{1}{2} \sum_{i<j} 
\left[ m_{i\alpha}^{\gamma} q_{j \gamma}^{A} - \frac{1}{2} (d_i +c_i) q_{j\alpha}^{A} + p_{i \alpha}^{\dot{\beta}} \bar{s}^{A}_{j \beta} + q_{i \alpha}^{B} r_{j B}^{A} - (i\leftrightarrow j) \right] \, ,
\end{align}
and indeed takes the form  \eqref{J1explicitdef}, with the ``densities''
$J^{(0)}_{ia}$ appearing quadratically along with 
 a trivial evaluation representation $J^{(1)}_{ia}=0$. A similar
 structure emerges for $K$~\cite{Drummond:2009fd}.

The Yangian is a hidden symmetry of tree-level superamplitudes, that is for any generator $J {\in} Y( \alg{psu}(2,2|4))$ one finds 
$ J \, \mathcal{A} {= }0$   up to contact terms 
 related to collinear kinematic configurations
  \cite{Bargheer:2009qu}.
In fact the Yangian symmetry  also 
constrains the structure of planar loop integrands, however infrared divergences break the symmetry at the integrated level \cite{Beisert:2010gn,Caron-Huot:2011dec,Bullimore:2011kg}. Being an infinite-dimensional
symmetry algebra,  the Yangian points to a hidden integrability of planar
$\mathcal{N}\!=\!4$ SYM, see \cite{Beisert:2010jr} for a review.

Finally,   the Yangian generators have a particularly simple form when re-expressed in the supertwistor variables
of \eqref{supertwistors}:
\begin{align}
J^{(0)}{}^{M}{}_{N} = \sum_{i} \mathcal{Z}_{i}^{M} \frac{\partial}{\partial \mathcal{Z}_{i}^{N}}\,, \qquad 
J^{(1)}{}^{M}{}_{N} = \sum_{i>j} \left[ \mathcal{Z}_{i}^{M} \mathcal{Z}_{j}^{O}  \frac{\partial}{\partial \mathcal{Z}_{i}^{O}} \frac{\partial}{\partial \mathcal{Z}_{j}^{N}}  - (i \leftrightarrow j) \right] \, .
\end{align}
Written in these variables the  Yangian symmetry of the scattering amplitudes can be made most manifest.

\section{Loops from unitarity cuts}
\label{sec:10}

\subsection{Basic ideas}

The fundamental  tenet of the modern amplitudes programme 
\cite{Bern:1994zx,Bern:1994cg,Bern:2004cz,Britto:2004nc}
is to 
use gauge-invariant quantities  such as amplitudes or form factors as input in computations, avoiding the use of Feynman diagrams. In previous sections we have shown how this can be achieved at tree level, and the  next  question is how to extend this approach to    loop amplitudes.
As we will now  review,  we can efficiently recycle tree-level amplitudes to obtain  loops from trees.%
\footnote{A different incarnation of this can be recognised in the Feynman tree theorem \cite{Feynman:1963ax,Feynman-magic},   see \cite{Brandhuber:2005kd} for a discussion and comparison of this theorem to the unitarity approach.}
    
If we are tasked to stay away from Feynman rules we have to
go back to more fundamental principles of QFT -- the relevant ones for us are locality and unitarity. These  tell us
that at tree level the only allowed singularities are simple poles arising  from 
propagator factors
$
    \frac{i}{p^2-m^2 + i \varepsilon}
$,
and the residue at such poles is the product of smaller scattering amplitudes. These facts
underpin tree-level factorisation theorems discussed  in Section \ref{sec:der-BCFW}, which in turn lead to  the BCFW recursion relations. 

Unitarity  is the statement of conservation of probability, it means that if we scatter something the probability that something happens is one:
\begin{align}
\label{SdaggerS}
    S^\dagger S = 1 \ .
\end{align}
Now writing the $S$-matrix as a trivial (forward) piece plus a part that describes the non-trivial scattering  as $S= 1 + i T$ we find
\begin{align}
\label{unitarity}
    T^\dagger T = -i (T-T^\dagger) \ . 
\end{align}
The formal matrix product on the left-hand side implies a summation over all
possible intermediate (helicity) states and an on-shell phase-space integration 
$\int\!d^4 p_i \, \delta(p_i^2-m_i^2)$ for each intermediate particle.
Taking matrix element of \eqref{unitarity} between external states, one obtains a  product of amplitudes that equals the imaginary part (or discontinuity) of the full amplitude,
from which one  can in principle obtain the  amplitude from a dispersion integral of the form $\int\!ds'\,  \frac{ {\rm Im}A(s')}{s-s'}$ where $s$ is some Mandelstam variable.
This is conceptually deep and beautiful, but unfortunately not of much practical use in particular if we consider a process with more than four particles.%
\footnote{
 We also mention important applications of unitarity to the study of black hole scattering in general relativity 
 \cite{Neill:2013wsa,Bjerrum-Bohr:2013bxa,Bjerrum-Bohr:2014zsa, Bjerrum-Bohr:2016hpa, Bai:2016ivl, Chi:2019owc,Bern:2019nnu,Bern:2019crd,Parra-Martinez:2020dzs,DiVecchia:2021bdo,Bjerrum-Bohr:2021din,Brandhuber:2021eyq,Herrmann:2021lqe,Herrmann:2021tct, Bern:2021dqo},
and  in theories of modified gravity with higher-derivative interactions \cite{Brandhuber:2019qpg, Emond:2019crr, AccettulliHuber:2020oou,AccettulliHuber:2020dal,Carrillo-Gonzalez:2021mqj}. See Chapters~13 and~14 of this review  \cite{Bjerrum-Bohr:2022blt,Kosower:2022yvp} for more details. 
}

\subsection{General structure of one-loop amplitudes}

From now on we will  focus on planar one-loop amplitudes in gauge theories. At one-loop,  using \eqref{leading-colour-one-loop}, these can be written in terms of a single primitive  amplitude
$A^{(1)}_{n;1}$  (and for brevity we will henceforth call it $A^{(1)}_{n}$). 
It is well known  \cite{Bern:1994zx} that also the non-planar contributions 
can be expressed as linear combinations of the $A^{(1)}_n$, giving a further reason  to focus on the computation of the planar parts.

Of utmost importance is the fact that one-loop amplitudes can be decomposed in terms of scalar Feynman integrals which contain transcendental functions such as logarithms and dilogarithms, i.e.~functions that contain discontinuities, and rational parts. 
In general the answer will contain ultraviolet (UV) and infrared (IR) divergences which we regulate using dimensional regularisation. From now on  we  consider   massless gauge theories, which implies that tadpoles are absent. In this case one can write  the following ansatz for a general one-loop amplitude:
\begin{align}
\label{1loopansatz}
    A^{(1)}_n = \sum_i a_i I_{4,i} + 
    \sum_j b_j I_{3,j} + \sum_k c_k I_{2,k} + R_n \ ,
\end{align}
where we have introduced the scalar Feynman integrals
\begin{align}
    I_{n,i} = \int\!\frac{d^D \ell}{(2\pi)^D}\,
    \frac{1}{\ell^2 (\ell-K_{i,1})^2 \cdots (\ell-K_{i,n-1})^2} \ ,
\end{align}
where the $K_{i,j}$ correspond to appropriate sums of subsets of external momenta $p_i$. 
 In the presence of colour ordering, only adjacent momentum labels appear in the set. 
%
The $I_{4,i}$ and $I_{3,i}$ are called boxes and triangles, which are UV finite but contain IR divergences, and the $I_{2,i}$ are  UV-divergent bubble integrals. 
We can  motivate the ansatz \eqref{1loopansatz} as follows. Had we started from a gedanken Feynman integral computation of a  one-loop  $n$-gluon amplitude, we would have  found  many more and much more complicated integrals, the most complicated one being an $n$-gon
\begin{align}
    \int \frac{d^D \ell}{(2\pi)^D}
    \frac{P_n(\ell)}{\ell^2 (\ell-p_1)^2 (\ell-p_1-p_2)^2\cdots (\ell+p_n)^2} \ ,
\end{align}
where $P_n(\ell)$ is a polynomial of degree $n$ in the loop momenta, coming from the $n$ momentum-dependent three-gluon vertices. Such an integral is usually called a tensor integral. Thanks to a theorem by Passarino and Veltman (PV) \cite{Passarino:1978jh}, all such higher-point and tensor integrals can be PV-reduced to only scalar bubbles, triangles and boxes. Since all these integrals have been evaluated and are tabulated, the remaining non-trivial task is to find the coefficients $a_i$, $b_j$ and $c_k$. 

Before  moving to  concrete examples, we now  
 discuss how one  can put unitarity to work to determine  these coefficients algebraically without ever performing any integrals, following the groundbreaking 
work of \cite{Bern:1994zx,Bern:1994cg}.

\subsection{Unitarity at one loop: two-particle cuts}

\begin{figure}
\begin{center}
\scalebox{0.45}{\includegraphics{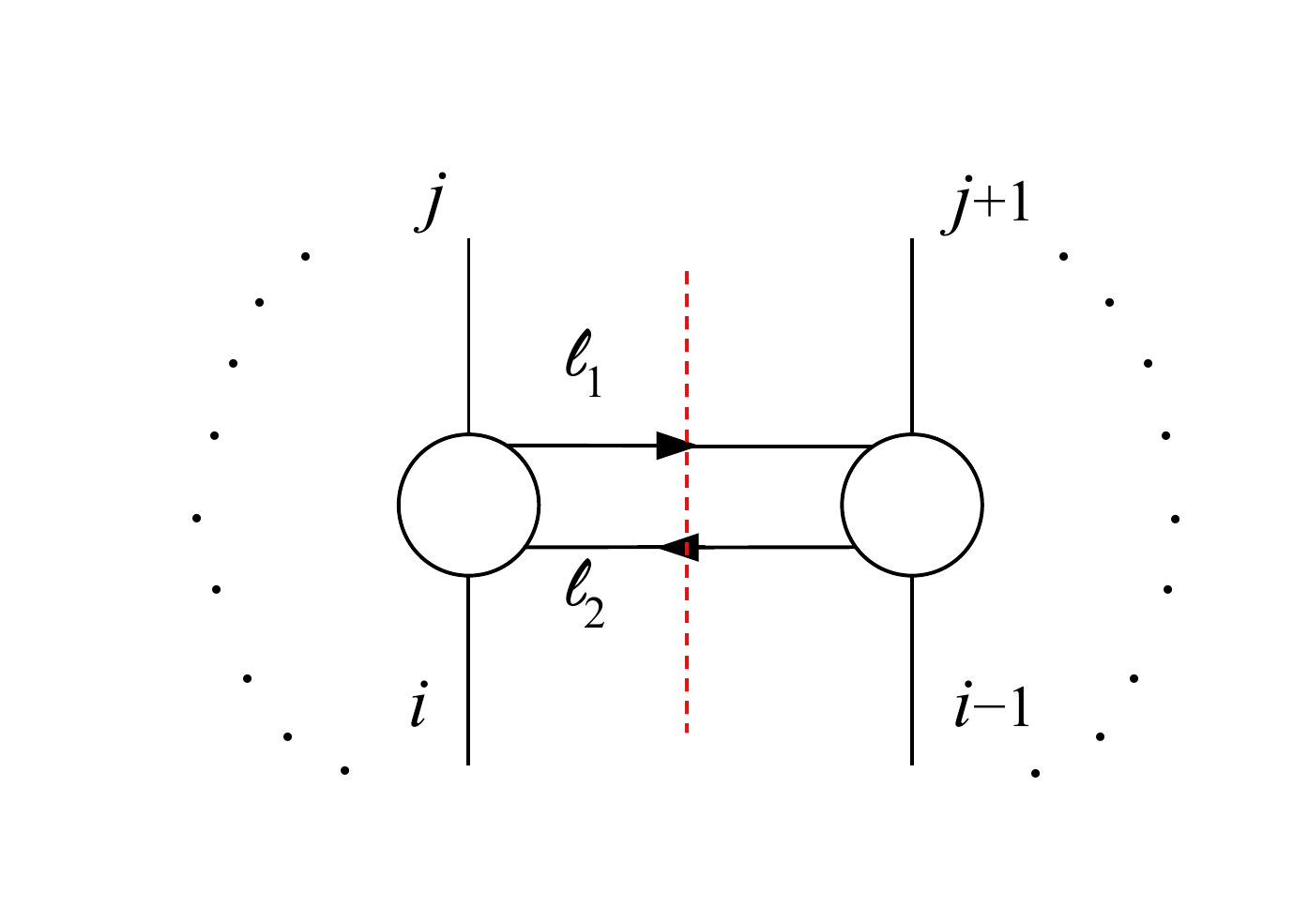}}
\end{center}
\vspace*{-10mm}
\caption{\it 
A cut diagram evaluating the discontinuity in the $s_{i\ldots j}$-channel.
}
\label{fig:generic-cut}
\end{figure}

The main idea is 
 to compute the discontinuities (or imaginary parts)
of the left-hand side and the right-hand side of \eqref{1loopansatz}:
\begin{align}
\label{1loopansatz-cut}
    {\rm Disc}(s_{i \ldots j}) A^{(1)}_n = \sum_i a_i {\rm Disc}(s_{i\ldots j}) I_{4,i} + 
    \sum_j b_j {\rm Disc}(s_{i \ldots j}) I_{3,j} + \sum_k c_k {\rm Disc}(s_{i\ldots j}) I_{2,k} \ ,
\end{align}
in all (two- or multi-particle) kinematic channels,  with   $s_{i \ldots j} {=} (p_i + p_{i+1} + \cdots + p_j)^2$. 
The left-hand side of \eqref{unitarity} is then evaluated as 
a product of two tree  amplitudes convoluted with a two-particle phase-space integral corresponding to two internal on-shell states, and we have to sum over all   internal helicities. This procedure is usually called a {\it two-particle cut} because two off-shell propagators are put on shell:
\begin{align}
\label{cut}
    \frac{i}{\ell_{1,2}^2 + i \varepsilon} \to 2 \pi \delta(\ell^2_{1,2}) \ ,
\end{align}
and the discontinuity of the one-loop amplitude in the channel $s_{i \ldots j}$
is then given by
\begin{align}
\label{cutintegral}
\begin{split}
{\rm Disc}(s_{i \ldots j}) A^{(1)}_n = 
 \sum_{h_1,h_2} \int \frac{d^D\ell_1}{(2 \pi)^{D-2}}
\delta(\ell_1^2) \delta(\ell_2^2) \  & A^{\rm tree}(-\ell_2^{-h_2},i\ldots j, \ell_1^{h_1}) \times \\
& A^{\rm tree}(-\ell_1^{-h_1},j+1,\ldots, i-1, \ell_2^{h_2}) \ ,
\end{split}
\end{align}
with $\ell_2=\ell_1 + p_i + p_{i+1} + \cdots + p_j$ (see Figure~\ref{fig:generic-cut}).
One could  perform this phase-space integral to obtain the discontinuity of the amplitude in this channel, however a more useful approach, advocated in \cite{Bern:1994cg}, 
is to observe that if we uplift this integral to a full Feynman integral by undoing \eqref{cut}, 
we obtain a Feynman integral
\begin{align}
\label{uncutintegral}
\begin{split}
&\sum_{h_1,h_2} \int \frac{d^D\ell_1}{(2 \pi)^{D}}
\frac{i}{\ell_1^2} \frac{i}{(\ell_2^2)}
A^{\rm tree}(-\ell_2^{-h_2},i\ldots j, \ell_1^{h_1}) \times 
A^{\rm tree}(-\ell_1^{-h_1},j+1,\ldots, i-1, \ell_2^{h_2}) \ ,
\end{split}
\end{align}
that has the correct discontinuity of the amplitude in this particular channel. Hence the {\it integrand} thus produced must be part of the complete answer, and by going through all kinematic channels we have enough constraints to fix the  integrand for the amplitude.
The key advantage is  that we  can 
simplify the cut integrand \eqref{cutintegral} as much as possible using on-shell conditions and powerful spinor-helicity techniques before lifting  it back 
to a full Feynman loop integrand \eqref{uncutintegral}.
Once we have combined the information from all cuts, we can PV-reduce the resulting integrand (which is an algebraic process) and read off the coefficients $a_i$, $b_j$ and $c_k$.

A  comment on the  rational terms $R_n$ in \eqref{1loopansatz}  is   in order.
In  \cite{Bern:1994zx,Bern:1994cg} it was shown that 
such  terms vanish at one loop  in supersymmetric theories, and  in computing  unitarity cuts
it is enough to use tree amplitudes valid  strictly in four dimensions. This allows us to use powerful  spinor-helicity techniques.
However, if we work in pure Yang-Mills or QCD we obtain only part of the answer --
the  four-dimensional cut-constructible pieces,  missing  further  rational terms. 
In order to get these we must  perform unitarity cuts in $D{=}4 {-}2 \epsilon$ dimensions \cite{vanNeerven:1985xr,Bern:1995db},  which requires amplitudes where at least the cut legs are  in $D$ dimensions. External momenta can be kept in four dimensions if, as we do, we use the four-dimensional helicity scheme 
\cite{Bern:1991aq,Bern:2002tk}. We will return to this  in Section~\ref{sec:nonsusy}.

\subsection{Example: four-gluon amplitude in 
$\mathcal{N}\!=\!4$ 
SYM from two-particle cuts}
\label{sec:four-g-oneloop}

\begin{figure}
\begin{center}
\scalebox{0.45}{\includegraphics{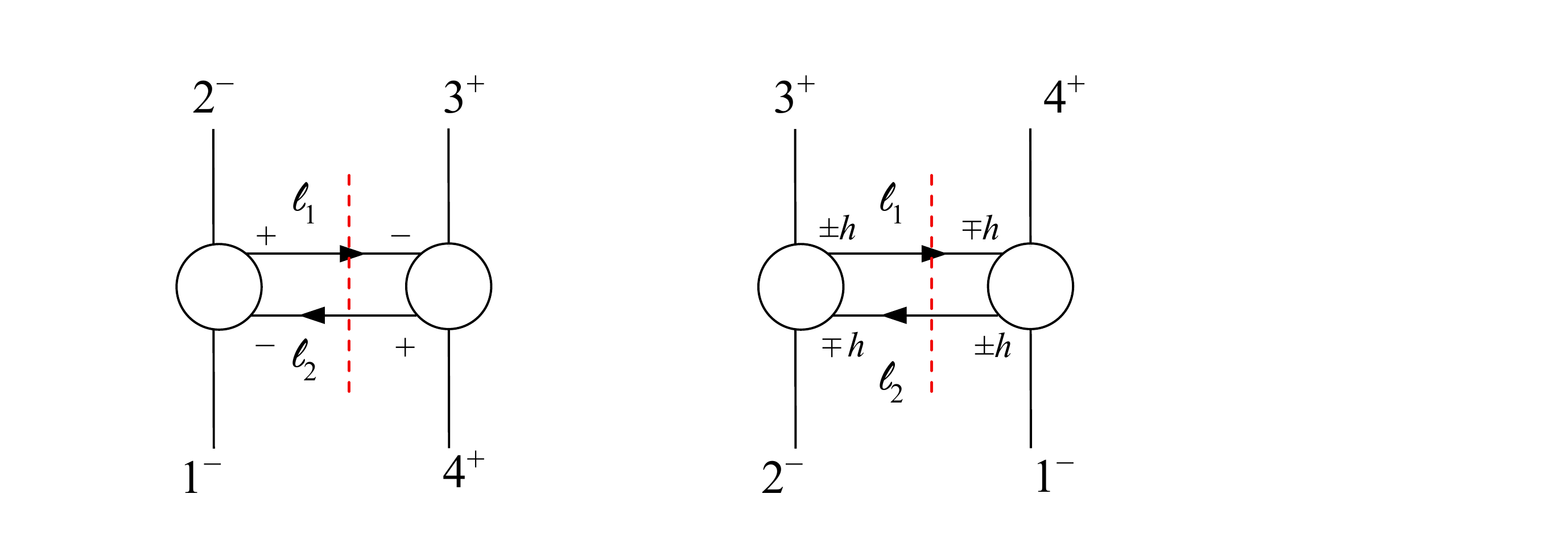}}
\end{center}
\vspace*{-10mm}
\caption{\it 
The $s$- and $t$-channel cut diagrams contributing to the one-loop MHV amplitude in $\cN{=}4$ SYM.
}
\label{fig:one-loop=N=4}
\end{figure}

We will now illustrate the previous discussion by computing the one-loop four-gluon amplitude $A^{(1)}(1^- 2^- 3^+ 4^+)$ from two-particle cuts. There are two channels to  consider, namely the $s$-channel and the $t$-channel, corresponding to  the Mandelstam invariants $s{=}(p_1+p_2)^2$ and 
$t{=}(p_2+p_3)^2$.
In the $s$-channel, the internal states can only be gluons, and the amplitudes entering the cut are (see Figure~\ref{fig:one-loop=N=4}):%
\footnote{In this section we drop powers of $g$, which can easily be reinstated at the end.}
\begin{align}
\begin{split}
    A((-\ell_2)^+, 1^-,2^-,\ell_1^+) = 
    i \frac{\langle 12\rangle^4}{\langle -\ell_2 1\rangle \langle 12 \rangle \langle 2 \ell_1\rangle \langle \ell_1 \, -\ell_2 \rangle}\, ,  \\
    A(-\ell_1^-, 3^+,4^+,\ell_2^-) = 
    i \frac{\langle \ell_2 \, -\ell_1\rangle^4}{\langle -\ell_1 3\rangle \langle 34 \rangle \langle 4 \ell_2\rangle \langle \ell_2 \, -\ell_1 \rangle}\, .
    \end{split}
\end{align}
Multiplying the product of these two amplitudes with $\frac{i}{\ell_1^2} \frac{i}{\ell_2^2}$ we find
the cut integrand
\begin{align}
{\rm Disc}(s) A^{(1)}(1^- 2^- 3^+ 4^+) = 
    A^{\rm tree}_4 \times \int \left. \frac{d^D \ell_1}{(2 \pi)^D}
    \frac{i}{\ell_1^2 \ell_2^2} 
    \frac{\langle 23 \rangle \langle 41 \rangle \langle \ell_1 \ell_2 \rangle^2}{\langle \ell_2 1\rangle \langle 2 \ell_1\rangle \langle \ell_1 3\rangle \langle 4 \ell_2 \rangle} \right|_{s-{\rm cut}} \ ,
\end{align}
where we have pulled out the tree amplitude $i\frac{\langle 12\rangle^4}{\langle 12 \rangle \langle 23 \rangle \langle 34 \rangle \langle 41 \rangle}$ and used our convention $\lambda_{-p} = i \lambda_p$.
By rationalising two of the denominator factors using 
$\langle 2 \ell_1 \rangle [\ell_1 2] = (\ell_1+p_2)^2$
and $\langle \ell_1 3 \rangle [3 \ell_1] = -(\ell_1-p_3)^2$
we can further massage the integrand to find
\begin{align}
\begin{split}
  &  A^{\rm tree}_4 \times \frac{i}{\ell_1^2 \ell_2^2 (\ell_1+p_2)^2 (\ell_1-p_3)^2} \frac{\langle 23 \rangle \langle 41 \rangle \langle \ell_2 \overbrace{\ell_1 \rangle [\ell_1}^{=\ell_2-p_1-p_2} 2] [3 \overbrace{\ell_1] \langle \ell_1}^{=\ell_2+p_3+p_4} \ell_2 \rangle}{\langle \ell_2 1\rangle \langle 4 \ell_2 \rangle} \\
  &=   A^{\rm tree}_4 \times \frac{i}{\ell_1^2 \ell_2^2 (\ell_1+p_2)^2 (\ell_1-p_3)^2} (-\langle 23 \rangle \langle 41 \rangle [12][34]) \\
  &=i s t A^{\rm tree}_4 \times \frac{1}{\ell_1^2 \ell_2^2 (\ell_1+p_2)^2 (\ell_1-p_3)^2} \ ,
    \end{split}
\end{align}
from which we see  that $i s t A_4^{\rm tree}$ is the coefficient of the  zero-mass box function \cite{Bern:1993kr}
\begin{align}
\label{0mass}
\begin{split}
    I_4^{\rm 0m}(s,t) &= \int \frac{d^D \ell}{(2 \pi)^D} \frac{1}{\ell^2 (\ell+p_1+p_2)^2 (\ell+p_2)^2 (\ell-p_3)^2} \\
  & =  -i \frac{2c_{\Gamma}}{st}\Big\{-\frac{1}{\eps^2}\Big[(-s)^{-\eps}+(-t)^{-\eps} \Big]  +\frac{1}{2}\log^2\Big(\frac{s}{t}\Big) + \frac{\pi^2}{2}\Big\} \ ,
    \end{split}
\end{align}
with
\begin{align}
c_{\Gamma} \,=\, \frac{\Gamma(1+\epsilon)\Gamma(1-\epsilon)^2}{(4\pi)^{2-\eps} \Gamma(1-2\epsilon)}\,.
\end{align}
Since we have only considered the $s$-channel cut, we know that $i s t A_4^{\rm tree} I_{\rm 0m}(s,t)$ must be part of the full answer  but we can only trust terms that have discontinuities in $s$.  
To complete the computation we need to consider also  the $t$-channel, also shown in Figure~\ref{fig:one-loop=N=4}. Initially, this looks more complicated because on both sides of the cuts the external legs are one positive and one negative helicity gluon and this allows all possible states of $\cN{=}4$ SYM to appear as internal states: $h{=}-1,-1/2,0,1/2,1$, with multiplicities $1,4,6,4,1$.
The corresponding tree amplitudes entering the cut are
\begin{align}
\begin{split}
    A((-\ell_2)^{-h}, 2^-,3^+,\ell_1^h) = 
    i \frac{\langle -\ell_2 2\rangle^{2+2h} \langle \ell_1 2 \rangle^{2-2h}}{\langle -\ell_2 2\rangle \langle 23 \rangle \langle 3 \ell_1\rangle \langle \ell_1 \, -\ell_2 \rangle}\, ,  \\
    A(-\ell_1^{-h}, 4^+,1^-,\ell_2^h) = 
    i \frac{\langle  -\ell_1 1 \rangle^{2+2h} \langle \ell_2 1 \rangle^{2-2h}}{\langle -\ell_1 4\rangle \langle 41 \rangle \langle 1 \ell_2\rangle \langle \ell_2 \, -\ell_1 \rangle} \ ,
    \end{split}
\end{align}
and the $t$-channel cut is given by the product of these amplitudes and summing over $h$. Focusing on the product of numerators,
the sum with  the correct multiplicities gives
\begin{align}
\begin{split}
   & \sum_h 
   \binom{4}{2+2h}
   \langle -\ell_2 2\rangle^{2+2h} \langle \ell_1 2 \rangle^{2-2h} \langle  -\ell_1 1 \rangle^{2+2h} \langle \ell_2 1 \rangle^{2-2h} \\
   & = \big( \langle \ell_1 2 \rangle \langle \ell_2 1 \rangle - \langle \ell_1 1 \rangle \langle \ell_2 2 \rangle\big)^4 
   = \langle 12 \rangle^4 \langle \ell_1 \ell_2 \rangle^4 \ .
    \end{split}
\end{align}
The numerator is then the same as in   the $s$-channel  and  denominator factors are obtained from a cyclic relabelling of the external legs. Thus from the $t$-channel cut we~get 
\begin{align}
i s t A^{\rm tree}_4 \times \frac{1}{\ell_1^2 \ell_2^2 (\ell_1+p_3)^2 (\ell_1-p_4)^2} \, , 
\end{align}
which  is proportional to the integrand of the box function $I_4^{\rm 0m}(s,t)$ up to a trivial shift of the loop momentum.
Summarising, the unique answer consistent with both cuts is
\begin{align}
    A_4^{(1)}(1^-,2^-,3^+,4^+) = i s\,  t\,  
     A_4^{\rm tree}(1^-,2^-,3^+,4^+) \, I_4^{\rm 0m}(s,t) \ .
\end{align}
One remarkable outcome of this computation is that it does not lead to any bubble or triangle integrals, and is consistent with the general
fact that one-loop amplitudes in $\cN{=}4$ SYM only contain boxes
\cite{Bern:1994zx,Bern:1994cg}. This can be linked to the improved power-counting behaviour of this theory, and in fact is a property of all one-loop amplitudes in the theory.  It can  also be related  to dual (pseudo)conformal symmetry  of the box functions \cite{Drummond:2006rz}, as anticipated in Section~\ref{sec:DSI}. 
Also note that $\cN{=}4$ SYM is UV-finite to all orders, and hence bubble integrals must be absent.


\setcounter{footnote}{0}


In the expression of  one-loop amplitudes involving massless particles 
one encounters IR divergences  which are known to be universal.  For  colour-ordered one-loop amplitudes, these take the form \cite{Giele:1991vf, Kunszt:1994np}   
\begin{align}
\label{IRoneloop}
 -A^{\rm tree}_n \times  c_\Gamma  \sum_{i=1}^n  \frac{(-s_{i,i+1})^{-\eps}}{\eps^2} \ .
\end{align}
One can see this in our four-point example by noticing that a factor of $st$ in the coefficient cancels the $1/(st)$ factor in $I_{\rm 0m}$ of \eqref{0mass}. 
It is also known \cite{Mueller:1979ih, Magnea:1990zb, Sterman:2002qn} that IR divergences are governed by 
Sudakov form factors,  we will return to this in Section~\ref{sec:FFSuda}.

\subsection{Generalised unitarity}


A key observation in our discussion so far is that any one-loop amplitude
of massless particles can be expressed in terms of a linear combination of a complete basis of scalar integrals functions: bubbles, triangles and boxes.
Performing a two-particle unitarity cut as above amounts to picking
two internal propagators with momenta $\ell_1$ and $\ell_2$ and putting them on-shell. In this factorisation limit we obtained a product of two tree amplitudes providing the cut integrand, while at the level of the ansatz in terms of integral functions this selects a particular set of integral functions that have these two propagators in common. This picks  a unique bubble, which only has two propagators, but allows in general a number of triangles and boxes. 
Considering all possible two-particle cuts gives us sufficient constraints to fix all the coefficients of the integral functions, but the information is entangled between the various  cut constraints.

A natural question is then if we can find a procedure, or rather projection, that directly selects a particular integral function and allows us to tackle individual integral coefficients directly. 
The loop momentum $\ell_1^\mu$ has four independent components and the two-particle cut only constrains two via $\delta(\ell_1^2) = \delta(\ell_2^2) =0 $; in principle we can impose up to two additional constraints 
$\delta(\ell_3^2) =0$ and/or  $\delta(\ell_4^2) =0$. Such {\it generalised cuts} \cite{Bern:2004cz,Britto:2004nc}
are called triple cuts and quadruple cuts, respectively, where the latter is also know as a maximal cut or  leading singularity.

In the case of a triple cut,   the integrand is  a product of
three tree amplitudes \cite{Bern:2004cz}, 
\begin{align}
    \sum_{h_{\ell_1},h_{\ell_2},h_{\ell_3}} A(-\ell_1, i,\ldots, j-1,\ell_2) \times A(-\ell_2, j,\ldots, k-1,\ell_3)  \times
    A(-\ell_3, k,\ldots i-1,\ell_1) \ ,
\end{align}
with $\ell_2 {=} \ell_1 {-} p_i {-} \cdots {-} p_{j-1}$ and $\ell_3 {=} \ell_1 {+} p_k {+} \cdots {+} p_{i-1}$ and $\ell_{1,2,3}^2{=}0$, where  a sum over internal helicity
states is implied. 
Such a cut will select a unique triangle and a number of boxes that share the same three propagators. Notice  that this cut does not detect contributions from bubbles, since they have only
two propagators. Furthermore,  there is a one-dimensional phase-space integration left.

\begin{center}
\begin{figure}[t]
\begin{center}
\scalebox{0.65}{\includegraphics[height=7cm]{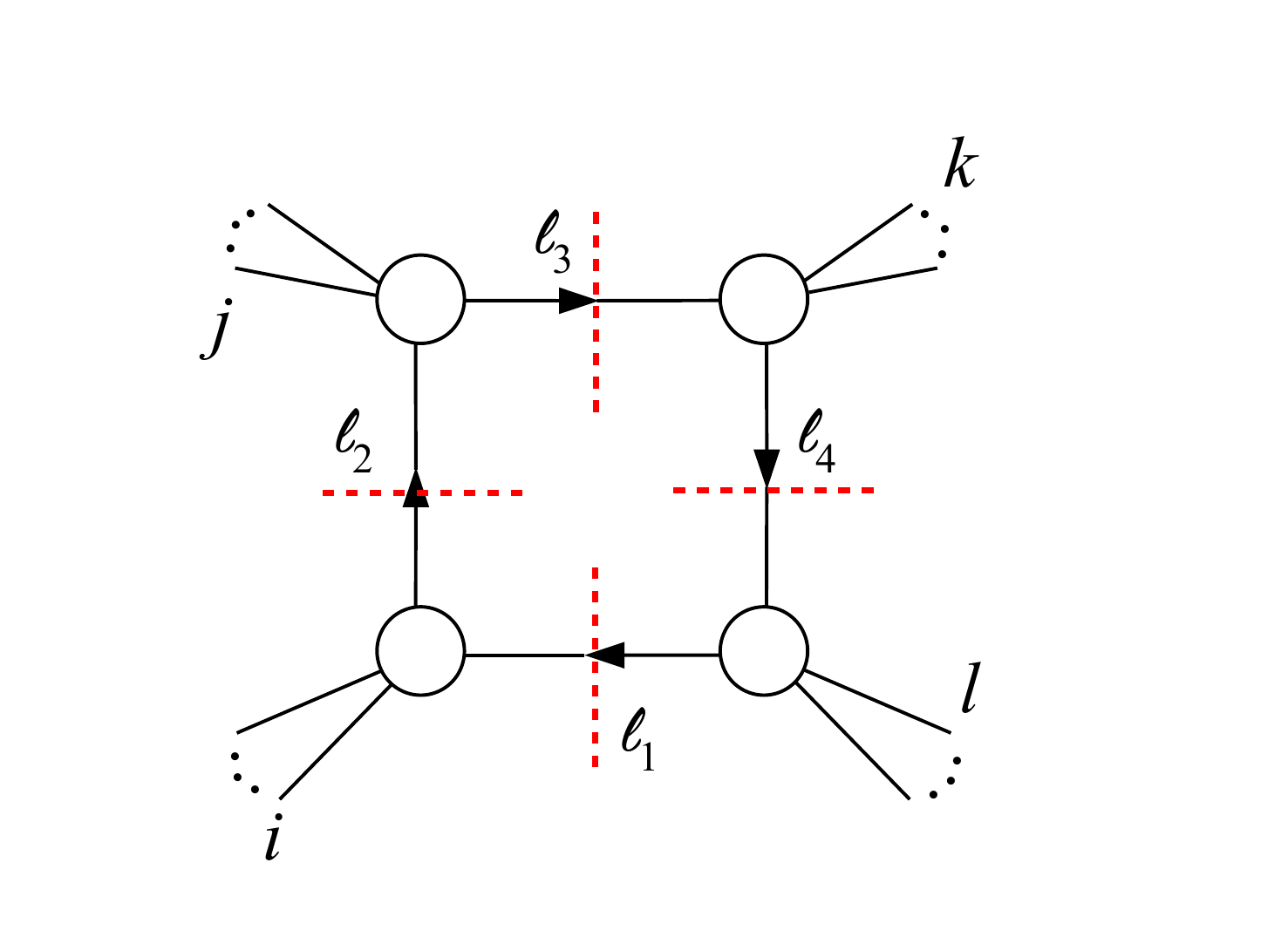} }
\end{center}
\vspace*{-5mm}
\caption{\it Generic quadruple cut of a one-loop amplitude. } 
\label{1loop-generic-qc}
\end{figure}
\end{center}

We now consider a  quadruple,  or maximal cut, shown in Figure~\ref{1loop-generic-qc}. Cutting four momenta  collapses  the loop integration to a sum over a set of solutions that in general is two dimensional. Indeed 
the   cut conditions $\ell_{1,2,3,4}^2{=}0$, with   $\ell_2 {=} \ell_1 {-} p_i {-} \cdots {-} p_{j-1}$, 
$\ell_3 {=} \ell_2 {-} p_j {-} \cdots {-} p_{k-1}$ and $\ell_4 {=} \ell_1 {+} p_l {+} \cdots {+} p_{i-1}$, 
are equivalent to 
\begin{align}
\label{qceqs}
\ell_1^2 = 0 \ , \ \ \ell_2^2-\ell_1^2 = 0
\ , \ \ \ell_3^2-\ell_1^2 = 0 \ , \ \ \ell_4^2-\ell_1^2 = 0 \ . 
\end{align}
This is one quadratic equation and three linear ones, hence there are
two solutions.
%
%

Consider now  our ansatz \eqref{1loopansatz}. The quadruple cut of the left-hand side is a product of four tree amplitudes, shown in Figure \ref{1loop-generic-qc}. 
As for the right-hand side, the quadruple cut picks a unique box function, times its coefficient. 
After integration, the quadruple cut of a scalar box, with  all four propagators replaced by delta functions, simply gives $1$  times a Jacobian.
However this Jacobian appears  on both sides and can then be dropped. 
Thus we arrive at the important result that the box coefficient is
equal  to \cite{Britto:2004nc}
\begin{align}
\begin{split}
    a_{i,j,k,l} = \frac{1}{2} \sum%
    & A(-\ell_1, i,\ldots, j-1,\ell_2) \times A(-\ell_2, j,\ldots, k-1,\ell_3) \\
   \times &  A(-\ell_3, k,\ldots , l-1,\ell_4) \times A(-\ell_4, l,\ldots , i-1,\ell_1) \ , 
\end{split}
\end{align}
where the sum is over the solution set of \eqref{qceqs} and the helicities 
of the four cut~legs, and the factor of $1/2$ is due to the averaging over the two solutions.  
As mentioned in Section~\ref{sec:four-g-oneloop},  bubble and triangle integrals are absent in  $\cN{=}4$ SYM, hence   one-loop amplitudes in this theory are completely determined by quadruple cuts.

\subsection{Example:  one-loop MHV superamplitude in $\cN{=}4$ SYM from quadruple cuts}

\begin{center}
\begin{figure}[t]
\begin{center}
\scalebox{0.65}{\includegraphics[height=7cm]{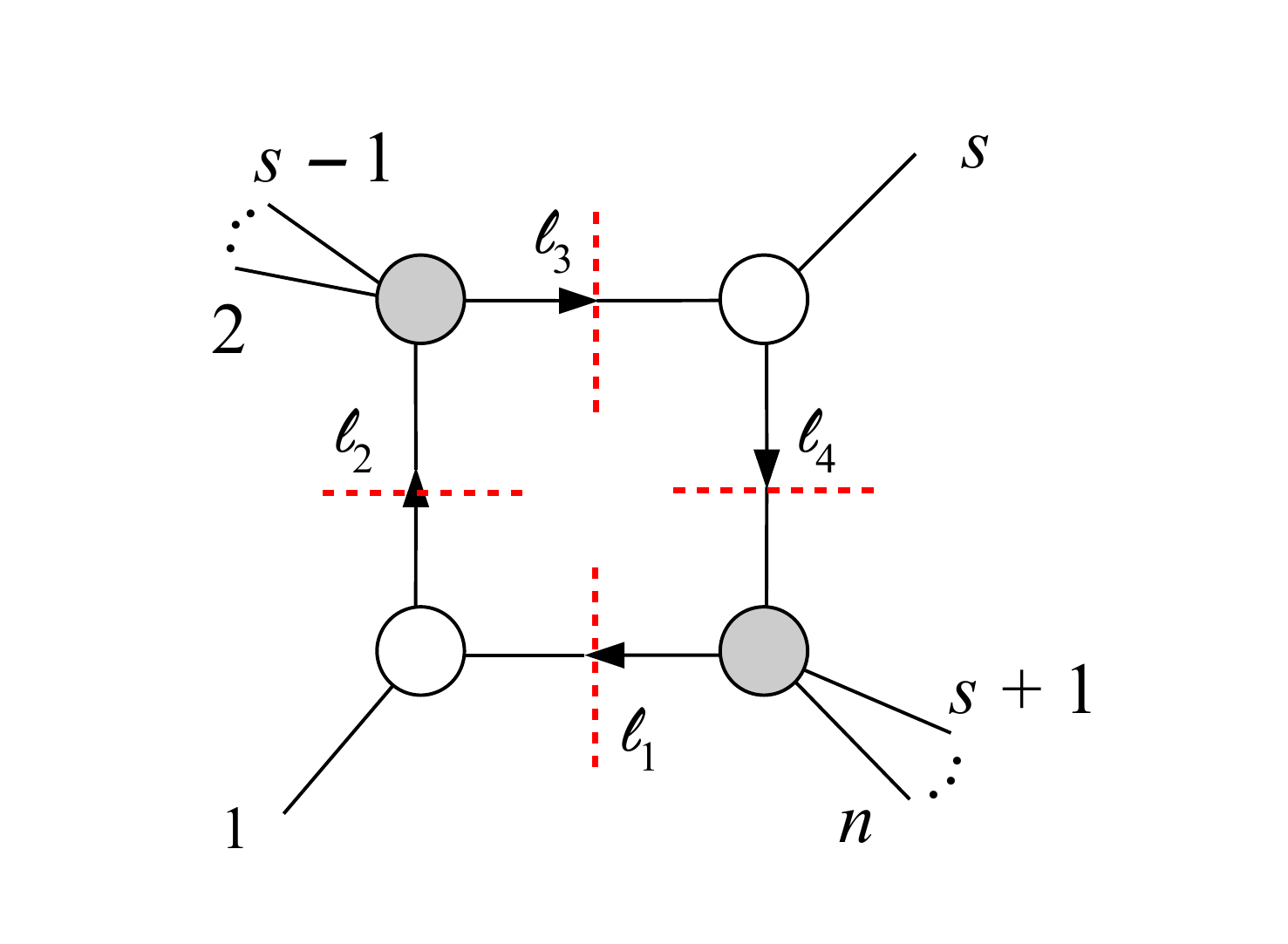} }
\end{center}
\vspace*{-5mm}
\caption{\it Quadruple cut diagram contributing to a one-loop MHV amplitude. The white amplitudes are $\overline{\rm MHV}$ while the grey ones MHV.} 
\label{1loopMHV2me}
\end{figure}
\end{center}

The simplest  application of generalised unitarity is to  the computation of the one-loop MHV amplitude in $\cN{=}4$ SYM, which we will now perform  using superamplitudes
\cite{Drummond:2008bq}. 
It is easy to see that the only non-vanishing quadruple cut  has the  ``two-mass easy'' configuration shown in Figure~\ref{1loopMHV2me}, where two massless legs sit on opposite three-point $\overline{\rm MHV}$ superamplitudes, while  the remaining two   are MHV. The two solutions to the cut equations can be found e.g.~in the Appendix of  \cite{RisagerLarsen:2007wtf}. The first one~is
\begin{align}
\label{sol2me}
    \ell_1 = \frac{|1\ran \lan s|Q }{\lan 1 s\ran}\, ,  \quad 
    \ell_2 = \frac{|1\ran \lan s|P }{\lan s 1\ran}\, ,  \quad
    \ell_3 = \frac{|s\ran \lan 1|P }{\lan s 1\ran}\, ,  \quad
    \ell_4 = \frac{|s\ran \lan 1|Q }{\lan 1s\ran}\, ,  
\end{align}
while the second can be obtained by exchanging $| \bullet \ran \leftrightarrow | \bullet ]$. Note that in  \eqref{sol2me} one has 
\begin{align}
\label{prop2me}
    \lambda_{\ell_1}\!\sim\!\lambda_{\ell_2}\!\sim \lambda_1\, , \qquad  
\lambda_{\ell_3}\!\sim\!\lambda_{\ell_4}\!\sim \lambda_s\, . 
\end{align}
In the case at hand only the first solution contributes since the two three-point amplitudes are $\overline{\rm MHV}$, and hence vanish when evaluated on the second solution, which has $\lt_{\ell_1}\!\sim\!\lt_{\ell_2}\!\sim \lt_1$ and $\lt_{\ell_3}\!\sim\!\lt_{\ell_4}\!\sim \lt_s$.

The   supercoefficient corresponding to the quadruple cut in Figure~\ref{1loopMHV2me} is then 
\begin{align}
\begin{split}
&\cC(1, P, s, Q)  \!=\!\frac{1}{2}\int\!\prod_{i=1}^4 d^4\eta_{\ell_i} \\ &
\cA_3^{\overline{\text{MHV}}} (-\ell_1, 1, \ell_2)
\cA^{\rm MHV} (-\ell_2, 2, \ldots, s{-}1, \ell_3)
\cA_3^{\overline{\text{MHV}}} (-\ell_3, s, \ell_4)
\cA^{\rm MHV} (-\ell_4, s{+}1, \ldots, n, \ell_1) , 
\end{split}
\end{align}
where the spinors are evaluated on the solutions  in \eqref{sol2me}, and the integral  over  the four internal Gra{\ss}mann variables elegantly takes care of the state sums. We also set  $P{:=}\sum_{i=2}^{s-1} p_i  $ and $Q{:= }\sum_{i=s+1}^{n} p_i $.
The relevant amplitudes are: 
\begin{align}
\begin{split}
\label{faa}
\cA_3^{\overline{\text{MHV}}} (-\ell_1, 1, \ell_2) &=-i \frac{\delta^{(4)} (\eta_{-\ell_1} [1\,  \ell_2] + \eta_1 [\ell_2 \, -\ell_1] + \eta_{\ell_2} [ -\ell_1 \, 1])}{[1\,  \ell_2][l_2 \, -\ell_1][ -\ell_1 \, 1]}
\, , 
\\
\cA_3^{\overline{\text{MHV}}} (-\ell_3, s, \ell_4) &=-i \frac{\delta^{(4)} (\eta_{-\ell_3} [s\,  \ell_4] + \eta_s [\ell_4 \, -\ell_3] + \eta_{\ell_4} [ -\ell_3 \, s])}{[s\,  \ell_4][\ell_4 \, -\ell_3][ -\ell_3 \, s]}\, , \\
\cA^{\rm MHV} (-\ell_2, 2, \ldots, s-1, \ell_3) & = i 
\frac{\delta^{(8)} (\lambda_{\ell_3} \eta_{\ell_3} -\lambda_{\ell_2} \eta_{\ell_2} +   \sum_{i=2}^{s-1} \lambda_i \eta_i)}{\lan -\ell_2\,  2\ran\cdots \lan s-1 \, \ell_3\ran \lan   \ell_3 \, -\ell_2\ran }\, , \\
\cA^{\rm MHV} (-\ell_4, s+1, \ldots, n, \ell_1) & =i
\frac{\delta^{(8)} (-\lambda_{\ell_4} \eta_{\ell_4} + \lambda_{\ell_1} \eta_{\ell_1} + \sum_{i=s+1}^{n} \lambda_i \eta_i)}{\lan -\ell_4\,  s+1\ran\cdots \lan n \, \ell_1\ran \lan   \ell_1 \, -\ell_4\ran }\, .
\end{split}
\end{align}
Next we perform the Gra{\ss}mann integrations. This task is simplified by noticing that by supermomentum conservation, we expect to find a result proportional to $\delta^{(8)} (\sum_{i=1}^n \lambda_i \eta_i)$;  we can then simply replace, for instance, the $\delta^{(8)}$ in the last amplitude in \eqref{faa} by this overall supermomentum conservation delta function. Then the integration over $\eta_{\ell_1}$ and  $\eta_{\ell_4}$ must be done using the $\delta^{(4)}$ in the two $\overline{\rm MHV}$ superamplitudes,  giving a factor of $[1\ell_2]^4 [\ell_3s]^4$;  integrating over $\eta_{\ell_2}$ and  $\eta_{\ell_3}$ using the remaining $\delta^{(8)}$ gives a factor of $\lan \ell_2 \ell_3\ran^4$. 

The quadruple-cut integrand can then be simplified by using momentum conservation and 
\eqref{prop2me}. Factoring out $\cA_{\rm MHV} (1, \ldots , n)$, one easily arrives at the result 
%
%
%
\begin{align}
\begin{split}
    \cC (1, P, s, Q)&= - \frac{1}{2}\cA_n^{\rm MHV}  \
     [s \ell_3] \lan \ell_3 \ell_2\ran [\ell_2 1] \lan 1 s\ran =  -\frac{1}{2} \cA_n^{\rm MHV}  \ \text{Tr}_{+}\big( s \ell_3 \ell_2 1 \big)  \ . 
   \end{split}
\end{align}
The  evaluation of the last trace can be  carried out using \eqref{sol2me}. One finds
\begin{align}
\label{factor2me}
    \text{Tr}_{+}\big( s \ell_3 \ell_2 1 \big) = \lan s | P | 1] \, \lan 1 | P | s]= P^2 Q^2- (P+p_1)^2 (Q+p_s)^2\ , 
\end{align}
 so that in conclusion 
 \begin{align}
 \label{coeff2me}
      \cC (1, P, s, Q)&= \frac{1}{2}\cA_n^{\rm MHV} \ \big[ (P+p_1)^2 (Q+p_s)^2-P^2 Q^2\big]\, .
 \end{align}
 This is the coefficient of the two-mass easy box function%
 \footnote{We mention that there is an  alternative expression for this function containing only four polylogarithms \cite{Brandhuber:2004yw}, related to this one by an application of Mantel's nine-dilogarithm identity~\cite{Mantel}.} %
 \begin{align}
\label{2me-box-f}
\begin{split}
    I^{\rm 2me}(p, q, P, Q) &= \int \frac{d^D \ell}{(2 \pi)^D} \frac{1}{\ell^2 (\ell-p)^2 (\ell-p - P)^2 (\ell+Q)^2} \\
  & =  -i \frac{2c_{\Gamma}}{st-P^2 Q^2}\Big\{-\frac{1}{\eps^2}\Big[(-s)^{-\eps}+(-t )^{-\eps}  - (-P^2)^{-\eps}-(-Q^2 )^{-\eps}\Big]  \\
  & +\text{Li}_2 \left( 1 - \frac{P^2}{s}\right) + 
  \text{Li}_2 \left( 1 - \frac{P^2}{t}\right) 
  \text{Li}_2 \left( 1 - \frac{Q^2}{s}\right) + 
  \text{Li}_2 \left( 1 - \frac{Q^2}{t}\right) \\ & - 
  \text{Li}_2 \left( 1 - \frac{P^2 Q^2 }{st}\right)
  + 
  \frac{1}{2}\log^2\Big(\frac{s}{t}\Big) \Big\} \ ,
    \end{split}
\end{align}
 where for generality we relabeled $p_1{\to} p$, $p_s {\to} q$ and set $s{=}(P+p)^2$ and $t{=} (P+q)^2$. 
 Note that the last factor in \eqref{coeff2me} cancels a corresponding one in the expression for $I^{\rm 2me}$ . 

\subsection{Supersymmetric  decomposition and rational terms}
\label{sec:nonsusy}

As discussed earlier, one-loop amplitudes in supersymmetric theories are
 special in that the only  rational terms that appear are  tied
to terms which have discontinuities in four dimensions,  allowing  for the use of spinor-helicity methods.

In non-supersymmetric theories, amplitudes can still
be reconstructed from their cuts, but this requires us to work in 
$4{-}2 \epsilon$ dimensions, with $\epsilon \neq 0$
\cite{vanNeerven:1985xr,Bern:1995db}. 
While this is important conceptually, it also implies that we 
have to work with
gluon amplitudes in  
away from four  dimensions and  the elegance of the spinor-helicity formalism.

A crucial  simplification comes from the following observation, known as the 
supersymmetric decomposition of one-loop gluon amplitudes
in pure Yang-Mills:
a one-loop amplitude $\cA_{\rm g}$
with gluons running in the loop can 
be re-written  as \cite{Bern:1994zx,Bern:1994cg}
\begin{align}
\label{susydec}
A^{(1)}_{\rm g} =\underbrace{(A^{(1)}_{\rm g} \, + \, 4 A^{(1)}_{\rm f} \, + \,
3 A^{(1)}_{\rm s})}_{A^{(1)}_{\cN=4}} \  - \
4\underbrace{( A^{(1)}_{\rm f}  + A^{(1)}_{\rm s})}_{A^{(1)}_{\cN=1}} \ + \  A^{(1)}_{\rm s}
\ , 
\end{align}
where $A^{(1)}_{\rm f}$ ($A^{(1)}_{\rm s}$) is an  amplitude with the same
external particles as  before  but with a Weyl fermion
(complex scalar) in the adjoint  of the gauge group circulating
in the loop.
This decomposition is powerful since we recognise that 
the first two terms on the right-hand side of
\eqref{susydec} come from  an $\cN \! = \! 4$ multiplet and
(minus four times) a chiral $\cN \! = \! 1$ multiplet,
respectively; therefore, they  are
four-dimensional cut-constructible -- an observation that 
 simplifies their calculation considerably.
The last term in \eqref{susydec}, $A^{(1)}_{\rm s}$,
is the contribution due to a complex scalar  in the loop, which  in $D$ dimensions    is much easier to compute than having a gluon in the loop.

An instructive  example is  all-plus four-point amplitude in pure Yang-Mills $A^{(1)}(1^+2^+3^+4^+)$ produced by gluons running in the loop. Using \eqref{susydec} we can immediately  relate this to the situation where  a scalar is  running in the loop since   the first two contributions in \eqref{susydec} vanish for this helicity configuration in any supersymmetric theory. 
Thus, the entire   contribution
comes from the last term in~\eqref{susydec}.
\begin{center}
\begin{figure}[h]
\begin{center}
\scalebox{0.65}{\includegraphics[height=7cm]{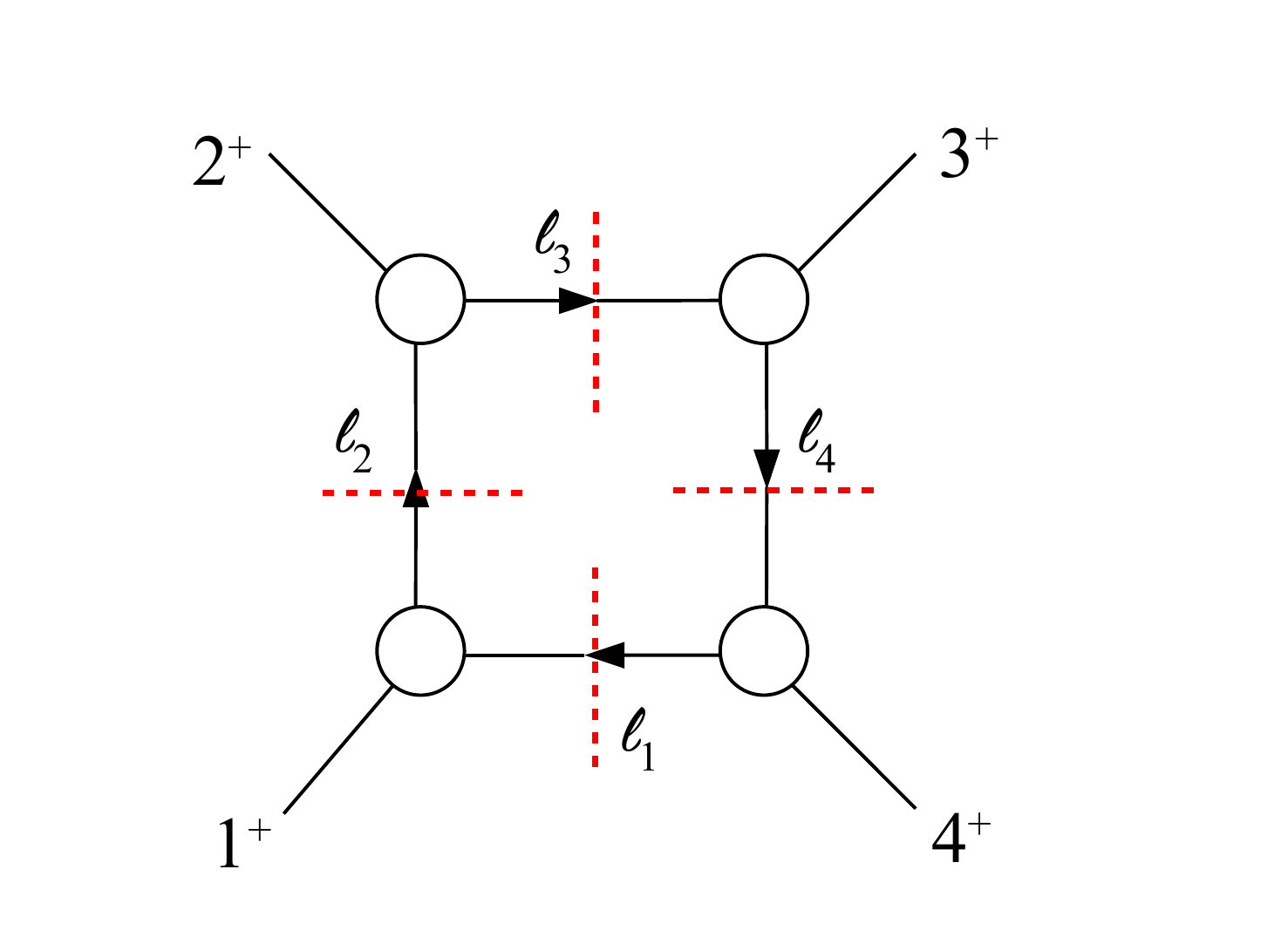} }
\end{center}
\vspace*{-5mm}
\caption{\it Quadruple cut diagram contributing to the all-plus one-loop amplitude. The internal loop is a real scalar field.} 

\label{allplus}
\end{figure}
\end{center}
In order to compute it,  we can perform  a $D$-dimensional quadruple cut
\cite{Brandhuber:2005jw}, by gluing four three-point amplitudes involving two scalars of mass $\mu^2$ and one gluon. Such amplitudes have  the form
\begin{align}
\label{threepoint}
A (\ell_1 , p_1^+ , \ell_2 ) =
A (\ell_1 , p_1^+ , \ell_2 )=
\frac{\lan \xi  | \ell_1 | p_1 ]}{\lan \xi  p_1 \ran}
\ ,
\end{align}
where $\ell_1 + \ell_2 + p_1 {=}0$ and $|\xi\ran$ is an arbitrary reference spinor.
The $D$-dimensional quadruple
cut integrand is then given by
\begin{align} \label{4vertices}
\frac{\lan \xi_1\vert \ell_1 \vert 1]}{\lan \xi_11\ran }\
\frac{\lan \xi_2\vert \ell_2 \vert 2]}{\lan \xi_22\ran }\
\frac{\lan \xi_3\vert \ell_3 \vert 3]}{\lan \xi_33\ran }\
\frac{\lan \xi_4\vert \ell_4 \vert 4]}{\lan \xi_44\ran }
\ ,
\end{align}
which, using  the $D$-dimensional on-shell condition $(\ell^{(D)}_i)^2 {=}0 {=}  (\ell^{(4)}_i)^2 {-} \mu^2$ and momentum conservation, evaluates to 
$
\mu^4 \, \frac{[12][34]}{\lan 12\ran\lan 34\ran}$.
We also have used here the standard trick that a massless scalar in $D$ dimensions can be viewed as a massive scalar in $D=4$ with mass $\mu^2$ coming from the loop momentum components in the extra $(-2\epsilon)$ dimensions, which we have to integrate over.

Finally, it can also be seen \cite{Brandhuber:2005jw} that two-particle and three-particle cuts  do not give any
new contributions, and hence we arrive at the result (after replacing the four delta functions by propagators):
\begin{align}
    A^{(1)}(1^+2^+3^+4^+)
    = 2 \frac{[12][34]}{\lan 12\ran\lan 34\ran} I^{\rm 0m}_4[\mu^4] \ ,
\end{align}
where $I^{\rm 0m}_4[\mu^4] {= -}\frac{i}{ (4 \pi)^2} \frac{1}{ 6} {+} \mathcal{O}(\epsilon)$ is the scalar box integral with an insertion
of $\mu^4$ 
\cite{Bern:1995db}, and the factor of two comes from the two real scalars running in the loop.

\subsection{Beyond unitarity and higher loops}

In the discussion above we only gave a flavour of the power of unitarity methods at one loop. In recent years these methods have been extended in many fruitful directions,  in particular generalised unitarity has been adapted to the computation of higher-loop amplitudes in essentially any theory as long as all internal propagators are massless. These include analytic approaches,  and several numerical implementations  for amplitudes in QCD up to two loops. In parallel, tremendous progress has been made in the evaluation of the required two- and higher-loop Feynman integrals,
and some of this progress is reviewed in detail in Chapters~3 \cite{Abreu:2022mfk} and~4 \cite{Blumlein:2022zkr} of this review.  

Furthermore, for highly symmetric theories such as  $\mathcal{N}{=}4$ supersymmetric Yang-Mills,
even more advanced methods have been developed.  These employ bootstrap ideas that completely avoid the (separate) determination of coefficients of the basis of integral functions and the problem of evaluating the integrals themselves. These are reviewed in Chapter~5 \cite{Papathanasiou:2022lan} of this review, and are based on a vastly improved understanding of the mathematical properties of complete amplitudes and the relevant function spaces, and include:  transcendental functions and their associated symbols, the relation between singularities of amplitudes and cluster algebras, and the Steinmann relations.  
Recently, these developments have been
exploited in \cite{Dixon:2022rse} for an unprecedented eight-loop computation of form factors
(see Section~\ref{sec:11} for more on form factors).
Readers interested in these exciting topics are invited to consult 
Chapter~5  of this review \cite{Papathanasiou:2022lan}.

\section{BPS and non-BPS form factors. Applications to Higgs amplitudes}

\label{sec:11}

\subsection{General properties}

Form factors  appear in several important contexts  in gauge theory. 
The form factor of  (gauge-invariant)
 $\mathcal{O}(x)$ between the vacuum and an $n$-particle state is defined as  
\begin{align}
\label{eq:Fourier-FF}
\begin{split}
F_\mathcal{O} (1,\dots,n;q):=\, &\int\!d^4x\, e^{-iq\cdot x}\langle 1\ldots n|\mathcal{O}(x)|0\rangle\,=\,(2\pi)^4 \ \delta^{(4)}\big(q-\sum_{i=1}^n p_i\big)\langle 1\ldots n|\mathcal{O}(0)|0\rangle\ , 
\end{split}
\end{align}
where the momentum conserving $\delta$-function follows from translation invariance. All legs are on shell except that   corresponding to the operator, since in general $q^2\neq 0$, hence form factors fall in between correlators (fully off shell) and amplitudes (fully on shell).

In addition to conceptual reasons, form factors are important because of their role in  several contexts. Notable  examples  include the form factor of the electromagnetic current, which computes the electron  $g{-}2$, and that of the 
 hadronic electromagnetic current with an external hadronic state, which appears in the study of deep inelastic  scattering and 
 $e^+ e^-\to$ hadrons. 
 
 Form factors also play a prominent role  in 
the study of  scattering processes in QCD  involving the Higgs boson and many gluons. 
  At one loop, the coupling of the Higgs  to the gluons is induced by  a quark loop,  with the top   giving the most important contribution.
  These gluon-fusion processes  can   be described using an    effective field theory approach, with  the   quark loop traded  for  a set of local interactions of increasing  dimension. The leading interaction in  the limit $m_H\!\ll\!m_t$, where  $m_H$  and $m_t$ are the masses of the Higgs and the top quark, is the dimension-5 operator $
\mathcal{L}_5 \!\sim\!H \, {\rm Tr} (F^2)
$,
 where $H$ denotes the Higgs boson and $F$  the gluon field strength \cite{Wilczek:1977zn,Shifman:1979eb,Dawson:1990zj}. It follows that  the  amplitude of a Higgs and a $n$ gluons    in the limit of infinite top mass  is the   form factor   
$\langle g_1\ldots g_n | \, {\rm Tr} (F^2 ) \, | 0 \rangle$  of  the  operator ${\rm Tr} ( F^2)$. 

Form factors share some of the beautiful properties of amplitudes, including their simplicity. For instance, at tree level one has \cite{Dixon:2004za}
\begin{align}
    \langle 1^+, \ldots, i^-, \ldots , j^-, \ldots  n^+ | \, {\rm Tr} (F_{\rm SD} ^2 ) \, | 0 \rangle \sim 
\frac{ \vev{i j }^4}{\vev{12}\vev{23} \cdots \vev{n1}}   \, ,
    \end{align}
with $q{=}p_1 + \cdots p_n$, and where  $F_{\rm SD}$ denotes the self-dual part of the field strength. It follows from the discussion above that this is the leading Higgs plus multi-gluon MHV amplitude at tree level. 
A systematic study of form factors was initiated in 
\cite{Brandhuber:2010ad}, and quickly extended to supersymmetric form factors \cite{Brandhuber:2011tv}. Among the various results, they satisfy BCFW recursion relations \cite{Brandhuber:2010ad}, also at loop level \cite{Bianchi:2018peu},  can be computed using (generalised) unitarity, and are invariant under a form of  dual conformal symmetry \cite{Bianchi:2018rrj},  which is broken at loop level.
Analytic non-supersymmetric form factors were recently computed at one loop \cite{AccettulliHuber:2019abj} using dimensional reconstruction \cite{Giele:2008ve,Ellis:2008ir,Bern:2010qa,Davies:2011vt}. 

A recent line of research has investigated supersymmetric form factors and  possible patterns or similarities with non-supersymmetric, phenomenologically relevant ones. While supersymmetrising the state is straightforward, some thoughts have to be devoted to which operators it may be worthwhile to consider. In this respect, one can observe that   
the operator ${\rm Tr} (F_{\rm SD} ^2 )$ discussed earlier is the first term in the on-shell Lagrangian of $\cN{=}4$ SYM,  which has the schematic form 
\begin{align}
\label{Lonshell}
    \mathcal{L}_\text{on-shell} \sim {\rm Tr} (F_{\rm SD}^2)
+ g\, {\rm Tr} (\psi \psi \phi) + g^2 \,{\rm Tr}( [\phi,\phi]^2)
\, .
\end{align}
 It is a descendant of the half-BPS  operator ${\rm Tr}(X^2)$, with  $X$ being  any  of the (complex) scalars in $\cN{=}4$ SYM, and is obtained by  acting on it with four supersymmetry charges.  Both ${\rm Tr} (X^2)$ and $\mathcal{L}_\text{on-shell} $  belong to the chiral part of the  stress-tensor multiplet  $\mathcal{T}_2$ \cite{Eden:2011yp} and, because they are protected, their form factors are free of ultraviolet divergences. 
Such form factors were studied vigorously in several works  \cite{Brandhuber:2011tv,Brandhuber:2012vm,Penante:2014sza,Brandhuber:2014ica,Dixon:2020bbt,Dixon:2021tdw}. This study was later extended to non-protected operators \cite{Brandhuber:2016fni,Brandhuber:2017bkg,Brandhuber:2018xzk,Brandhuber:2018kqb} such as the Konishi multiplet.  

Form factors are a source of many surprises, and we would like to list some of the most unexpected ones. To begin with, it was found at two loops \cite{Brandhuber:2012vm} and later confirmed at higher loops \cite{Dixon:2020bbt,Dixon:2021tdw,Dixon:2022rse}, that  the form factor of the stress-tensor multiplet in $\cN{=}4$ SYM  with three external particles is  maximally transcendental,  similarly to amplitudes.%
\footnote{The precise statement is  for certain finite remainders \cite{Bern:2005iz} of the form factors obtained by subtracting universal infrared-divergent terms.} 
Even more surprisingly, it was found in  \cite{Brandhuber:2012vm} that the two-loop remainder is identical to the maximally transcendental part of that of the  form factor $\lan g^+ g^+ g^{\pm} | {\rm Tr}\, F_{\rm SD}^2 | 0 \rangle$ \cite{Gehrmann:2011aa} in QCD -- the first occurrence of the  principle of maximal transcendentality  \cite{Kotikov:2004er} in a kinematic-dependent quantity. These connections do not stop at protected operators: the maximally transcendental part of the remainder of the minimal form factor of the operator ${\rm  Tr} ( X[Y, Z])$, that is $\lan \bar{X}\bar{Y} \bar{Z} | {\rm Tr} ( X[Y, Z]) |0\ran $ is identical to that of $\lan \bar{X}\bar{X} \bar{X} | {\rm Tr} (X^3) |0\ran $ \cite{Brandhuber:2016fni}; and finally, this equality extends to the form factor of the operator ${\rm Tr}\, (F_{\rm SD}^3)$  \cite{Brandhuber:2017bkg,Brandhuber:2018xzk,Brandhuber:2018kqb}, which describes higher-derivative corrections to the Higgs effective theory \cite{Buchmuller:1985jz,Neill:2009tn,Neill:2009mz,Harlander:2013oja,Dawson:2014ora}. 
In \cite{Guo:2022pdw} a proof of the principle of maximal transcendentality for
two-loop form factors involving ${\rm Tr}\, (F^2)$ and ${\rm Tr}\, (F^3)$ was presented.
As a result, it seems that the maximally transcendental part of  Higgs plus multi-gluon processes could be equivalently computed in $\cN{=}4$ SYM!

 We also mention the intriguing  connections   between the  (infrared-finite) remainder functions of the three-point form factor $\lan \bar{X}\bar{X} \bar{X} | {\rm Tr} (X^3) |0\ran $ and the remainder for the six-point MHV amplitude \cite{Brandhuber:2012vm,Dixon:2020bbt,Dixon:2021tdw,Dixon:2022rse}, which recently have been explained by an antipodal duality that relates the discontinuities of the form factor to the derivatives of the amplitude. The remainder functions for these two quantities can be expressed in terms of three dimensionless variables,  $(u, v, w)$, representing  ratios of Mandelstam variables satisfying  $u+v+w{=}1$ in the former case, and unconstrained dual-conformal invariant cross-ratios in the latter. This remarkable duality then connects the form factor remainder function and the parity-even part of the amplitude remainder function on the surface $u+v+w=1$.  
\black 

Finally, form factors  found  an application in \cite{Zwiebel:2011bx,Wilhelm:2014qua} to the study of the complete one-loop dilatation operator of $\cN{=}4$ SYM \cite{Beisert:2003tq,Beisert:2003jj,Beisert:2003yb}, with the Yangian invariance of the latter \cite{Dolan:2003uh}
being a direct consequence \cite{Brandhuber:2015dta} of that of the $\cN{=}4$ SYM $S$-matrix \cite{Drummond:2009fd}. We also mention recent applications in effective field theories of the Standard Model, e.g.~in classifying marginal operators and studying the mixing problem \cite{Caron-Huot:2016cwu,EliasMiro:2020tdv,Baratella:2020lzz,Bern:2020ikv,AccettulliHuber:2021uoa}.

\subsection{Example: one-loop Sudakov form factor}
\label{sec:FFSuda}

To have a taste of  form factors,  we now  compute that  of the on-shell Lagrangian \eqref{Lonshell} with an external state of two positive-helicity gluons, known as the Sudakov form factor.  Only the field-strength part of $\cL_\text{on-shell}$ contributes, and at tree level we can normalise the operator   to~have  
$
    F^{\rm tree}_{{\rm Tr}(F^2)}(1^{+} 2^{+}) = [12]^2 $.
This form factor depends on a single kinematic invariant   $s=(p_1 + p_2)^2$, and at one loop we only have   a cut in this channel. 
This gives
\begin{align}
\begin{split}
F^{(1)}_{{\rm Tr}(F^2)}(1^{+} 2^{+}) |_{s-{\rm cut}} & =
F^{\rm tree}_{{\rm Tr}(F^2)}(-\ell_1^{+} -\ell_2^{+}) A^{(0)}_4(\ell_1^-,1^+,2^+,\ell_2^-) \\
& = 2 [-\ell_1 \, {-}\ell_2]^2 \times
i \frac{\langle \ell_2 \ell_1 \rangle^3}{\langle \ell_1 1 \rangle \langle 12 \rangle \langle 2 \ell_2 \rangle} 
= \frac{2 i (\ell_1+\ell_2)^2 \langle \ell_2  \ell_1 \rangle [\ell_2 \ell_1]}{\langle 12 \rangle \langle 2 \underbrace{\ell_2 \rangle [\ell_2}_{\ell_2 = -\ell_1 -p_1-p_2} \ell_1] \langle \ell_1 1 \rangle} \\
& = \frac{- 2 i (p_1+p_2)^3}{\langle 12 \rangle^2 (\ell_1+p_1)^2 }  = \frac{- 2 i s [12]^2}{ (\ell_1+p_1)^2} \ .
\end{split}
\end{align}
In order to obtain the uplifted integrand, following the strategy described in Section~\ref{sec:four-g-oneloop},  we have to multiply this by  $\frac{i}{\ell_1^2} \frac{i}{\ell_2^2}$, and further integrating we find 
 $   2 i s [12]^2 \times I_3^{\rm 1m}(s)$,
  where the one-mass triangle is given by
\begin{align}
I_3^{\rm 1m}(s) = \int \frac{d^{4-2\epsilon} \ell}{(2\pi)^{4-2\epsilon}}\dfrac{1}{\ell^2(\ell+p_1)^2(\ell+p_1+p_2)^2}= -i \frac{c_\Gamma}{\epsilon^2}
(-s)^{-1-\epsilon} \ .
\end{align}
In conclusion, we find that 
\begin{align}
    F^{(1)}_{{\rm Tr}(F^2)}(1^{+} 2^{+})/F^{\rm tree}_{{\rm Tr}(F^2)}(1^{+} 2^{+}) = -\frac{2 c_\Gamma}{\eps^2} (-s)^{-\eps} \ .
\end{align}
Note  that the Sudakov form factor is equal
to twice the IR-divergent term of the contribution of a given two-particle invariant $s_{i,i+1}$ to the one-loop amplitude computed in  \eqref{IRoneloop}. Interestingly this relation holds for general one-loop amplitudes (see for instance \cite{Bena:2004xu} for a unitarity-based proof), and also beyond one loop  \cite{Magnea:1990zb, Sterman:2002qn}.

\newpage

\section*{Acknowledgments}

It is a pleasure to thank   James Bedford, Lorenzo Bianchi, Gang Chen, James Drummond, \"{O}mer G\"{u}rdo\u{g}an, Johannes Henn, 
Paul Heslop, Edward Hughes, Dimitrios Korres, Martyna Kostaci\'{n}ska Jones, Simon McNamara, Robert Mooney, Rodolfo Panerai,   Brenda Penante, Bill Spence, Congkao Wen, Gang Yang and Donovan Young for enjoyable collaborations on topics related to this  article. Many thanks to Manuel Accettulli Huber, Stefano De Angelis and Shun-Qing Zhang   for their careful reading of our article and for comments. 
This work  was supported  by the European Union's Horizon 2020 research and innovation programme under the Marie Sk\l{}odowska-Curie grant agreement No.~764850 {\it ``\href{https://sagex.org}{SAGEX}''}. We also acknowledge support from 
 the Science and Technology Facilities Council (STFC) Consolidated Grant ST/T000686/1 \textit{``Amplitudes, strings  \& duality''}.  \\

\appendix

\section{Conventions and   Lorentz transformations of spinor variables}
\label{app:1}

\noindent
{\it Conventions}. 
Spinor indices are raised and lowered using the Levi-Civita tensor as 
\begin{align}
\begin{split}
   \lambda^{\alpha} &= \eps^{\alpha \beta}\lambda_\beta \ , \qquad
   \lambda_{\alpha} = \eps_{\alpha \beta}\lambda^\beta
    \ , \\
    \lt^{\dot\alpha} &= \eps^{\dot{\alpha} \dot{\beta}}\lt_{\dot\beta}
 \ , \qquad
 \lt_{\dot\alpha} = \eps_{\dot{\alpha} \dot{\beta}}\lt^{\dot\beta}
    \,  ,
\end{split}
\end{align}
with $\eps^{\alpha \gamma}\, \eps_{\gamma\beta} = \delta^\alpha_{\, \beta}$, and $\eps^{\dot\alpha \dot\gamma}\, \eps_{\dot\gamma\dot\beta} = \delta^{\dot\alpha}_{\, \dot\beta}$. 
We also define \begin{align}
    \sigma_{\mu \alpha\dot\alpha} = (\mathbb{1}, \vec{\sigma})\ , \qquad 
 \bar{\sigma}_\mu^{\dot\alpha\alpha} =(\mathbb{1},-\vec{\sigma})\ , 
 \end{align}
 where   
$\vec{\sigma}$ are  the Pauli matrices. They are related as
\begin{align}
    \sigma_{\mu\, \alpha\dot\alpha}= \epsilon_{\alpha\beta}\,  \epsilon_{\dot\alpha\dot\beta}\, 
\bar{\sigma}_\mu^{\dot\beta\beta} \, .
\end{align}
We also note the completeness relations
\begin{align}
\sigma_{\mu\, \alpha\dot\alpha}\, \sigma^{\mu}_{\beta \dot\beta} = 2 \, \epsilon_{\alpha \beta}\, \epsilon_{\dot\alpha \dot\beta}\ , \qquad
\bar{\sigma}_\mu^{\dot\alpha\alpha}\, \bar{\sigma}^{\mu\, \dot\beta\beta} = 2 \, \epsilon^{\alpha \beta}\, \epsilon^{\dot\alpha \dot\beta}\ , 
\end{align}
and the normalisation condition \begin{align}
    {\rm Tr}\, (\bar\sigma^\mu \sigma^\nu) = 2 \, \eta^{\mu \nu}\ , 
\end{align}
with $\eta^{\mu \nu} = (1, -1, -1, -1)$. \\

\noindent
{\it Transformations under the Lorentz group}. 
 The complexified four-dimensional Lorentz group $SO(3,1)$ is locally isomorphic to $SL(2, \mathbb{C}) {\times} SL(2, \mathbb{C})$, with the isomorphism being realised as in \eqref{isomorphism}. Its  representations are then labeled  as $(m,n)$, where $m, n {\in} \frac{1}{2}\mathbb{Z}$. 
In real Minkowski space, the second
$SL(2, \mathbb{C})$  has to be identified with the complex conjugate of the first.%
\footnote{Recall from Section~\ref{sec:2.2} that in real Minkowski space   $(\la^{\alpha})^{\ast}{=}\pm \tla^{\da}$. For completeness we also note that in $(++--)$ signature the Lorentz group $SO(2,2)$ is locally isomorphic to   $SL(2, \mathbb{R}) {\times} SL(2, \mathbb{R})$, with the two $SL(2, \mathbb{R})$ factors being independent; in this case the spinor representations are real.}
Our helicity spinors $\lambda$ and $\tilde{\lambda}$ then transform  in the 
$(1/2, 0)$ and $(0, 1/2)$ representations, respectively, that~is 
\begin{align}
    \lambda_{\alpha}\to M_{\alpha}^{\ \beta}\,  \lambda_\beta\ , \qquad \lt_{\dot{\alpha}} \to (M^{\ast})_{\dot{\alpha}}^{\ \dot{\beta}}\, \lt_{\dot{\beta}}= \lt_{\dot{\beta}} \, (M^\dagger)^{\dot{\beta}}_{\ 
    \dot\alpha}\, , 
\end{align}
and the momentum $p_{\alpha \dot{\alpha}}$  in the $(1/2,1/2)$,  
i.e.
\begin{align}
    p_{\alpha \dot{\alpha}}\to M_{\alpha}^{\ \beta} \, p_{\beta \dot{\beta}}\, (M^{\ast})_{\dot{\alpha}}^{\ \dot{\beta}} =
    (M p M^\dagger)_{\alpha \dot{\alpha}} \, . 
    \end{align}
Here $M{\in} SL(2, \mathbb{C})$, so that $\text{det}\, M {=} 1$. The Levi-Civita symbols $\epsilon_{\alpha \beta}$ and $\epsilon_{\dot{\alpha}\dot{ \beta}}$
are  invariant tensors: for  instance,  
 $   \epsilon_{\alpha \beta}{\to} M_\alpha^{\ \alpha^{\prime}}  M_\beta^{\ \beta^{\prime}} \epsilon_{\alpha^\prime \beta^\prime}{=} \text{det} M \, \epsilon_{\alpha \beta}{ =} \epsilon_{\alpha \beta}\ . 
$
We can write this transformation   in matrix form as $M\epsilon M^T = \epsilon$, with a similar relation  $M^\ast \epsilon M^\dagger = \epsilon$ for $\epsilon_{\dot{\alpha}\dot{\beta}}$. 
It is also important to work out  the transformations  of 
$\lambda^\alpha := \epsilon^{\alpha \beta}\lambda_\beta$ and 
$\lt^{\dot{\alpha}} := \epsilon^{\dot{\alpha} \dot{\beta}}\lt_{\dot{\beta}}$. Calling $\epsilon$ and $\tilde{\epsilon}$ the matrices whose elements are $\epsilon_{\alpha \beta}$ and~$\epsilon^{\alpha \beta}$,
\begin{align}
    \lambda^\alpha \to \epsilon^{\alpha \beta}M_{\beta}^{ \ \gamma} \epsilon_{\gamma \delta} \lambda^\delta = (\tilde{\epsilon} M \epsilon \lambda)^\alpha = \big(\tilde{\epsilon} \epsilon ( M^T)^{-1}  \lambda\big)^\alpha = \lambda^\beta (M^{-1})_\beta^{\ \alpha} \ , 
\end{align}
where we used $M \epsilon = \epsilon (M^T)^{-1}$.
Similarly
$\lt^{\dot\alpha}:= \epsilon^{\dot{\alpha} \dot{\beta}}\lt_{\dot{\beta}}$ transforms as 
\begin{align}
\lt^{\dot\alpha}\to \epsilon^{\dot{\alpha}\dot{\beta}} (M^\ast)_{\dot{\beta}}^{\ \dot{\gamma}}\, \epsilon_{\dot{\gamma}\dot{\rho}} \, \lt^{\dot{\rho}} = (\tilde{\epsilon} \epsilon (M^\dagger)^{-1}\lt)^{\dot{\alpha}}= ((M^\dagger)^{-1})^{\dot{\alpha}}_{\ \dot{\beta}}\, \lt^{\dot{ \beta}} 
\ , 
\end{align}
where we used $M^\ast \epsilon (M^\ast)^T = \epsilon$, from which it follows that $M^\ast \epsilon = \epsilon (M^\dagger)^{-1}$, and we also called $\eps$ and $\tilde{\eps}$ the matrices with elements $\eps_{\dot{\alpha} \dot{\beta}}$ and $\eps^{\dot{\alpha} \dot{\beta}}$.  Summarising, we have that 
\begin{align}
\begin{split}
\lambda_{\alpha}&\to M_{\alpha}^{\ \beta}\,   \lambda_\beta\ , \qquad\qquad  \lt_{\dot{\alpha}} \to \lt_{\dot{\beta}} \, (M^\dagger)^{\dot{\beta}}_{\ 
    \dot\alpha}\, , \\ 
  \lambda^\alpha & \to  \lambda^\beta (M^{-1})_\beta^{\ \alpha} \ , \ \, \qquad 
  \lt^{\dot\alpha}\to  ((M^\dagger)^{-1})^{\dot{\alpha}}_{\ \dot{\beta}}\, \lt^{\dot{ \beta}} 
\ . 
\end{split}
\end{align}
As a result, the brackets  $\lan i \, j\ran {:=} \lambda_i^\alpha \lambda_{j\alpha }$ and 
$[i \, j] {:= }\lt_{i\dot{\alpha}} {\tilde{\lambda}}^{j\dot{\alpha}}$ are manifestly Lorentz invariant.

 \newpage

\bibliographystyle{iopart-num}
\newcommand{\eprint}[2][]{\href{https://arxiv.org/abs/#2}{\tt{#2}}}
\addcontentsline{toc}{section}{References}
\section*{References}
\bibliography{remainder}

\providecommand{\newblock}{}
\begin{thebibliography}{100}
\expandafter\ifx\csname url\endcsname\relax
  \def\url#1{{\tt #1}}\fi
\expandafter\ifx\csname urlprefix\endcsname\relax\def\urlprefix{URL }\fi
\providecommand{\eprint}[2][]{\url{#2}}

\bibitem{Mangano:1990by}
Mangano M~L and Parke S~J 1991 {\em Phys. Rept.\/} {\bf 200} 301--367
  (\textit{Preprint} \eprint{hep-th/0509223})

\bibitem{Parke:1986gb}
Parke S~J and Taylor T~R 1986 {\em Phys. Rev. Lett.\/} {\bf 56} 2459

\bibitem{Mangano:1987xk}
Mangano M~L, Parke S~J and Xu Z 1988 {\em Nucl. Phys. B\/} {\bf 298} 653--672

\bibitem{Witten:2003nn}
Witten E 2004 {\em Commun. Math. Phys.\/} {\bf 252} 189--258 (\textit{Preprint}
  \eprint{hep-th/0312171})

\bibitem{Penrose:1967wn}
Penrose R 1967 {\em J. Math. Phys.\/} {\bf 8} 345

\bibitem{Penrose:1972ia}
Penrose R and MacCallum M~A~H 1972 {\em Phys. Rept.\/} {\bf 6} 241--316

\bibitem{Roiban:2004vt}
Roiban R, Spradlin M and Volovich A 2004 {\em JHEP\/} {\bf 04} 012
  (\textit{Preprint} \eprint{hep-th/0402016})

\bibitem{Berkovits:2004hg}
Berkovits N 2004 {\em Phys. Rev. Lett.\/} {\bf 93} 011601 (\textit{Preprint}
  \eprint{hep-th/0402045})

\bibitem{Roiban:2004yf}
Roiban R, Spradlin M and Volovich A 2004 {\em Phys. Rev. D\/} {\bf 70} 026009
  (\textit{Preprint} \eprint{hep-th/0403190})

\bibitem{Cachazo:2004kj}
Cachazo F, Svrcek P and Witten E 2004 {\em JHEP\/} {\bf 09} 006
  (\textit{Preprint} \eprint{hep-th/0403047})

\bibitem{Lance-talk}
Dixon L~J Future challenges for {(N)NLO} accuracy in {QCD}
  \urlprefix\url{https://indico.cern.ch/event/93790/contributions/1281096/attachments/1103796/1574776/LDTrentoFuture.pdf}

\bibitem{Xu:1986xb}
Xu Z, Zhang D~H and Chang L 1987 {\em Nucl. Phys. B\/} {\bf 291} 392--428

\bibitem{Gunion:1985vca}
Gunion J~F and Kunszt Z 1985 {\em Phys. Lett. B\/} {\bf 161} 333

\bibitem{Kleiss:1985yh}
Kleiss R and Stirling W~J 1985 {\em Nucl. Phys. B\/} {\bf 262} 235--262

\bibitem{Bjorken:1966kh}
Bjorken J~D and Chen M~C 1966 {\em Phys. Rev.\/} {\bf 154} 1335--1337

\bibitem{Henry:1967jm}
Henry G~R 1967 {\em Phys. Rev.\/} {\bf 154} 1534--1536

\bibitem{DeCausmaecker:1981wzb}
De~Causmaecker P, Gastmans R, Troost W and Wu T~T 1981 {\em Phys. Lett. B\/}
  {\bf 105} 215

\bibitem{DeCausmaecker:1981jtq}
De~Causmaecker P, Gastmans R, Troost W and Wu T~T 1982 {\em Nucl. Phys. B\/}
  {\bf 206} 53--60

\bibitem{Berends:1981uq}
Berends F~A, Kleiss R, De~Causmaecker P, Gastmans R, Troost W and Wu T~T 1982
  {\em Nucl. Phys. B\/} {\bf 206} 61--89

\bibitem{Berends:1983ez}
Berends F~A, Kleiss R, de~Causmaecker P, Gastmans R, Troost W and Wu T~T
  (CALKUL) 1984 {\em Nucl. Phys. B\/} {\bf 239} 382--394

\bibitem{Berends:1983ey}
Berends F~A, Kleiss R, de~Causmaecker P, Gastmans R, Troost W and Wu T~T
  (CALKUL) 1984 {\em Nucl. Phys. B\/} {\bf 239} 395--409

\bibitem{Berends:1984qe}
Berends F~A, De~Causmaecker P, Gastmans R, Kleiss R, Troost W and Wu T~T
  (CALKUL) 1986 {\em Nucl. Phys. B\/} {\bf 264} 243

\bibitem{Berends:1984qf}
Berends F~A, De~Causmaecker P, Gastmans R, Kleiss R, Troost W and Wu T~T
  (CALKUL) 1986 {\em Nucl. Phys. B\/} {\bf 264} 265--276

\bibitem{wigner1}
Wigner E~P 1939 {\em Annals Math.\/} {\bf 40} 149--204

\bibitem{wigner2}
Bargmann V and Wigner E~P 1948 {\em Proc. Nat. Acad. Sci.\/} {\bf 34} 211

\bibitem{Arkani-Hamed:2017jhn}
Arkani-Hamed N, Huang T~C and Huang Y~t 2021 {\em JHEP\/} {\bf 11} 070
  (\textit{Preprint} \eprint{1709.04891})

\bibitem{Ita:2011ar}
Ita H and Ozeren K 2012 {\em JHEP\/} {\bf 02} 118 (\textit{Preprint}
  \eprint{1111.4193})

\bibitem{Reuschle:2013qna}
Reuschle C and Weinzierl S 2013 {\em Phys. Rev. D\/} {\bf 88} 105020
  (\textit{Preprint} \eprint{1310.0413})

\bibitem{Bern:1990ux}
Bern Z and Kosower D~A 1991 {\em Nucl. Phys. B\/} {\bf 362} 389--448

\bibitem{DelDuca:1999rs}
Del~Duca V, Dixon L~J and Maltoni F 2000 {\em Nucl. Phys. B\/} {\bf 571} 51--70
  (\textit{Preprint} \eprint{hep-ph/9910563})

\bibitem{Kleiss:1988ne}
Kleiss R and Kuijf H 1989 {\em Nucl. Phys. B\/} {\bf 312} 616--644

\bibitem{Bern:2008qj}
Bern Z, Carrasco J~J~M and Johansson H 2008 {\em Phys. Rev. D\/} {\bf 78}
  085011 (\textit{Preprint} \eprint{0805.3993})

\bibitem{Bern:2010ue}
Bern Z, Carrasco J~J~M and Johansson H 2010 {\em Phys. Rev. Lett.\/} {\bf 105}
  061602 (\textit{Preprint} \eprint{1004.0476})

\bibitem{Bern:2022wqg}
Bern Z, Carrasco J~J, Chiodaroli M, Johansson H and Roiban R 2022 {\em J. Phys.
  A\/} {\bf 55} 443003 (\textit{Preprint} \eprint{2203.13013})

\bibitem{Johansson:2015oia}
Johansson H and Ochirov A 2016 {\em JHEP\/} {\bf 01} 170 (\textit{Preprint}
  \eprint{1507.00332})

\bibitem{Melia:2015ika}
Melia T 2015 {\em JHEP\/} {\bf 12} 107 (\textit{Preprint} \eprint{1509.03297})

\bibitem{Dixon:1993xd}
Dixon L~J and Shadmi Y 1994 {\em Nucl. Phys. B\/} {\bf 423} 3--32 [Erratum:
  Nucl.Phys.B 452, 724--724 (1995)] (\textit{Preprint} \eprint{hep-ph/9312363})

\bibitem{Dixon:2004za}
Dixon L~J, Glover E~W~N and Khoze V~V 2004 {\em JHEP\/} {\bf 12} 015
  (\textit{Preprint} \eprint{hep-th/0411092})

\bibitem{Broedel:2012rc}
Broedel J and Dixon L~J 2012 {\em JHEP\/} {\bf 10} 091 (\textit{Preprint}
  \eprint{1208.0876})

\bibitem{DeWitt:1967uc}
DeWitt B~S 1967 {\em Phys. Rev.\/} {\bf 162} 1239--1256

\bibitem{Eden:1966dnq}
Eden R~J, Landshoff P~V, Olive D~I and Polkinghorne J~C 1966 {\em {The analytic
  S-matrix}\/} (Cambridge: Cambridge Univ. Press)

\bibitem{Britto:2004ap}
Britto R, Cachazo F and Feng B 2005 {\em Nucl. Phys. B\/} {\bf 715} 499--522
  (\textit{Preprint} \eprint{hep-th/0412308})

\bibitem{Britto:2005fq}
Britto R, Cachazo F, Feng B and Witten E 2005 {\em Phys. Rev. Lett.\/} {\bf 94}
  181602 (\textit{Preprint} \eprint{hep-th/0501052})

\bibitem{Risager:2005vk}
Risager K 2005 {\em JHEP\/} {\bf 12} 003 (\textit{Preprint}
  \eprint{hep-th/0508206})

\bibitem{Elvang:2008vz}
Elvang H, Freedman D~Z and Kiermaier M 2009 {\em JHEP\/} {\bf 06} 068
  (\textit{Preprint} \eprint{0811.3624})

\bibitem{Arkani-Hamed:2008bsc}
Arkani-Hamed N and Kaplan J 2008 {\em JHEP\/} {\bf 04} 076 (\textit{Preprint}
  \eprint{0801.2385})

\bibitem{Henn:2014yza}
Henn J~M and Plefka J~C 2014 {\em {Scattering Amplitudes in Gauge Theories}\/}
  vol 883 (Berlin: Springer) ISBN 978-3-642-54021-9

\bibitem{Bedford:2005yy}
Bedford J, Brandhuber A, Spence B~J and Travaglini G 2005 {\em Nucl. Phys. B\/}
  {\bf 721} 98--110 (\textit{Preprint} \eprint{hep-th/0502146})

\bibitem{Cachazo:2005ca}
Cachazo F and Svrcek P 2005  (\textit{Preprint} \eprint{hep-th/0502160})

\bibitem{Badger:2005zh}
Badger S~D, Glover E~W~N, Khoze V~V and Svrcek P 2005 {\em JHEP\/} {\bf 07} 025
  (\textit{Preprint} \eprint{hep-th/0504159})

\bibitem{Badger:2005jv}
Badger S~D, Glover E~W~N and Khoze V~V 2006 {\em JHEP\/} {\bf 01} 066
  (\textit{Preprint} \eprint{hep-th/0507161})

\bibitem{Bern:2005hs}
Bern Z, Dixon L~J and Kosower D~A 2005 {\em Phys. Rev. D\/} {\bf 71} 105013
  (\textit{Preprint} \eprint{hep-th/0501240})

\bibitem{Bern:2005ji}
Bern Z, Dixon L~J and Kosower D~A 2005 {\em Phys. Rev. D\/} {\bf 72} 125003
  (\textit{Preprint} \eprint{hep-ph/0505055})

\bibitem{Brandhuber:2007up}
Brandhuber A, McNamara S, Spence B and Travaglini G 2007 {\em JHEP\/} {\bf 03}
  029 (\textit{Preprint} \eprint{hep-th/0701187})

\bibitem{Dunbar:2010xk}
Dunbar D~C, Ettle J~H and Perkins W~B 2010 {\em JHEP\/} {\bf 06} 027
  (\textit{Preprint} \eprint{1003.3398})

\bibitem{Alston:2015gea}
Alston S~D, Dunbar D~C and Perkins W~B 2015 {\em Phys. Rev. D\/} {\bf 92}
  065024 (\textit{Preprint} \eprint{1507.08882})

\bibitem{Brandhuber:2010ad}
Brandhuber A, Spence B, Travaglini G and Yang G 2011 {\em JHEP\/} {\bf 01} 134
  (\textit{Preprint} \eprint{1011.1899})

\bibitem{Brandhuber:2011tv}
Brandhuber A, Gurdogan O, Mooney R, Travaglini G and Yang G 2011 {\em JHEP\/}
  {\bf 10} 046 (\textit{Preprint} \eprint{1107.5067})

\bibitem{Kampf:2012fn}
Kampf K, Novotny J and Trnka J 2013 {\em Phys. Rev. D\/} {\bf 87} 081701
  (\textit{Preprint} \eprint{1212.5224})

\bibitem{Cheung:2014dqa}
Cheung C, Kampf K, Novotny J and Trnka J 2015 {\em Phys. Rev. Lett.\/} {\bf
  114} 221602 (\textit{Preprint} \eprint{1412.4095})

\bibitem{Cheung:2015ota}
Cheung C, Kampf K, Novotny J, Shen C~H and Trnka J 2016 {\em Phys. Rev.
  Lett.\/} {\bf 116} 041601 (\textit{Preprint} \eprint{1509.03309})

\bibitem{Mojahed:2021sxy}
Mojahed M~A and Brauner T 2021 {\em Phys. Lett. B\/} {\bf 822} 136705
  (\textit{Preprint} \eprint{2108.03189})

\bibitem{Brandhuber:2008pf}
Brandhuber A, Heslop P and Travaglini G 2008 {\em Phys. Rev. D\/} {\bf 78}
  125005 (\textit{Preprint} \eprint{0807.4097})

\bibitem{Arkani-Hamed:2008owk}
Arkani-Hamed N, Cachazo F and Kaplan J 2010 {\em JHEP\/} {\bf 09} 016
  (\textit{Preprint} \eprint{0808.1446})

\bibitem{Arkani-Hamed:2012zlh}
Arkani-Hamed N, Bourjaily J~L, Cachazo F, Goncharov A~B, Postnikov A and Trnka
  J 2016 {\em {Grassmannian Geometry of Scattering Amplitudes}\/} (Cambridge
  University Press) ISBN 978-1-107-08658-6, 978-1-316-57296-2
  (\textit{Preprint} \eprint{1212.5605})

\bibitem{Herrmann:2022nkh}
Herrmann E and Trnka J 2022 {\em J. Phys. A\/} {\bf 55} 443008
  (\textit{Preprint} \eprint{2203.13018})

\bibitem{Cheung:2015cba}
Cheung C, Shen C~H and Trnka J 2015 {\em JHEP\/} {\bf 06} 118
  (\textit{Preprint} \eprint{1502.05057})

\bibitem{Cohen:2010mi}
Cohen T, Elvang H and Kiermaier M 2011 {\em JHEP\/} {\bf 04} 053
  (\textit{Preprint} \eprint{1010.0257})

\bibitem{Brandhuber:2004yw}
Brandhuber A, Spence B~J and Travaglini G 2005 {\em Nucl. Phys. B\/} {\bf 706}
  150--180 (\textit{Preprint} \eprint{hep-th/0407214})

\bibitem{Brandhuber:2005kd}
Brandhuber A, Spence B and Travaglini G 2006 {\em JHEP\/} {\bf 01} 142
  (\textit{Preprint} \eprint{hep-th/0510253})

\bibitem{Brandhuber:2011ke}
Brandhuber A, Spence B and Travaglini G 2011 {\em J. Phys. A\/} {\bf 44} 454002
  (\textit{Preprint} \eprint{1103.3477})

\bibitem{Kosower:1999rx}
Kosower D~A and Uwer P 1999 {\em Nucl. Phys. B\/} {\bf 563} 477--505
  (\textit{Preprint} \eprint{hep-ph/9903515})

\bibitem{Bern:1999ry}
Bern Z, Del~Duca V, Kilgore W~B and Schmidt C~R 1999 {\em Phys. Rev. D\/} {\bf
  60} 116001 (\textit{Preprint} \eprint{hep-ph/9903516})

\bibitem{Bedford:2004py}
Bedford J, Brandhuber A, Spence B~J and Travaglini G 2005 {\em Nucl. Phys. B\/}
  {\bf 706} 100--126 (\textit{Preprint} \eprint{hep-th/0410280})

\bibitem{Bedford:2004nh}
Bedford J, Brandhuber A, Spence B~J and Travaglini G 2005 {\em Nucl. Phys. B\/}
  {\bf 712} 59--85 (\textit{Preprint} \eprint{hep-th/0412108})

\bibitem{Quigley:2004pw}
Quigley C and Rozali M 2005 {\em JHEP\/} {\bf 01} 053 (\textit{Preprint}
  \eprint{hep-th/0410278})

\bibitem{McLoughlin:2022ljp}
McLoughlin T, Puhm A and Raclariu A~M 2022 {\em J. Phys. A\/} {\bf 55} 443012
  (\textit{Preprint} \eprint{2203.13022})

\bibitem{Cachazo:2014fwa}
Cachazo F and Strominger A 2014  (\textit{Preprint} \eprint{1404.4091})

\bibitem{Casali:2014xpa}
Casali E 2014 {\em JHEP\/} {\bf 08} 077 (\textit{Preprint} \eprint{1404.5551})

\bibitem{Bern:1998sv}
Bern Z, Dixon L~J, Perelstein M and Rozowsky J~S 1999 {\em Nucl. Phys. B\/}
  {\bf 546} 423--479 (\textit{Preprint} \eprint{hep-th/9811140})

\bibitem{Weinberg:1965nx}
Weinberg S 1965 {\em Phys. Rev.\/} {\bf 140} B516--B524

\bibitem{Laddha:2017ygw}
Laddha A and Sen A 2017 {\em JHEP\/} {\bf 10} 065 (\textit{Preprint}
  \eprint{1706.00759})

\bibitem{Broedel:2014fsa}
Broedel J, de~Leeuw M, Plefka J and Rosso M 2014 {\em Phys. Rev. D\/} {\bf 90}
  065024 (\textit{Preprint} \eprint{1406.6574})

\bibitem{Bern:2014vva}
Bern Z, Davies S, Di~Vecchia P and Nohle J 2014 {\em Phys. Rev. D\/} {\bf 90}
  084035 (\textit{Preprint} \eprint{1406.6987})

\bibitem{Klose:2015xoa}
Klose T, McLoughlin T, Nandan D, Plefka J and Travaglini G 2015 {\em JHEP\/}
  {\bf 07} 135 (\textit{Preprint} \eprint{1504.05558})

\bibitem{Volovich:2015yoa}
Volovich A, Wen C and Zlotnikov M 2015 {\em JHEP\/} {\bf 07} 095
  (\textit{Preprint} \eprint{1504.05559})

\bibitem{DHoker:2002nbb}
D'Hoker E and Freedman D~Z 2002 {Supersymmetric gauge theories and the AdS /
  CFT correspondence} {\em {Theoretical Advanced Study Institute in Elementary
  Particle Physics (TASI 2001): Strings, Branes and Extra Dimensions}\/} pp
  3--158 (\textit{Preprint} \eprint{hep-th/0201253})

\bibitem{Nair:1988bq}
Nair V~P 1988 {\em Phys. Lett. B\/} {\bf 214} 215--218

\bibitem{Ferber:1977qx}
Ferber A 1978 {\em Nucl. Phys. B\/} {\bf 132} 55--64

\bibitem{Drummond:2008vq}
Drummond J~M, Henn J, Korchemsky G~P and Sokatchev E 2010 {\em Nucl. Phys. B\/}
  {\bf 828} 317--374 (\textit{Preprint} \eprint{0807.1095})

\bibitem{Drummond:2008bq}
Drummond J~M, Henn J, Korchemsky G~P and Sokatchev E 2013 {\em Nucl. Phys. B\/}
  {\bf 869} 452--492 (\textit{Preprint} \eprint{0808.0491})

\bibitem{Bern:2004bt}
Bern Z, Dixon L~J and Kosower D~A 2005 {\em Phys. Rev. D\/} {\bf 72} 045014
  (\textit{Preprint} \eprint{hep-th/0412210})

\bibitem{Drummond:2008cr}
Drummond J~M and Henn J~M 2009 {\em JHEP\/} {\bf 04} 018 (\textit{Preprint}
  \eprint{0808.2475})

\bibitem{Dixon:1996wi}
Dixon L~J 1996 {Calculating scattering amplitudes efficiently} {\em
  {Theoretical Advanced Study Institute in Elementary Particle Physics (TASI
  95): QCD and Beyond}\/} pp 539--584 (\textit{Preprint}
  \eprint{hep-ph/9601359})

\bibitem{Mandelstam:1982cb}
Mandelstam S 1983 {\em Nucl. Phys. B\/} {\bf 213} 149--168

\bibitem{Drummond:2006rz}
Drummond J~M, Henn J, Smirnov V~A and Sokatchev E 2007 {\em JHEP\/} {\bf 01}
  064 (\textit{Preprint} \eprint{hep-th/0607160})

\bibitem{Bern:1994zx}
Bern Z, Dixon L~J, Dunbar D~C and Kosower D~A 1994 {\em Nucl. Phys. B\/} {\bf
  425} 217--260 (\textit{Preprint} \eprint{hep-ph/9403226})

\bibitem{tHooft:1978jhc}
't~Hooft G and Veltman M~J~G 1979 {\em Nucl. Phys. B\/} {\bf 153} 365--401

\bibitem{Denner:1991qq}
Denner A, Nierste U and Scharf R 1991 {\em Nucl. Phys. B\/} {\bf 367} 637--656

\bibitem{Bourjaily:2019jrk}
Bourjaily J~L, Dulat F and Panzer E 2019 {\em Nucl. Phys. B\/} {\bf 942}
  251--302 (\textit{Preprint} \eprint{1901.02887})

\bibitem{Duhr:2019tlz}
Duhr C and Dulat F 2019 {\em JHEP\/} {\bf 08} 135 (\textit{Preprint}
  \eprint{1904.07279})

\bibitem{Brandhuber:2009xz}
Brandhuber A, Heslop P and Travaglini G 2009 {\em JHEP\/} {\bf 08} 095
  (\textit{Preprint} \eprint{0905.4377})

\bibitem{Brandhuber:2009kh}
Brandhuber A, Heslop P and Travaglini G 2009 {\em JHEP\/} {\bf 10} 063
  (\textit{Preprint} \eprint{0906.3552})

\bibitem{Drummond:2007cf}
Drummond J~M, Henn J, Korchemsky G~P and Sokatchev E 2008 {\em Nucl. Phys. B\/}
  {\bf 795} 52--68 (\textit{Preprint} \eprint{0709.2368})

\bibitem{Drummond:2007au}
Drummond J~M, Henn J, Korchemsky G~P and Sokatchev E 2010 {\em Nucl. Phys. B\/}
  {\bf 826} 337--364 (\textit{Preprint} \eprint{0712.1223})

\bibitem{Alday:2007hr}
Alday L~F and Maldacena J~M 2007 {\em JHEP\/} {\bf 06} 064 (\textit{Preprint}
  \eprint{0705.0303})

\bibitem{Drummond:2007aua}
Drummond J~M, Korchemsky G~P and Sokatchev E 2008 {\em Nucl. Phys. B\/} {\bf
  795} 385--408 (\textit{Preprint} \eprint{0707.0243})

\bibitem{Brandhuber:2007yx}
Brandhuber A, Heslop P and Travaglini G 2008 {\em Nucl. Phys. B\/} {\bf 794}
  231--243 (\textit{Preprint} \eprint{0707.1153})

\bibitem{Elvang:2009ya}
Elvang H, Freedman D~Z and Kiermaier M 2010 {\em JHEP\/} {\bf 03} 075
  (\textit{Preprint} \eprint{0905.4379})

\bibitem{Bern:2007ct}
Bern Z, Carrasco J~J~M, Johansson H and Kosower D~A 2007 {\em Phys. Rev. D\/}
  {\bf 76} 125020 (\textit{Preprint} \eprint{0705.1864})

\bibitem{Bourjaily:2011hi}
Bourjaily J~L, DiRe A, Shaikh A, Spradlin M and Volovich A 2012 {\em JHEP\/}
  {\bf 03} 032 (\textit{Preprint} \eprint{1112.6432})

\bibitem{Bern:2005iz}
Bern Z, Dixon L~J and Smirnov V~A 2005 {\em Phys. Rev.\/} {\bf D72} 085001
  (\textit{Preprint} \eprint{hep-th/0505205})

\bibitem{Bern:2006ew}
Bern Z, Czakon M, Dixon L~J, Kosower D~A and Smirnov V~A 2007 {\em Phys. Rev.
  D\/} {\bf 75} 085010 (\textit{Preprint} \eprint{hep-th/0610248})

\bibitem{Cachazo:2006az}
Cachazo F, Spradlin M and Volovich A 2007 {\em Phys. Rev. D\/} {\bf 75} 105011
  (\textit{Preprint} \eprint{hep-th/0612309})

\bibitem{Hodges:2009hk}
Hodges A 2013 {\em JHEP\/} {\bf 05} 135 (\textit{Preprint} \eprint{0905.1473})

\bibitem{Mason:2009qx}
Mason L~J and Skinner D 2009 {\em JHEP\/} {\bf 11} 045 (\textit{Preprint}
  \eprint{0909.0250})

\bibitem{Drummond:2009fd}
Drummond J~M, Henn J~M and Plefka J 2009 {\em JHEP\/} {\bf 05} 046
  (\textit{Preprint} \eprint{0902.2987})

\bibitem{Drinfeld:1985rx}
Drinfeld V~G 1985 {\em Sov. Math. Dokl.\/} {\bf 32} 254--258

\bibitem{Drinfeld:1986in}
Drinfeld V~G 1986 {\em Zap. Nauchn. Semin.\/} {\bf 155} 18--49

\bibitem{Bargheer:2009qu}
Bargheer T, Beisert N, Galleas W, Loebbert F and McLoughlin T 2009 {\em JHEP\/}
  {\bf 11} 056 (\textit{Preprint} \eprint{0905.3738})

\bibitem{Beisert:2010gn}
Beisert N, Henn J, McLoughlin T and Plefka J 2010 {\em JHEP\/} {\bf 04} 085
  (\textit{Preprint} \eprint{1002.1733})

\bibitem{Caron-Huot:2011dec}
Caron-Huot S and He S 2012 {\em JHEP\/} {\bf 07} 174 (\textit{Preprint}
  \eprint{1112.1060})

\bibitem{Bullimore:2011kg}
Bullimore M and Skinner D 2011  (\textit{Preprint} \eprint{1112.1056})

\bibitem{Beisert:2010jr}
Beisert N {\em et~al.\/} 2012 {\em Lett. Math. Phys.\/} {\bf 99} 3--32
  (\textit{Preprint} \eprint{1012.3982})

\bibitem{Bern:1994cg}
Bern Z, Dixon L~J, Dunbar D~C and Kosower D~A 1995 {\em Nucl. Phys. B\/} {\bf
  435} 59--101 (\textit{Preprint} \eprint{hep-ph/9409265})

\bibitem{Bern:2004cz}
Bern Z, Dixon L~J and Kosower D~A 2004 {\em JHEP\/} {\bf 08} 012
  (\textit{Preprint} \eprint{hep-ph/0404293})

\bibitem{Britto:2004nc}
Britto R, Cachazo F and Feng B 2005 {\em Nucl. Phys. B\/} {\bf 725} 275--305
  (\textit{Preprint} \eprint{hep-th/0412103})

\bibitem{Feynman:1963ax}
Feynman R~P 1963 {\em Acta Phys. Polon.\/} {\bf 24} 697--722

\bibitem{Feynman-magic}
Feynman R~P 1972 {\em {Closed loop and tree diagrams, in J.~R.~Klauder, Magic
  without Magic}\/} (W.~H.~Freeman and Company, San Francisco) ISBN
  978-0716703372

\bibitem{Neill:2013wsa}
Neill D and Rothstein I~Z 2013 {\em Nucl. Phys.\/} {\bf B877} 177--189
  (\textit{Preprint} \eprint{1304.7263})

\bibitem{Bjerrum-Bohr:2013bxa}
Bjerrum-Bohr N~E~J, Donoghue J~F and Vanhove P 2014 {\em JHEP\/} {\bf 02} 111
  (\textit{Preprint} \eprint{1309.0804})

\bibitem{Bjerrum-Bohr:2014zsa}
Bjerrum-Bohr N~E~J, Donoghue J~F, Holstein B~R, Plante L and Vanhove P 2015
  {\em Phys. Rev. Lett.\/} {\bf 114} 061301 (\textit{Preprint}
  \eprint{1410.7590})

\bibitem{Bjerrum-Bohr:2016hpa}
Bjerrum-Bohr N~E~J, Donoghue J~F, Holstein B~R, Plante L and Vanhove P 2016
  {\em JHEP\/} {\bf 11} 117 (\textit{Preprint} \eprint{1609.07477})

\bibitem{Bai:2016ivl}
Bai D and Huang Y 2017 {\em Phys. Rev.\/} {\bf D95} 064045 (\textit{Preprint}
  \eprint{1612.07629})

\bibitem{Chi:2019owc}
Chi H~H 2019 {\em Phys. Rev.\/} {\bf D99} 126008 (\textit{Preprint}
  \eprint{1903.07944})

\bibitem{Bern:2019nnu}
Bern Z, Cheung C, Roiban R, Shen C~H, Solon M~P and Zeng M 2019 {\em Phys. Rev.
  Lett.\/} {\bf 122} 201603 (\textit{Preprint} \eprint{1901.04424})

\bibitem{Bern:2019crd}
Bern Z, Cheung C, Roiban R, Shen C~H, Solon M~P and Zeng M 2019 {\em JHEP\/}
  {\bf 10} 206 (\textit{Preprint} \eprint{1908.01493})

\bibitem{Parra-Martinez:2020dzs}
Parra-Martinez J, Ruf M~S and Zeng M 2020 {\em JHEP\/} {\bf 11} 023
  (\textit{Preprint} \eprint{2005.04236})

\bibitem{DiVecchia:2021bdo}
Di~Vecchia P, Heissenberg C, Russo R and Veneziano G 2021 {\em JHEP\/} {\bf 07}
  169 (\textit{Preprint} \eprint{2104.03256})

\bibitem{Bjerrum-Bohr:2021din}
Bjerrum-Bohr N~E~J, Damgaard P~H, Plant\'e L and Vanhove P 2021 {\em JHEP\/}
  {\bf 08} 172 (\textit{Preprint} \eprint{2105.05218})

\bibitem{Brandhuber:2021eyq}
Brandhuber A, Chen G, Travaglini G and Wen C 2021 {\em JHEP\/} {\bf 10} 118
  (\textit{Preprint} \eprint{2108.04216})

\bibitem{Herrmann:2021lqe}
Herrmann E, Parra-Martinez J, Ruf M~S and Zeng M 2021 {\em Phys. Rev. Lett.\/}
  {\bf 126} 201602 (\textit{Preprint} \eprint{2101.07255})

\bibitem{Herrmann:2021tct}
Herrmann E, Parra-Martinez J, Ruf M~S and Zeng M 2021 {\em JHEP\/} {\bf 10} 148
  (\textit{Preprint} \eprint{2104.03957})

\bibitem{Bern:2021dqo}
Bern Z, Parra-Martinez J, Roiban R, Ruf M~S, Shen C~H, Solon M~P and Zeng M
  2021 {\em Phys. Rev. Lett.\/} {\bf 126} 171601 (\textit{Preprint}
  \eprint{2101.07254})

\bibitem{Brandhuber:2019qpg}
Brandhuber A and Travaglini G 2020 {\em JHEP\/} {\bf 01} 010 (\textit{Preprint}
  \eprint{1905.05657})

\bibitem{Emond:2019crr}
Emond W~T and Moynihan N 2019 {\em JHEP\/} {\bf 12} 019 (\textit{Preprint}
  \eprint{1905.08213})

\bibitem{AccettulliHuber:2020oou}
Accettulli~Huber M, Brandhuber A, De~Angelis S and Travaglini G 2020 {\em Phys.
  Rev. D\/} {\bf 102} 046014 (\textit{Preprint} \eprint{2006.02375})

\bibitem{AccettulliHuber:2020dal}
Accettulli~Huber M, Brandhuber A, De~Angelis S and Travaglini G 2021 {\em Phys.
  Rev. D\/} {\bf 103} 045015 (\textit{Preprint} \eprint{2012.06548})

\bibitem{Carrillo-Gonzalez:2021mqj}
Carrillo-Gonz\'alez M, de~Rham C and Tolley A~J 2021 {\em JHEP\/} {\bf 11} 087
  (\textit{Preprint} \eprint{2107.11384})

\bibitem{Bjerrum-Bohr:2022blt}
Bjerrum-Bohr N~E~J, Damgaard P~H, Plante L and Vanhove P 2022 {\em J. Phys.
  A\/} {\bf 55} 443014 (\textit{Preprint} \eprint{2203.13024})

\bibitem{Kosower:2022yvp}
Kosower D~A, Monteiro R and O'Connell D 2022 {\em J. Phys. A\/} {\bf 55} 443015
  (\textit{Preprint} \eprint{2203.13025})

\bibitem{Passarino:1978jh}
Passarino G and Veltman M~J~G 1979 {\em Nucl. Phys. B\/} {\bf 160} 151--207

\bibitem{vanNeerven:1985xr}
van Neerven W~L 1986 {\em Nucl. Phys. B\/} {\bf 268} 453--488

\bibitem{Bern:1995db}
Bern Z and Morgan A~G 1996 {\em Nucl. Phys. B\/} {\bf 467} 479--509
  (\textit{Preprint} \eprint{hep-ph/9511336})

\bibitem{Bern:1991aq}
Bern Z and Kosower D~A 1992 {\em Nucl. Phys. B\/} {\bf 379} 451--561

\bibitem{Bern:2002tk}
Bern Z, De~Freitas A and Dixon L~J 2002 {\em JHEP\/} {\bf 03} 018
  (\textit{Preprint} \eprint{hep-ph/0201161})

\bibitem{Bern:1993kr}
Bern Z, Dixon L~J and Kosower D~A 1994 {\em Nucl. Phys. B\/} {\bf 412} 751--816
  (\textit{Preprint} \eprint{hep-ph/9306240})

\bibitem{Giele:1991vf}
Giele W~T and Glover E~W~N 1992 {\em Phys. Rev. D\/} {\bf 46} 1980--2010

\bibitem{Kunszt:1994np}
Kunszt Z, Signer A and Trocsanyi Z 1994 {\em Nucl. Phys. B\/} {\bf 420}
  550--564 (\textit{Preprint} \eprint{hep-ph/9401294})

\bibitem{Mueller:1979ih}
Mueller A~H 1979 {\em Phys. Rev. D\/} {\bf 20} 2037

\bibitem{Magnea:1990zb}
Magnea L and Sterman G~F 1990 {\em Phys. Rev. D\/} {\bf 42} 4222--4227

\bibitem{Sterman:2002qn}
Sterman G~F and Tejeda-Yeomans M~E 2003 {\em Phys. Lett. B\/} {\bf 552} 48--56
  (\textit{Preprint} \eprint{hep-ph/0210130})

\bibitem{RisagerLarsen:2007wtf}
Risager~Larsen K 2007 {\em {Unitarity and On-Shell Recursion Methods for
  Scattering Amplitudes}\/} Ph.D. thesis Copenhagen U. (\textit{Preprint}
  \eprint{0804.3310})

\bibitem{Mantel}
Mantel W 1898 {\em Nieuw Archief\/} {\bf (2)3} 292

\bibitem{Brandhuber:2005jw}
Brandhuber A, McNamara S, Spence B~J and Travaglini G 2005 {\em JHEP\/} {\bf
  10} 011 (\textit{Preprint} \eprint{hep-th/0506068})

\bibitem{Abreu:2022mfk}
Abreu S, Britto R and Duhr C 2022 {\em J. Phys. A\/} {\bf 55} 443004
  (\textit{Preprint} \eprint{2203.13014})

\bibitem{Blumlein:2022zkr}
Bl\"umlein J and Schneider C 2022 {\em J. Phys. A\/} {\bf 55} 443005
  (\textit{Preprint} \eprint{2203.13015})

\bibitem{Papathanasiou:2022lan}
Papathanasiou G 2022 {\em J. Phys. A\/} {\bf 55} 443006 (\textit{Preprint}
  \eprint{2203.13016})

\bibitem{Dixon:2022rse}
Dixon L~J, Gurdogan O, McLeod A~J and Wilhelm M 2022  (\textit{Preprint}
  \eprint{2204.11901})

\bibitem{Wilczek:1977zn}
Wilczek F 1977 {\em Phys. Rev. Lett.\/} {\bf 39} 1304

\bibitem{Shifman:1979eb}
Shifman M~A, Vainshtein A~I, Voloshin M~B and Zakharov V~I 1979 {\em Sov. J.
  Nucl. Phys.\/} {\bf 30} 711--716

\bibitem{Dawson:1990zj}
Dawson S 1991 {\em Nucl. Phys. B\/} {\bf 359} 283--300

\bibitem{Bianchi:2018peu}
Bianchi L, Brandhuber A, Panerai R and Travaglini G 2019 {\em JHEP\/} {\bf 02}
  182 (\textit{Preprint} \eprint{1812.09001})

\bibitem{Bianchi:2018rrj}
Bianchi L, Brandhuber A, Panerai R and Travaglini G 2019 {\em JHEP\/} {\bf 02}
  134 (\textit{Preprint} \eprint{1812.10468})

\bibitem{AccettulliHuber:2019abj}
Accettulli~Huber M, Brandhuber A, De~Angelis S and Travaglini G 2020 {\em Phys.
  Rev. D\/} {\bf 101} 026004 (\textit{Preprint} \eprint{1910.04772})

\bibitem{Giele:2008ve}
Giele W~T, Kunszt Z and Melnikov K 2008 {\em JHEP\/} {\bf 04} 049
  (\textit{Preprint} \eprint{0801.2237})

\bibitem{Ellis:2008ir}
Ellis R~K, Giele W~T, Kunszt Z and Melnikov K 2009 {\em Nucl. Phys. B\/} {\bf
  822} 270--282 (\textit{Preprint} \eprint{0806.3467})

\bibitem{Bern:2010qa}
Bern Z, Carrasco J~J, Dennen T, Huang Y~t and Ita H 2011 {\em Phys. Rev. D\/}
  {\bf 83} 085022 (\textit{Preprint} \eprint{1010.0494})

\bibitem{Davies:2011vt}
Davies S 2011 {\em Phys. Rev. D\/} {\bf 84} 094016 (\textit{Preprint}
  \eprint{1108.0398})

\bibitem{Eden:2011yp}
Eden B, Heslop P, Korchemsky G~P and Sokatchev E 2013 {\em Nucl. Phys. B\/}
  {\bf 869} 329--377 (\textit{Preprint} \eprint{1103.3714})

\bibitem{Brandhuber:2012vm}
Brandhuber A, Travaglini G and Yang G 2012 {\em JHEP\/} {\bf 05} 082
  (\textit{Preprint} \eprint{1201.4170})

\bibitem{Penante:2014sza}
Penante B, Spence B, Travaglini G and Wen C 2014 {\em JHEP\/} {\bf 04} 083
  (\textit{Preprint} \eprint{1402.1300})

\bibitem{Brandhuber:2014ica}
Brandhuber A, Penante B, Travaglini G and Wen C 2014 {\em JHEP\/} {\bf 08} 100
  (\textit{Preprint} \eprint{1406.1443})

\bibitem{Dixon:2020bbt}
Dixon L~J, McLeod A~J and Wilhelm M 2021 {\em JHEP\/} {\bf 04} 147
  (\textit{Preprint} \eprint{2012.12286})

\bibitem{Dixon:2021tdw}
Dixon L~J, Gurdogan O, McLeod A~J and Wilhelm M 2022 {\em Phys. Rev. Lett.\/}
  {\bf 128} 111602 (\textit{Preprint} \eprint{2112.06243})

\bibitem{Brandhuber:2016fni}
Brandhuber A, Kostacinska M, Penante B, Travaglini G and Young D 2016 {\em
  JHEP\/} {\bf 08} 134 (\textit{Preprint} \eprint{1606.08682})

\bibitem{Brandhuber:2017bkg}
Brandhuber A, Kostacinska M, Penante B and Travaglini G 2017 {\em Phys. Rev.
  Lett.\/} {\bf 119} 161601 (\textit{Preprint} \eprint{1707.09897})

\bibitem{Brandhuber:2018xzk}
Brandhuber A, Kostacinska M, Penante B and Travaglini G 2018 {\em JHEP\/} {\bf
  12} 076 (\textit{Preprint} \eprint{1804.05703})

\bibitem{Brandhuber:2018kqb}
Brandhuber A, Kostacinska M, Penante B and Travaglini G 2018 {\em JHEP\/} {\bf
  12} 077 (\textit{Preprint} \eprint{1804.05828})

\bibitem{Gehrmann:2011aa}
Gehrmann T, Jaquier M, Glover E~W~N and Koukoutsakis A 2012 {\em JHEP\/} {\bf
  02} 056 (\textit{Preprint} \eprint{1112.3554})

\bibitem{Kotikov:2004er}
Kotikov A~V, Lipatov L~N, Onishchenko A~I and Velizhanin V~N 2004 {\em Phys.
  Lett. B\/} {\bf 595} 521--529 [Erratum: Phys.Lett.B 632, 754--756 (2006)]
  (\textit{Preprint} \eprint{hep-th/0404092})

\bibitem{Buchmuller:1985jz}
Buchmuller W and Wyler D 1986 {\em Nucl. Phys.\/} {\bf B268} 621--653

\bibitem{Neill:2009tn}
Neill D 2009  (\textit{Preprint} \eprint{0908.1573})

\bibitem{Neill:2009mz}
Neill D 2009  (\textit{Preprint} \eprint{0911.2707})

\bibitem{Harlander:2013oja}
Harlander R~V and Neumann T 2013 {\em Phys. Rev.\/} {\bf D88} 074015
  (\textit{Preprint} \eprint{1308.2225})

\bibitem{Dawson:2014ora}
Dawson S, Lewis I~M and Zeng M 2014 {\em Phys. Rev. D\/} {\bf 90} 093007
  (\textit{Preprint} \eprint{1409.6299})

\bibitem{Guo:2022pdw}
Guo Y, Jin Q, Wang L and Yang G 2022  (\textit{Preprint} \eprint{2205.12969})

\bibitem{Zwiebel:2011bx}
Zwiebel B~I 2012 {\em J. Phys. A\/} {\bf 45} 115401 (\textit{Preprint}
  \eprint{1111.0083})

\bibitem{Wilhelm:2014qua}
Wilhelm M 2015 {\em JHEP\/} {\bf 02} 149 (\textit{Preprint} \eprint{1410.6309})

\bibitem{Beisert:2003tq}
Beisert N, Kristjansen C and Staudacher M 2003 {\em Nucl. Phys. B\/} {\bf 664}
  131--184 (\textit{Preprint} \eprint{hep-th/0303060})

\bibitem{Beisert:2003jj}
Beisert N 2004 {\em Nucl. Phys. B\/} {\bf 676} 3--42 (\textit{Preprint}
  \eprint{hep-th/0307015})

\bibitem{Beisert:2003yb}
Beisert N and Staudacher M 2003 {\em Nucl. Phys. B\/} {\bf 670} 439--463
  (\textit{Preprint} \eprint{hep-th/0307042})

\bibitem{Dolan:2003uh}
Dolan L, Nappi C~R and Witten E 2003 {\em JHEP\/} {\bf 10} 017
  (\textit{Preprint} \eprint{hep-th/0308089})

\bibitem{Brandhuber:2015dta}
Brandhuber A, Heslop P, Travaglini G and Young D 2015 {\em Phys. Rev. Lett.\/}
  {\bf 115} 141602 (\textit{Preprint} \eprint{1507.01504})

\bibitem{Caron-Huot:2016cwu}
Caron-Huot S and Wilhelm M 2016 {\em JHEP\/} {\bf 12} 010 (\textit{Preprint}
  \eprint{1607.06448})

\bibitem{EliasMiro:2020tdv}
Elias~Mir\'o J, Ingoldby J and Riembau M 2020 {\em JHEP\/} {\bf 09} 163
  (\textit{Preprint} \eprint{2005.06983})

\bibitem{Baratella:2020lzz}
Baratella P, Fernandez C and Pomarol A 2020 {\em Nucl. Phys. B\/} {\bf 959}
  115155 (\textit{Preprint} \eprint{2005.07129})

\bibitem{Bern:2020ikv}
Bern Z, Parra-Martinez J and Sawyer E 2020 {\em JHEP\/} {\bf 10} 211
  (\textit{Preprint} \eprint{2005.12917})

\bibitem{AccettulliHuber:2021uoa}
Accettulli~Huber M and De~Angelis S 2021 {\em JHEP\/} {\bf 11} 221
  (\textit{Preprint} \eprint{2108.03669})

\bibitem{Bena:2004xu}
Bena I, Bern Z, Kosower D~A and Roiban R 2005 {\em Phys. Rev. D\/} {\bf 71}
  106010 (\textit{Preprint} \eprint{hep-th/0410054})

\end{thebibliography}

\end{document}